\documentclass[structabstract]{aa}  

\usepackage{graphicx}
\usepackage{txfonts}
\usepackage{natbib}
\usepackage{lscape}
\usepackage{longtable}
\usepackage{supertabular}
\usepackage{epsfig}
\usepackage{ifthen}
\usepackage[figuresright]{rotating}

\usepackage{color}
\newcommand{\micron}{$\mu$m}
\newcommand{\GB}{Galactic bulge}
\newcommand{\Myr}{M$_{\odot}$\,yr$^{-1}$}
\newcommand{\Msun}{M$_{\odot}$}
\newcommand{\Mdot}{$\dot{\rm{M}}$}
\newcommand{\Lsun}{L$_{\odot}$}
\newcommand{\Mms}{M$_{\rm{MS}}$}
\newcommand{\mic}{$\mu$m}
\newcommand{\arcdeg}{$^{\circ}$}
\newcommand{\kms}{km\,s$^{-1}$}

\begin{document}

   \title{Study of extremely reddened AGB stars in the Galactic bulge} 

   \titlerunning{Study of extremely reddened AGB stars in the GB}

   \author{F. M. Jim\'enez-Esteban\inst{1,2,3}
     \and
     D. Engels\inst{4}
}

   \institute{
     Centro de Astrobiolog\'{\i}a (INTA-CSIC), Departamento de
     Astrof\'{\i}sica, PO Box 78, E-28691, Villanueva de la Ca\~nada,
     Madrid, Spain\\
     \email{fran.jimenez-esteban@cab.inta-csic.es}
     \and
     Spanish Virtual Observatory, Spain
     \and
     Suffolk University, Madrid Campus, C/ Valle de la Viña 3, 28003, Madrid, Spain
     \and 
     Hamburger Sternwarte, Gojenbergsweg 112, D-21029 Hamburg, Germany
}

   \date{Received 15 July 2014 / Accepted 30 March 2015}

 
  \abstract
   {Extremely reddened asymptotic giant branch stars (AGB) lose mass at
    high rates of $>$\,10$^{-5}$ \Myr. This is the very last stage of
    AGB evolution, in which stars in the mass range $\sim$2.0\,--\,4.0
    \Msun\ (for solar metallicity) should have been converted to
    C stars already. The extremely reddened AGB stars in the Galactic
    bulge are however predominantly O-rich, implying that they 
    might be either low-mass stars or stars at the upper end of
    the AGB mass range.}
  {Our goal is to determine the mass range of the most reddened AGB
    stars in the Galactic bulge.}
   {Using Virtual Observatory tools, we constructed spectral energy
     distributions of a sample of 37 evolved stars in the \GB\ with
     extremely red IRAS colours. We fitted DUSTY models to the
     observational data to infer the bolometric fluxes. Applying
     individual corrections for interstellar extinction and adopting a
     common distance, we determined luminosities and mass-loss rates,
     and inferred the progenitor mass range from comparisons with AGB
     evolutionary models.}
   {The observed spectral energy distributions are consistent with a
     classification as reddened AGB stars, except for two stars, which
     are proto-planetary nebula candidates. For the AGB stars, we found
     luminosities in the range $\sim$\,3000\,--\,30,000 \Lsun\ and
     mass-loss rates $\sim$\,10$^{-5}$\,--\,$3\times10^{-4}$ \Myr. The
     corresponding mass range is $\sim$\,1.1\,--\,6.0 \Msun\ assuming
     solar metallicity.}
   {Contrary to the predictions of the evolutionary models, the
     luminosity distribution is continuous, with many O-rich AGB stars
     in the mass range in which they should have been converted into C
     stars already. We suspect that bulge AGB stars have higher than
     solar metallicity and therefore may avoid the conversion to
     C-rich. The presence of low-mass stars in the sample shows that
     their termination of the AGB evolution also occurs during a final
     phase of very high mass-loss rate, leading to optically thick
     circumstellar shells.}


   \maketitle

%

\section{Introduction}
\label{introduction}
Low- and intermediate-mass stars pass along the asymptotic giant
branch (AGB) towards the end of their evolution. Stellar evolution
models predict that stars approaching the tip of the AGB experience
thermal pulses. In this so-called `thermal pulses AGB' (TP-AGB) phase,
stars steadily increase in luminosities and mass-loss rates. This
increase is interrupted on timescales of several ten thousand years by
short-term variations due to the pulses. The heavy mass loss leads to
the formation of circumstellar envelopes (CSE) of gas and dust. If the
mass-loss rate surpasses \Mdot\,$\ge$\,10$^{-6}$\,\Myr, the dust shell
eventually becomes opaque to visible light \citep{Habing96}. In
addition, the stars may change their photospheric chemistry from O- to
C-rich as a consequence of the so-called third dredge-up following a
thermal pulse. The range of initial masses, in which the star will end
as C-rich, depends on the metallicity. This range, defined on the main
sequence, is 2\,$\la$\,\Mms\,$\la$\,4 \Msun\ for solar metallicities
\citep{Marigo07}. At lower metallicities, as for example in the
Magellanic Clouds, this mass range is extended to lower masses,
significant increasing the fraction of C stars in the AGB population
at the end of the TP-AGB phase \citep{Blum06}. The stars ending the
TP-AGB O-rich must have developed either from low-mass stars with
main-sequence masses \Mms\,$\la$\,2 M$_{\odot}$ \citep{Bertelli08},
for which the TP-AGB phase is terminated before the conversion is
completed, or from intermediate-mass stars with \Mms\,$\ga$\,4 \Msun,
in which hot bottom burning (HBB) is active at the base of the
convective envelope, preventing the formation of carbon.

Low-mass stars are expected to experience only moderate mass-loss
rates during most of their AGB lifetime, leading to the formation of
optically thin CSE. Low-mass AGB stars have been discovered
predominantly in the optical and near-infrared wavelength
range. Classical Mira variables belong to these kind of stars. AGB
stars descending from progenitors with larger main-sequence masses are
expected to be more luminous and to have higher mass-loss rates. They
probably spend a larger fraction of their TP-AGB life in an optically
obscured state \citep{Mouhcine02}, which constrains their detection to
the infrared or radio wavelength range. OH/IR stars, originally
discovered by their radio OH maser emission and later identified in
the infrared are an example of these kind of stars. OH maser emission
is however also widespread among Mira variables, so that the term
`OH/IR star' encompasses low- and intermediate-mass stars.

If low-mass stars develop high mass-loss rates at the end of their
evolution on the TP-AGB, they will form optically thick CSEs at lower
mass-loss rates than intermediate-mass stars because they have more
compact shells with smaller outflow velocities. Also, because of their
frequency they should outnumber the intermediate-mass stars even among
the obscured stars. However, in the Galactic disk, \cite{Chen01} and
\cite{Likkel89} found that the reddest (e.g. with highest mass-loss
rates) OH/IR stars are dominated by progenies of main-sequence
intermediate-mass stars. A dearth of low-mass stars among infrared
bright sources in clusters of the Magellanic Clouds was also noted by
\cite{vanLoon05}. Therefore, low-mass stars may never evolve into the
high mass-loss regime observed for the very red OH/IR stars, or it
might be a very brief phase because the envelope mass above the core
is rapidly depleted.

These stars should have relatively low luminosities, L\,$<$\,8000
\Lsun\ for \Mms\,$\la$\,2 M$_{\odot}$ \citep{Bertelli08}.
Because of the uncertain distances in the Galactic disk and the strong
variability, mean luminosities for AGB stars are difficult to
determine and objects with relatively low luminosities are difficult
to single out. A way to overcome this pitfall is to study AGB stars in
the \GB, where a common distance can be used \citep{vanderVeen90}. AGB
stars are luminous enough to be detectable there, especially in the
mid-infrared. During their heavily obscured stage, they emit most of
their radiation in the 10\,--\,30\,\micron\ range, which is less
affected by the significant interstellar extinction towards many parts
of the \GB.

The \GB\ is the central part of the Milky Way galaxy, consisting of a
dominant old population of stars (t\,$>$\,8~Gyr;
\citealt{Zoccali03,Vanhollebeke09}) and a smaller population of
younger stars with no well constrained ages of $\sim$\,200~Myr --
7~Gyr \citep{vanLoon03} or 1\,--\,3 Gyr \citep{Groenewegen05}, to
which the AGB stars in the \GB\ belong. In Baade's window two
metallicity populations have been found, of which the metal-poor
population is associated with the old stellar population and the
metal-rich population is consistent with a younger stellar population
supporting a bar \citep{Hill11}. A metallicity gradient was noted
along the minor axis of the bulge \citep{Zoccali08} with the
metallicity decreasing towards higher latitudes. No luminous C-rich
AGB stars are known in the \GB\ \citep{Blanco89}, suggesting that the
younger stellar population in the \GB\ may be restricted to low-mass
stars with AGB evolutionary timescales too short to experience enough
dredge-ups to convert them into C stars \citep{Buell13}. Then, the
\GB\ OH/IR stars should also exclusively be descendants from low-mass
stars, including the extremely red stars. This was implied by the
findings of \cite{vanderVeen90}, who derived low progenitor masses of
1.0\,--\,1.4 \Msun\ for a sample of these red sources from the IRAS
survey.

Although subsequent studies of AGB stars in the \GB\ also concluded
that their samples originate from low-mass stars, luminosity
distributions are not necessarily compatible with an exclusive origin
from low-mass stars. For example, \cite{Glass01} studied the
variability properties of long-period, large-amplitude variable stars
near the Galactic centre, and found a period-luminosity (P-L) relation
that is compatible with the existence of a young population with
initial masses in the range 2.5\,--\,3.0 \Msun. In a later study on
Mira variables in the optical gravitational lensing experiment (OGLE)
bulge fields, \cite{Groenewegen05} found a different P-L relation,
which they explained with a population with low initial masses
(1.5\,--\,2.0 \Msun). However, possibly more massive stars with
P\,$>$\,700 days, such as obscured OH/IR stars, might have been missed
in the I-band of the OGLE survey.  \cite{Blommaert06} studied a small
sample of Galactic bulge AGB stars with near-infrared spectroscopy,
and found a luminosity range from 1700 to 7700 \Lsun\ with a mean of
4100 \Lsun, as expected for a population of AGB stars with low initial
masses. Again, their study was restricted to objects with relatively
low mass-loss rates, and may therefore have excluded more massive
stars. This can be inferred from \cite{Ojha07}, who focussed their
study on high mass-loss AGB stars, and found a wider luminosity
distribution with a peak at 8000 \Lsun\ and an appreciable number of
sources with a luminosity exceeding 20,000\,\Lsun.

A number of studies have been specifically focussed on OH/IR stars.
\cite{Blommaert98} and \cite{Wood98b} found that the OH/IR stars of
the Galactic centre do not fit the classical Mira P-L relations
\citep{Feast89,Whitelock91}. They systematically have lower luminosity
and higher expansion velocity for a given period. OH/IR stars with
periods as long as 1000 days, expected to have higher initial masses,
were found to have low luminosities corresponding to Miras with masses
below 1.5 \Msun. These findings were explained by \cite{Wood98b} with
a higher metallicity in the Galactic centre compared to the rest of
the bulge. This fits into the general view that at low latitudes the
metal-rich stellar population of the disk merges with the generally
more metal-poor population of the bulge prevailing at larger
latitudes. The origin from low-mass stars was confirmed by
\cite{Ortiz02} for other OH/IR stars observed by the ISOGAL
survey. They concluded that the small number of objects more luminous
than M$_{bol}$\,=\,--5.5 (L\,=\,12,000 \Lsun) indicates that the
initial masses of stars in the sample are rarely higher than $\sim$\,3
\Msun, but
instead the masses are about 1\,--\,2 \Msun, on average.

All AGB star samples studied in the \GB\ were dominated by objects
with blue or moderate reddened colours. This may have biased the
studies towards low-mass objects, which would explain the scarcity of
high-luminosity sources. In this paper, we study a sample of AGB stars
of the \GB\ with extremely reddened colours. These stars are expected
to be at the very end of the TP-AGB evolution. If low-mass stars
evolve into the high mass-loss regime, the sample should include
objects with rather low luminosities. If not, they must have
originated from intermediate mass stars with correspondingly higher
luminosities.

Currently, with new infrared surveys available, the spectral energy
distributions (SED) of heavily obscured objects are covered with
unprecedented coverage. Together with theoretical models of the CSEs
of AGB stars, reliable bolometric fluxes can be derived without
invoking uncertain bolometric corrections. Finally, it is also
possible to obtain individual extinctions for each source leading to
more reliable luminosities.

The sample of extremely reddened AGB stars is presented in
Sect.\,\ref{thesample}. Section\,\ref{analysis} describes the
construction of the spectral energy distributions (SED), their fitting
with DUSTY models, and the estimation of luminosities and mass-loss
rates. The nature of the sources in the sample is discussed in
Sect.\,\ref{discussion}, and the conclusions are presented in
Sect.\,\ref{conclusions}.


\section{The sample}
\label{thesample}

Our sample of AGB stars of the \GB\ (hereafter: GB sample) was
selected from the sample of sources in \cite{Jimenez-Esteban06a}. They
studied around 100 IRAS sources distributed throughout the Milky Way
with very red IRAS colours ($[12]-[25]$\,$\ge$\,0.75\,mag), located in
a characteristic region of the IRAS two-colour diagram along the
sequence of colours predicted for O-rich AGB stars with increasing
mass loss \citep{Bedijn87}, and having a high (VAR\,$>$\,50) IRAS
variability index \citep{Beichman88}. The high value of this index
implies that all sources should be large-amplitude variables. By
comparison with the colours of the `Arecibo sample' of O-rich AGB
stars \citep{Jimenez-Esteban05}, they concluded that their sample of
extremely reddened IRAS sources was also mainly made up of O-rich AGB
stars highly obscured by their CSEs. To define the GB sample presented
here, we selected the extremely reddened IRAS sources from
\cite{Jimenez-Esteban06a} with angular distances $\le$\,10$^{\circ}$
from the Galactic centre.

Of the 39 sources selected in this region of the sky, two
(IRAS~17317--3331 and IRAS~17411--3154) do not belong to the
bulge. They are known to be foreground sources, located at a distance
of 3.2 and 1.2 kpc, respectively \citep{Cohen05,Yuasa99}. The
remaining 37 sources, presented in Table\,\ref{sample}, define our
sample of extremely reddened AGB stars in the \GB.

Although the sample is small, it is probably rather complete within
its selection boundaries. Its flux density distribution at
25~\micron\ peaks at $F_{25}$\,$\sim$\,7~Jy, while the weakest source
has $F_{25}$\,$\sim$\,4~Jy (Table\,\ref{phot}). AGB stars in the bulge
matching the colour selection criterion $[12]-[25]$\,$\ge$\,0.75\,mag
and $F_{25}$\,$>$\,7~Jy are probably all contained in the sample,
except those that are missing because of incompleteness of the IRAS
catalogue. Surpassing the $F_{25}$ flux limit requires a minimum
luminosity, which depends on the optical thickness of the CSE. This
minimum luminosity is the lowest for sources in which the bulk of the
emission is emitted in the 15--35~\micron\ wavelength range. This
minimum luminosity is also not biased by extinction, which is low at
these wavelengths. As we show below, the minimum luminosity required
to surpass the $F_{25}$ flux limit is low enough for high optical
thickness, so that low-mass stars will also be present in the sample,
if they develop the corresponding high mass-loss rates.

In accordance with the two foreground disk sources identified, we also
expect contamination of the sample by background disk sources. The
number of background disk sources can be roughly estimated as
follows. Considering the foreground disk sources as ``typical'' for
contaminating disk sources and using 4 Jy as the $F_{25}$ flux limit,
all such disk sources within the solar orbit around the centre of the
Milky Way should have been detected by IRAS. Assuming the same disk
source density in front and behind the Galactic bulge, the number of
foreground and background disk sources is coarsely proportional to the
surface area of the Galactic disk segments in front and behind the
bulge. Using a distance to the Galactic centre of
8.0\,($\pm$\,0.5)\,kpc \citep{Reid93}, the adopted 10$^{\circ}$
angular size of the \GB\ corresponds to a radius of
$\sim$\,1.4~kpc. Based on this consideration, we expect $\sim$4 times
more background sources than foreground sources. This yields about 8
sources out of 37, which could be background sources and whose
luminosities would be underestimated in the following.

We searched in the literature for information on the predominant
chemistry in the CSE of our objects. Six sources have IRAS Low
Resolution Spectra (LRS) spectral classifications by
\cite{Kwok97}. IRAS~17292--2727, IRAS~17316--3523, IRAS~17367--2722,
IRAS~17418--2713, and IRAS~17495--2534 were classified as LRS class A,
meaning that their spectra present the 9.7\mic\ silicate dust feature
in absorption. IRAS~18092--2347 was classified as LRS class I, meaning
that the LRS was noisy or incomplete. For the rest of the sources, we
searched for LRS spectra in the Calgary database\footnote{\tt
  http://www.iras.ucalgary.ca/saldatabase/index.html}. Only the LRS of
IRAS~17367--3633 was found and visually classified. It shows a clear
9.7\mic\ silicate dust feature in absorption. We also visually
inspected the LRS of IRAS~18092--2347, which also showed the silicate
dust feature in absorption. In addition, \cite{Golriz14} have recently
observed three of our sources that lack LRS classifications (IRAS~
17251--2821, IRAS~17276--2846, and IRAS~17323--2424) with the {\em
  Spitzer}-Infrared Spectrograph. All three Spitzer spectra show the
9.7\mic\ silicate in absorption. We classified these five sources as
LRS class A.


\begin{sidewaystable*}

\begin{minipage}[t][180mm]{\textwidth} 
\vspace{9cm}
  \caption[]{The GB sample. Observational parameters.}
  \label{sample}
  \begin{center}
  \begin{tabular}{l ccrr cccc ccc ccl}
    \hline
    \hline
    \noalign{\smallskip}
(1) & (2) & (3) & (4) & (5) & (6) & (7) & (8) & (9) & (10) & (11) & (12) & (13) & (14) & (15) \\ 
IRAS & \multicolumn{4}{c}{WISE coordinates} & \multicolumn{4}{c}{IRAS}  & \multicolumn{3}{c}{Radio data$^{e}$} & v$_{exp}$$^{f}$~~ & A$_{Ks}$ & Comment$^{g}$ \\
name & RA(J2000) & Dec(J2000) & Gal Long (\arcdeg) & Gal Lat (\arcdeg) & [12]-[25]$^{a}$ & [25]-[60]$^{a}$ & LRS$^{b}$ & Var$^{d}$ & OH & H$_{2}$O & SiO & [km\,s$^{-1}$]& mag &\\
    \noalign{\smallskip}                                                                                                                                                             
    \hline                                                                                                                                                                           
    \noalign{\smallskip}                                                                                                                                                             
16582--3059 & 17:01:29.85 & -31:04:19.1 & 352.843477 &  6.688608 & 0.89 & --0.86 &           &  57 & Y &   &   & 13.4 & 0.15 &  \\
17030--3053 & 17:06:14.06 & -30:57:38.2 & 353.547542 &  5.945143 & 0.77 & --0.73 &           &  99 & Y &   &   & 15.3 & 0.15 &  \\
17107--3330 & 17:14:05.05 & -33:33:57.9 & 352.421186 &  3.067547 & 0.86 & --0.96 &           &  99 & N &   &   &      & 0.40 &  \\
17128--3528 & 17:16:12.91 & -35:32:12.8 & 351.072561 &  1.563949 & 1.52 & ~~0.12 &           &  98 &   &   &   &      & 0.82 & NIR-Exc \\
17151--3642 & 17:18:29.53 & -36:46:01.9 & 350.333858 &  0.477947 & 0.99 & --0.30 &           &  99 & N &   &   &      & 1.63 &  \\
17171--2955 & 17:20:20.35 & -29:58:21.7 & 356.131283 &  4.050980 & 0.82 & --1.30 &           &  99 & N &   &   &      & 0.30 &  \\
17207--3632 & 17:24:07.27 & -36:35:40.6 & 351.118359 &--0.351526 & 0.95 & ~~0.46 &           &  99 & Y & N &   & 16.1 & 2.45 &  \\
17251--2821 & 17:28:18.61 & -28:24:00.5 & 358.413518 &  3.491327 & 0.94 & --0.75 &   A$^{c}$ &  99 & Y &   &   & 16.0 & 0.35 &  \\
17276--2846 & 17:30:48.32 & -28:49:02.0 & 358.366433 &  2.804559 & 1.25 & --0.12 &   A$^{c}$ &  79 & Y &   &   & 15.5 & 0.48 & PPN$^{c}$; DP \\
17292--2727 & 17:32:23.57 & -27:30:01.2 & 359.662488 &  3.230668 & 0.81 & --0.79 &   A       &  64 & Y &   & Y & 17.6 & 0.53 & FIR-Exc \\
17316--3523 & 17:34:57.50 & -35:25:52.5 & 353.298313 &--1.536651 & 1.02 & --0.72 &   A       &  99 & Y &   &   & 12.3 & 0.98 &  \\
17323--2424 & 17:35:25.92 & -24:26:30.5 & ~~2.611416 &  4.308932 & 0.99 & --0.73 &   A$^{c}$ &  97 & Y &   &   & 13.8 & 0.60 &  \\
17341--3529 & 17:37:30.29 & -35:31:04.4 & 353.504370 &--2.020286 & 0.88 & --1.26 &           &  97 & N &   & Y &      & 0.50 &  \\
17350--2413 & 17:38:08.86 & -24:14:49.2 & 3.107790   &  3.890090 & 0.85 & --1.50 &           &  99 & Sp&   &   &      & 0.32 &  \\
17351--3429 & 17:38:26.28 & -34:30:40.7 & 354.457640 &--1.643595 & 0.95 & --0.58 &           &  99 & Y &   &   & 20.0 & 0.56 &  \\
17361--2358 & 17:39:14.82 & -23:59:56.8 & 3.451800   &  3.809114 & 0.80 & --0.88 &           &  99 & N &   &   &      & 0.31 &  \\
17367--2722 & 17:39:52.43 & -27:23:31.7 & 0.646949   &  1.889337 & 0.98 & --0.47 &   A       &  97 & Y &   &   &  7.7 & 0.54 & FIR-Exc \\
17367--3633 & 17:40:07.62 & -36:34:41.4 & 352.888394 &--3.033409 & 0.88 & --0.75 &   A$^{c}$ &  92 & Y &   &   & 14.5 & 0.46 & FIR-Exc \\
17368--3515 & 17:40:12.98 & -35:16:40.8 & 354.002057 &--2.360174 & 0.77 & --0.71 &           &  99 & Y &   &   & 13.5 & 0.35 &  \\
17392--3020 & 17:42:30.54 & -30:22:07.4 & 358.425640 &--0.174669 & 1.09 & ~~0.81 &           &  54 & Y & Y,N&  & 21.1 & 1.80 &  \\
17418--2713 & 17:44:58.73 & -27:14:42.7 & 1.369081   &  1.002630 & 1.34 & --0.12 &   A       &  99 & Y & N &   & 15.2 & 0.76 &  \\
17428--2438 & 17:45:56.94 & -24:39:57.9 & 3.685349   &  2.159567 & 0.86 & --1.04 &           &  73 & Y &   &   & 13.8 & 0.58 &  \\
17495--2534 & 17:52:39.54 & -25:34:39.1 & 3.684492   &  0.388289 & 0.80 & --1.31 &   A       &  99 & N &   &   & ~16.0$^{f}$ & 1.28 &  \\
17504--3312 & 17:53:50.25 & -33:13:26.5 & 357.219129 &--3.708828 & 1.17 & ~~0.21 &           &  84 & Y &   &   & 15.7 & 0.26 & PPN$^{c}$; DP \\
17521--2938 & 17:55:21.80 & -29:39:13.0 & 0.471972   &--2.191825 & 1.07 & --0.17 &           &  98 & Y &   &   & 16.3 & 0.28 &  \\
17545--3056 & 17:57:48.41 & -30:56:25.8 & 359.621118 &--3.292745 & 1.36 & --0.04 &           &  89 & Y &   &   & 15.0 & 0.29 &  \\
17545--3317 & 17:57:49.20 & -33:17:47.6 & 357.573330 &--4.466135 & 0.86 & --0.91 &           &  99 & Y &   &   & 13.9 & 0.25 &  \\
17583--3346 & 18:01:39.29 & -33:46:00.5 & 357.556949 &--5.393255 & 0.76 & --0.99 &           &  99 & Y &   &   & 13.4 & 0.20 & FIR-Exc \\
17584--3147 & 18:01:42.04 & -31:47:54.7 & 359.284968 &--4.439273 & 1.04 & --0.53 &           &  99 & Y &   &   & 16.8 & 0.20 &  \\
18019--3121 & 18:05:12.26 & -31:21:43.2 & 0.032084   &--4.878452 & 0.99 & --1.00 &           &  99 & Sp&   &   &      & 0.21 &  \\
18040--2726 & 18:07:08.92 & -27:25:52.4 & 3.685892   &--3.345684 & 1.05 & --0.58 &           &  84 & Y &   &   & 16.3 & 0.19 &  \\
18040--2953 & 18:07:18.44 & -29:53:11.2 & 1.547809   &--4.562185 & 0.82 & --0.64 &           &  99 & N &   &   &      & 0.17 &  \\
18091--2437 & 18:12:16.15 & -24:36:42.9 & 6.715421   &--2.997292 & 0.80 & --0.60 &           &  99 & Y &   &   & 15.7 & 0.47 &  \\
18092--2347 & 18:12:20.44 & -23:46:55.9 & 7.452681   &--2.614618 & 0.94 & --0.15 & I-A$^{c}$ &  92 & Y & N &   & 17.4 & 0.45 &  \\
18092--2508 & 18:12:21.87 & -25:07:20.6 & 6.276710   &--3.260295 & 0.80 & --0.80 &           &  91 & Y &   &   & 14.5 & 0.40 & NIR-Exc \\
18195--2804 & 18:22:40.21 & -28:03:08.8 & 4.760139   &--6.667509 & 0.81 & --0.70 &           &  99 & Y &   &   & 16.3 & 0.13 &  \\
18201--2549 & 18:23:12.28 & -25:47:58.7 & 6.828182   &--5.738816 & 0.77 & --1.50 &           &  98 & N &   &   &      & 0.16 &  \\
  \noalign{\smallskip}
  \hline
  \end{tabular}
  \end{center}
   \begin{list}{}{}
      \item[Notes:] $^{a}$ IRAS colours as defined in
        \cite{Jimenez-Esteban05}; $^{b}$ IRAS Low Resolution Spectra;
        $^{c}$ Classification by this work (see Sections
        \ref{thesample} and \ref{SED});$^{d}$ IRAS variability
        index. $^{e}$ Radio data. Y: detection, N: no detection, Sp:
        single peak; $^{f}$ Expansion velocity from the OH maser
        velocity range, or from CO data for IRAS~17495--2534; $^{g}$
        SED peculiarities (see Section \ref{SED}). DP: double peaked,
        NIR-Exc: near-infrared excess, FIR-Exc: far-infrared excess,
        PPN: proto-planetary nebula candidate.
  \end{list}
\end{minipage}
\end{sidewaystable*}


In addition, all sources of our sample except IRAS~17128--3528 were
observed for OH maser emission
\citep{teLintel-Hekkert91a,David93,Sevenster97a}. Of these, eight were
not detected, two (IRAS~17350--2413 and IRAS~18019--3121) presented
single-peak OH emission, and the rest showed the typical double-peak
spectrum commonly shown by OH/IR stars. Expansion velocities of the
CSEs were derived for our objects from the velocity separations
between the two OH peaks. For IRAS~17495--2534, with no OH maser
emission detected, the expansion velocity was obtained from CO data
\citep{Loup93}. Four sources (IRAS~17207--3632, IRAS~17392--3020,
IRAS~17418--2713, and IRAS~18092--2347) were observed for H$_{2}$O
maser emission \citep{Deacon07,Suarez07}. Only IRAS~17392--3020 was
detected by \cite{Deacon07} in April 2003, although it was not
detected two years later, in September 2005, by
\cite{Suarez07}. Finally, two sources (IRAS~17292--2727 and
IRAS~17341--3529) were detected as SiO masers
\citep{Nyman98,Nakashima03}. Evidence for the O-rich nature of 30
sources is therefore provided by either detection of a maser and/or
the presence of the silicate dust feature.

In Table\,\ref{sample} we present the observational parameters for the
sources in our sample and summarise the above information. The IRAS
object name is listed in the first column. Columns 2 to 5 give the
equatorial and Galactic WISE coordinates (see
Sect\,\ref{SED}). Columns 6 and 7 give the IRAS colours. The IRAS LRS
spectral classification and the IRAS variability index are given in
the eight and ninth columns, respectively. In columns 10 to 12, we
summarise the detection status for OH, H$_{2}$O, and SiO maser
emission, respectively. The expansion velocity, if available, is
listed in column 13, and the estimated individual extinction in the
line of sight used to de-redden the observational photometry (see
Sect.\,\ref{obsflux}) is in column 14. Some comments on individual
SEDs, discussed below, are given in the last column.


\section{Analysis}
\label{analysis}

\subsection{Spectral energy distributions}
\label{SED}

To create the SEDs of the sources of our sample we collected,
when available, photometric information from astronomical
catalogues. To do that, we took advantage of the Virtual
Observatory\footnote{\tt http://www.ivoa.net} (VO). In particular, we
made use of the Multiple Cone Search utility of Topcat\footnote{\tt
  http://www.star.bris.ac.uk/$\sim$mbt/topcat/}, an interactive
graphical viewer and editor for tabular data that allows the user to
examine, analyse, combine, and edit astronomical tables. With Topcat,
we cross-matched our sample with several available astronomical
catalogues within the VO, using different search radii depending on
the particular catalogue.

This way, we populated the SEDs from the near- to the far-infrared,
wavelength range where these objects radiate the majority of their
energy. We collected data from the following catalogues: Two Micron
All Sky Survey (2MASS; \citealt{Cutri03}) and VISTA variables in the
V\'ia L\'actea catalogue (VVV-DR1; \citealt{Saito12}) at 1.25, 1.65,
and 2.2\,\mic\ (filter J, H, and K$_s$, respectively); GLIMPSE
\citep{Churchwell09} at 3.56, 4.51, 5.76, and 7.96\,\mic\ (filters I1,
I2, I3, and I4, respectively); WISE catalogue \citep{Wright10} at 3.4,
4.6, 11.6, and 22.1\,\mic\ (filters W1, W2, W3, and W4, respectively);
MSX6C Infrared Point Source Catalogue \citep{Egan03a} at 8.28, 12.13,
14.65, and 21.3\,\mic\ (filters A, C, D, and E, respectively); AKARI
Point Source Catalogue \citep{Ishihara10,Yamamura10} at 9, 18, 65, 90,
140, and 160\,\mic\ (filters S09, S18, S65, S90, S140, and S160,
respectively); and IRAS Point Sources Catalogue \citep{Beichman88} at
12, 25, 60 and 100\,\mic\ (filters 12, 25, 60, and 100,
respectively). Since WISE data were not available through the VO, they
were downloaded directly from the NASA/IPAC Infrared Science Archive.


We plotted the SED for each source and visually inspected them to
check the consistency of the photometric data. When deviating
photometric data were found, we used Aladin\footnote{\tt
  http://aladin.u-strasbg.fr/} \citep[][]{Bonnarel00}, another VO
tool, to visualise the astronomical images of the field and to check
the reliability of the selected counterpart. Aladin allows users to
visualise and analyse digitised astronomical images, and superimpose
entries from astronomical catalogues or databases available from the
VO services.

We found few inconsistencies, basically because of chance coincidences
with field stars within our search radii. In all these cases the
offending counterparts were removed. In the case of missing
counterparts both in 2MASS and VISTA catalogues, we used a nearby
($<$30\arcsec) faint star to define an upper flux limit, preferably in
the VISTA catalogue.

The photometric data are presented in Table~\ref{phot}, and the SEDs
are plotted in Fig.\,7. This figure shows the
observational data after correcting for interstellar extinction and
the DUSTY model that best fits the observations (see
Sect.\,\ref{obsflux}). Most of the sources present the typical SEDs of
an AGB star heavily obscured by an optically thick CSE, with the
maximum of their emission in the mid-IR. Note that most of the sources
were not detected in the near-IR, or only barely in the K$_s$ band.

Two sources (IRAS 17276--2846 and IRAS 17504--3312) show a peculiar
double-peaked SED (Fig.\,7). In the case of IRAS
17276--2846, the photospheric emission extends into the visual so that
we added VVVDR1 data for filters Z (0.88 \mic) and Y (1.02 \mic) to
the SED. These SEDs appear as soon as the mass-loss rates have greatly
decreased and the obscuration has substantially diminished. The bluer
peak at $\sim$\,2 \micron\ is then attributed to the reddened stellar
photosphere, and the redder peak corresponds to the emission at longer
wavelengths by the cool circumstellar dust \citep[e.g. see Fig.\,9
  in][]{SanchezContreras08}. This kind of drop in mass-loss rate can
be expected in the aftermath of a thermal pulse or when the stellar
envelope has nearly been lost at the end of AGB evolution. Stars with
this characteristic SED are generally considered to be proto-planetary
nebula (PPN) \citep{Oudmaijer96}, and therefore we assume that these
stars have left the AGB. Since post-AGB evolution is outside the scope
of the present paper, the two PPN candidates are not included in the
following analyses.



\begin{sidewaystable*}
\begin{minipage}[t][180mm]{\textwidth} 
\vspace{9cm}
  \caption[]{Photometric data collected.}
  \label{phot}
  \begin{center}
\tiny
  \begin{tabular}{l cccccccccccccc}
    \hline
    \hline
    \noalign{\smallskip} 
Filter              & J       & H      &$K_{\rm s}$& W1      & I1     & I2      & W2     & I3      & I4      & A      & S09     & W3     & 12     & C     \\
$\lambda$ ($\mu$m)  & 1.25    & 1.65   & 2.20     & 3.4     & 3.56   & 4.51    & 4.6    & 5.76    & 7.96    & 8.28   & 8.61    & 11.6   & 11.60  & 12.13  \\
Flux in Jy~$\times$ &$10^{-4}$&$10^{-4}$&$10^{-4}$  &$10^{-3}$&$10^{-3}$&$10^{-3}$&$10^{-2}$&$10^{-2}$&$10^{-2}$&$10^{-1}$&$10^{-1}$&$10^{-1}$&$10^{0}$&$10^{0}$ \\
    \noalign{\smallskip}    
    \hline   
    \noalign{\smallskip}        
16582--3059 & $^{a}$2.6          & $^{a}$3.9           & $^{a}$25.1$\pm$1.3   & 138$\pm$3    &               &              & 77.3$\pm$1.6  &              &            &              & 20.5$\pm$0.2     & 23.1$\pm$0.4  & 2.5$\pm$0.2   &               \\
17030--3053 & $^{a}$1.6          & $^{a}$1.8           & $^{a}$66.8$\pm$1.7   & 514$\pm$16   &               &              & 310$\pm$10    &              &            & 20.0$\pm$0.8 & 31.3$\pm$1.6     & 32.0$\pm$0.7  & 2.0$\pm$0.3   & 3.2           \\
17107--3330 & $^{a}$4.0          & $^{a}$29.3$\pm$1.5  & $^{a}$609$\pm$16     & 680$\pm$20   &               &              & 270$\pm$9     &              &            & 20.3$\pm$0.8 & 47$\pm$3         & 34.2$\pm$0.7  & 2.60$\pm$0.16 & 2.87$\pm$0.16 \\
17128--3528 & $^{a}$24           & $^{b}$1.52$\pm$0.14 & $^{b}$17.6$\pm$0.2   & 25.0$\pm$1.0 &               &              & 20.9$\pm$0.4  &              &            & 29.0$\pm$1.2 & 50$\pm$40        & 36.3$\pm$0.7  & 2.2$\pm$0.3   & 7.9$\pm$0.4   \\
17151--3642 & $^{b}$0.2          & $^{b}$3.5           & $^{b}$0.7            & 132$\pm$4    &               &              & 69.0$\pm$1.6  &              &            & 47$\pm$2     & 20.3$\pm$0.2     & 30.4$\pm$0.6  & 5$\pm$2       & 7.1$\pm$0.4   \\
17171--2955 & $^{b}$0.5          & $^{b}$20.2          & $^{b}$156.4$\pm$0.3  & 219$\pm$5    &               &              & 100$\pm$3     &              &            & 35.2$\pm$1.4 & 23.0$\pm$0.4     & 17.5$\pm$0.3  & 3.2$\pm$1.7   & 4.7$\pm$0.3   \\
17207--3632 & $^{b}$0.2          & $^{b}$1.39          & $^{b}$38.9           & 43.7$\pm$1.6 &               &              & 163$\pm$5     &              &            & 56$\pm$2     & 54.7$\pm$0.7     & 104.7$\pm$1.2 & 6.2$\pm$1.2   & 8.8$\pm$0.4   \\
17251--2821 & $^{a}$75           & $^{b}$12.1          & $^{b}$1.2            & 38.0$\pm$1.4 & 192$\pm$6     & 859$\pm$40   & 41.8$\pm$0.8  & 263$\pm$4    & 404$\pm$9  & 21.3$\pm$0.9 &                  & 24.2$\pm$0.5  & 3.6$\pm$0.8   & 3.7$\pm$0.2   \\
17276--2846$^{d}$ & $^{a}$700$\pm$30   & $^{a}$1700          & $^{b}$1441.0$\pm$1.3 & 135$\pm$4    & 138$\pm$4     & 261$\pm$7    & 32.7$\pm$0.7  & 137$\pm$3    & 340$\pm$20 & 21.3$\pm$0.9 &                  & 34.0$\pm$0.7  & 2.5$\pm$0.5   & 3.4$\pm$0.2   \\
17292--2727 & $^{a}$70           & $^{a}$64            & $^{a}$63             & 870$\pm$30   &               &              & 780$\pm$30    &              &            & 80$\pm$3     & 88$\pm$12        & 145.1$\pm$0.8 & 11.6$\pm$0.6  & 11.4$\pm$0.6  \\
17316--3523 & $^{b}$0.7          & $^{b}$1.1           & $^{a}$42$\pm$2       & 437$\pm$12   &               &              & 608$\pm$16    &              &            & 62$\pm$3     & 61$\pm$7         & 145.1$\pm$0.5 & 10$\pm$3      & 10.7$\pm$0.5  \\
           &                    &                     & $^{b}$24.4$\pm$0.3   &              &               &              &               &              &            &              &                  &               &               &               \\
17323--2424 & $^{b}$0.4          & $^{b}$3.8           & $^{b}$0.6            & 24.2$\pm$0.9 &               &              & 44.1$\pm$0.9  &              &            & 23.4$\pm$1.0 &                  & 30.6$\pm$0.6  & 3.4$\pm$0.5   & 2.9$\pm$0.2   \\
17341--3529 & $^{a}$22.6$\pm$1.3 & $^{a}$316$\pm$8     & $^{a}$1590$\pm$30    & 1140$\pm$40  &               &              & 409$\pm$14    &              &            & 83$\pm$3     & 41$\pm$4         & 63.7$\pm$0.8  & 6.3$\pm$1.6   & 12.5$\pm$0.6  \\
           & $^{b}$3.0$\pm$0.3  & $^{b}$67.0$\pm$0.4  & $^{b}$279.2$\pm$0.5  &              &               &              &               &              &            &              &                  &               &               &               \\
17350--2413 & $^{b}$0.7          & $^{b}$0.8           & $^{b}$175.7$\pm$0.3  & 178$\pm$4    &               &              & 63.8$\pm$1.3  &              &            & 14.8$\pm$0.6 &                  & 12.9$\pm$0.2  & 3.6$\pm$1.3   & 1.99$\pm$0.13 \\
17351--3429 & $^{b}$1.3          & $^{b}$1.8           & $^{a}$107$\pm$4      & 414$\pm$11   &               &              & 457$\pm$15    &              &            & 45$\pm$2     &                  & 58.5$\pm$0.7  & 3.2$\pm$0.7   & 6.1$\pm$0.3   \\
           &                    &                     & $^{b}$16.3$\pm$0.5   &              &               &              &               &              &            &              &                  &               &               &               \\
17361--2358 & $^{b}$0.5          & $^{b}$0.8           & $^{b}$19.2$\pm$0.3   & 253$\pm$7    &               &              & 253$\pm$7     &              &            & 19.6$\pm$0.8 & 63.5$\pm$0.8     & 49.2$\pm$0.4  & 3.6$\pm$0.8   & 2.70$\pm$0.15 \\
17367--2722 & $^{b}$1.0          & $^{b}$1.1           & $^{b}$1.1            & 29$\pm$2     &               &              & 50$\pm$1      &              &            & 24.1$\pm$1.0 &                  & 32.0$\pm$0.7  & 3.7$\pm$0.4   & 3.8$\pm$0.2   \\
17367--3633 & $^{b}$0.3          & $^{b}$8.4$\pm$0.2   & $^{a}$66$\pm$2       & 690$\pm$20   &               &              & 710$\pm$20    &              &            & 62$\pm$3     &                  & 147.2$\pm$1.0 & 8.8$\pm$1.1   & 8.2$\pm$0.4   \\
           &                    &                     & $^{b}$43.9$\pm$0.4   &              &               &              &               &              &            &              &                  &               &               &               \\
17368--3515 & $^{b}$0.2          & $^{b}$1.2           & $^{b}$0.8            & 27.9$\pm$1.6 &               &              & 60.6$\pm$1.3  &              &            & 31.4$\pm$1.3 &                  & 39.3$\pm$0.7  & 4.0$\pm$1.1   & 4.6$\pm$0.2   \\
17392--3020 & $^{b}$0.2          & $^{b}$0.3           & $^{b}$1.0            & 74$\pm$5     & 37.7$\pm$1.5  & 350$\pm$20   & 80.7$\pm$1.8  & 179$\pm$3    &            & 40.1$\pm$1.6 & 20.30$\pm$0.03   & 33.9$\pm$0.7  & 6.3$\pm$0.5   & 5.2$\pm$0.3   \\
17418--2713 & $^{b}$1.0          & $^{b}$1.3           & $^{b}$2.0            & 40$\pm$3     & 22.6$\pm$1.0  & 490$\pm$30   & 175$\pm$4     & 368$\pm$7    &            & 179$\pm$7    & 179.3$\pm$0.6    & 384.4$\pm$1.1 & 15$\pm$3      & 29.8$\pm$1.5  \\
17428--2438 & $^{b}$0.9          & $^{b}$1.1           & $^{b}$20.6$\pm$0.4   & 292$\pm$8    & 246$\pm$8     & 630$\pm$40   & 308$\pm$9     & 240$\pm$5    &            & 36.6$\pm$1.5 & 45$\pm$15        & 57.0$\pm$0.6  & 6.6$\pm$0.4   & 4.73$\pm$0.13 \\
17495--2534 & $^{b}$0.3          & $^{b}$0.3           & $^{b}$144.3$\pm$0.5  & 1108$\pm$49  & 370$\pm$20    & 8400$\pm$800 & 1140$\pm$30   & 436$\pm$11   &            & 85$\pm$4     &                  & 314.5$\pm$1.2 & 29$\pm$5      & 13.7$\pm$0.7  \\
17504--3312 & $^{b}$54.1$\pm$0.3 & $^{a}$70.3$\pm$1.4  & $^{a}$209$\pm$4      & 59$\pm$2     &               &              & 11.4$\pm$0.3  &              &            & 7.3$\pm$0.3  & 6.97$\pm$0.05    & 18.1$\pm$0.3  & 4.1$\pm$0.8   & 2.42$\pm$0.15 \\
           &                    & $^{b}$193.1$\pm$0.4 & $^{b}$425.8$\pm$0.4  &              &               &              &               &              &            &              &                  &               &               &               \\
17521--2938 & $^{a}$31           & $^{a}$44            & $^{b}$2.3            & 3.6$\pm$1.7  & 29.2$\pm$1.5  & 269$\pm$9    & 16.5$\pm$0.4  & 130$\pm$3    &            & 15.2$\pm$0.6 & 26$\pm$6         & 18.3$\pm$0.4  & 3.1$\pm$0.6   & 2.36$\pm$0.12 \\
17545--3056 & $^{b}$1.3          & $^{b}$1.4           & $^{b}$1.6            & 2.2          & 1.52$\pm$0.08 & 31.6$\pm$1.2 & 5.72$\pm$0.16 & 33.4$\pm$0.7 & 119$\pm$3  & 11.1$\pm$0.5 &                  & 22.4$\pm$0.4  & 2.9$\pm$0.4   & 2.53$\pm$0.13 \\
17545--3317 & $^{b}$0.8          & $^{b}$1.3           & $^{a}$45$\pm$3       & 204$\pm$5    &               &              & 140$\pm$4     &              &            & 46$\pm$2     &                  & 32.7$\pm$0.6  & 3.6$\pm$0.3   & 6.2$\pm$0.3   \\
           &                    &                     & $^{b}$11.4$\pm$0.3   &              &               &              &               &              &            &              &                  &               &               &               \\
17583--3346 & $^{b}$25.8$\pm$0.2 & $^{a}$39$\pm$2      & $^{a}$522$\pm$9      & 631$\pm$19   &               &              & 210$\pm$6     &              &            &              &                  & 36.2$\pm$0.7  & 2.00$\pm$0.14 &               \\
           &                    & $^{b}$265.1$\pm$0.2 & $^{b}$1082.0$\pm$1.0 &              &               &              &               &              &            &              &                  &               &               &               \\
17584--3147 & $^{b}$0.7          & $^{b}$0.7           & $^{b}$1.0            & 46.9$\pm$1.6 &               &              & 95$\pm$2      &              &            & 22.6$\pm$0.9 &                  & 67.7$\pm$0.8  & 6.9$\pm$1.0   & 3.4$\pm$0.2   \\
18019--3121 & $^{b}$0.5          & $^{b}$0.7           & $^{b}$0.8            & 25.2$\pm$0.9 &               &              & 50.0$\pm$1.0  &              &            & 24.5$\pm$1.0 & 28$\pm$4         & 29.3$\pm$0.6  & 3.5$\pm$1.1   & 3.0$\pm$0.2   \\
18040--2726 & $^{b}$1.1          & $^{b}$1.1           & $^{b}$0.8            & 39.1$\pm$1.3 &               &              & 97$\pm$2      &              &            & 24.2$\pm$1.0 & 51$\pm$15        & 59.4$\pm$0.9  & 6.5$\pm$0.5   & 3.5$\pm$0.2   \\
18040--2953 & $^{b}$0.9          & $^{b}$1.0           & $^{b}$8.6$\pm$0.3    & 345$\pm$9    &               &              & 492$\pm$17    &              &            & 28.3$\pm$1.2 & 30$\pm$4         & 84.4$\pm$0.6  & 5.0$\pm$0.8   & 3.7$\pm$0.2   \\
18091--2437 & $^{b}$0.8          & $^{b}$0.7           & $^{b}$10.6$\pm$0.3   & 177$\pm$4    & 236$\pm$8     &              & 150$\pm$4     & 212$\pm$5    &            & 16.3$\pm$0.7 & 20.2$\pm$0.9     & 32.8$\pm$0.7  & 4.6$\pm$0.7   & 2.30$\pm$0.14 \\
18092--2347 & $^{b}$0.5          & $^{b}$1.0           & $^{b}$1.0            & 4.3$\pm$1.1  & 13.9$\pm$0.6  & 279$\pm$13   & 36.5$\pm$0.7  & 230$\pm$5    &            & 50$\pm$2     & 55$\pm$3         & 121.2$\pm$0.9 & 10.9$\pm$0.8  & 9.8$\pm$0.5   \\
18092--2508 & $^{b}$1.5$\pm$0.2  & $^{b}$2.4$\pm$0.3   & $^{b}$20.2$\pm$0.3   & 311$\pm$9    &               &              & 352$\pm$13    &              &            & 49$\pm$2     & 51.8$\pm$7       & 58.4$\pm$0.5  & 5.6$\pm$0.7   & 6.6$\pm$0.3   \\
18195--2804 & $^{a}$0.4          & $^{b}$0.5           & $^{b}$21.76$\pm$0.16 & 362$\pm$8    &               &              & 341$\pm$10    &              &            &              & 20.6$\pm$6       & 49.2$\pm$0.5  & 3.6$\pm$0.8   &               \\
18201--2549 & $^{a}$11.1$\pm$0.7 & $^{a}$323$\pm$7     & $^{a}$3260$\pm$60    & 2340$\pm$140 &               &              & 678$\pm$16    &              &            &              & 93.627$\pm$0.003 & 95.0$\pm$0.5  & 5.0$\pm$0.7   &               \\
          & $^{b}$14.22$\pm$0.18& $^{b}$366.9$\pm$0.3 & $^{b}$2759$\pm$3     &              &               &              &               &              &            &              &                  &               &               &               \\
  \noalign{\smallskip}
  \hline
  \end{tabular}
  \end{center}
   \begin{list}{}{}
     \item[Notes:] Data without errors are upper limits; $^{a}$ 2MASS
       data; $^{b}$ VVV-DR1 data; $^{c}$ No confirmed detection;
       $^{d}$ VVV-DR1 fluxes in Jy\,$\times$\,10$^{-5}$:
       F$_{z}$=761$\pm$2 and F$_{Y}$=3017$\pm$4.
  \end{list}
\end{minipage}
\end{sidewaystable*}

\addtocounter{table}{-1} %
\begin{sidewaystable*}
\begin{minipage}[t][180mm]{\textwidth} 
\vspace{9cm}
  \caption[]{Photometric data collected. Continued}
  \begin{center}
\tiny
  \begin{tabular}{l ccccccccccc}
    \hline
    \hline
    \noalign{\smallskip} 
Filter ($\lambda$ ($\mu$m))   & D     & S18     & E      & W4     & 25    & 60     & S65   & S90     & 100    & S140   & S160 \\
$\lambda$ ($\mu$m)            & 14.65 & 18.39   & 21.3   & 22.1   & 23.88 & 60     & 65    & 90      & 100    & 140    & 160 \\
Flux in Jy~$\times$          &$10^{0}$&$10^{-1}$&$10^{0}$&$10^{-1}$&$10^{0}$&$10^{0}$&$10^{0}$&$10^{-1}$&$10^{0}$&$10^{0}$&$10^{0}$\\
    \noalign{\smallskip}  
    \hline              
    \noalign{\smallskip}          
16582--3059 &               &                &              & 51.7$\pm$0.9  & 5.8$\pm$0.6  & 2.6$\pm$0.3  & $^{c}$1.6         & 13$\pm$2       & 20   &                   & $^{c}$1.3      \\
17030--3053 & 3.5$\pm$0.2   &                & 5.5          & 58.0$\pm$0.8  & 4.1$\pm$0.7  & 2.1$\pm$0.2  &                   &                & 26   &                   &                \\
17107--3330 & 3.4$\pm$0.2   &                & 3.8$\pm$0.2  & 64.9$\pm$0.8  & 5.7$\pm$1.0  & 2.4$\pm$0.5  &                   &                & 20   &                   &                \\
17128--3528 & 13.0$\pm$0.8  & 93.0$\pm$0.3   & 13.8$\pm$0.8 & 114.9$\pm$1.5 & 9.1$\pm$1.4  & 10.1$\pm$0.8 &     17$\pm$5      & 86$\pm$8       & 55   & $^{c}$9.9         & $^{c}$7.3      \\
17151--3642 & 11.2$\pm$0.7  &                & 11.3$\pm$0.7 & 73.2$\pm$1.6  & 13$\pm$4     & 10.1$\pm$1.6 & $^{c}$6.8         & 71$\pm$5       & 320  & $^{c}$30$\pm$4    & $^{c}$82$\pm$7 \\
17171--2955 & 5.8$\pm$0.4   &                & 6.1$\pm$0.4  & 33.6$\pm$0.3  & 7$\pm$4      & 2.1$\pm$0.6  & $^{c}$1.1         & 16.5$\pm$0.2   & 18   &                   &                \\
17207--3632 & 15.4$\pm$0.9  &                & 12.6$\pm$0.8 & 194$\pm$3     & 15$\pm$3     & 23$\pm$3     &     16$\pm$4      & 113$\pm$6      & 1500 &                   & $^{c}$45$\pm$9 \\
17251--2821 & 4.9$\pm$0.3   & 47$\pm$8       & 5.8$\pm$0.4  & 58.8$\pm$0.7  & 8.5$\pm$1.6  & 4.2$\pm$0.4  & $^{c}$1.7$\pm$0.6 & 22$\pm$2       & 20   & $^{c}$1.8         &                \\
17276--2846$^{d}$ & 5.7$\pm$0.4   & 59.20$\pm$4    & 5.4$\pm$0.3  & 75.7$\pm$0.9  & 7.9$\pm$0.9  & 7.2$\pm$0.8  & $^{c}$6.2         & 37$\pm$3       & 30   & $^{c}$1.4         &                \\
17292--2727 & 16.1$\pm$1.0  & 153.3$\pm$0.7  & 19.5$\pm$1.2 & 213$\pm$6     & 24$\pm$2     & 11.8$\pm$1.1 &     8.9$\pm$0.4   & 71$\pm$9       & 12   &     4.6$\pm$1.2   & $^{c}$4.4      \\
17316--3523 & 16.4$\pm$1.0  &                & 19.6$\pm$1.2 & 246$\pm$4     & 25$\pm$4     & 13$\pm$2     &                   &                & 170  &                   &                \\
17323--2424 & 5.5$\pm$0.3   & 43.0$\pm$1.6   & 7.1$\pm$0.5  & 67.9$\pm$0.9  & 8.5$\pm$1.7  & 4.3$\pm$0.5  & $^{c}$3.8         & 19$\pm$4       & 13   &                   & $^{c}$0.66     \\
17341--3529 & 13.3$\pm$0.8  & 87$\pm$10      & 16.4$\pm$1.0 & 116.9$\pm$1.3 & 14$\pm$2     & 4.5$\pm$0.9  &                   &                & 100  &                   &                \\
17350--2413 & 2.76$\pm$0.18 &                & 3.1          & 27.6$\pm$0.3  & 7.8$\pm$0.6  & 2.0$\pm$0.2  & $^{c}$1.4         & 13.1$\pm$0.6   & 13   & $^{c}$0.61        &                \\
17351--3429 & 8.6$\pm$0.5   & 106.9$\pm$1.6  & 10.2$\pm$0.6 & 117.1$\pm$1.1 & 7.8$\pm$1.4  & 4.6$\pm$0.7  &     6.0$\pm$0.8   & 29$\pm$6       & 160  & $^{c}$2.0         & $^{c}$0.13     \\
17361--2358 & 3.7$\pm$0.2   &                & 4.1$\pm$0.3  & 85.9$\pm$1.7  & 7.5$\pm$1.4  & 3.3$\pm$0.4  &                   &                & 11   &                   &                \\
17367--2722 & 6.1$\pm$0.4   & 90$\pm$30      & 6.6$\pm$0.4  & 76.7$\pm$1.3  & 9.2$\pm$1.2  & 6.0$\pm$0.5  & $^{c}$4.0         & 44$\pm$7       & 36   & $^{c}$4.4$\pm$1.1 & $^{c}$4.7      \\
17367--3633 & 11.6$\pm$0.7  &                & 13.3$\pm$0.8 & 204$\pm$4     & 20$\pm$2     & 9.9$\pm$1.0  &     11$\pm$5      & 69.6$\pm$0.4   & 10   & $^{c}$6$\pm$4     & $^{c}$0.92     \\
17368--3515 & 7.3$\pm$0.5   & 74.3$\pm$1.3   & 7.5$\pm$0.5  & 85.0$\pm$1.0  & 8$\pm$2      & 4.2$\pm$1.0  & $^{c}$5.9$\pm$1.5 & 21$\pm$3       & 13   &                   &                \\
17392--3020 & 9.0$\pm$0.5   & 90$\pm$30      & 9.4$\pm$0.6  & 91.6$\pm$1.6  & 17.3$\pm$1.6 & 36$\pm$7     &                   &                & 210  &                   &                \\
17418--2713 & 58$\pm$4      &                & 68$\pm$4     & 720$\pm$9     & 51$\pm$7     & 46$\pm$5     &                   &                & 500  &                   &                \\
17428--2438 & 6.3$\pm$0.4   & 103.5$\pm$1.1  & 7.2$\pm$0.4  & 110.5$\pm$1.7 & 14.6$\pm$1.2 & 5.6$\pm$0.6  &                   &                & 47   &                   &                \\
17495--2534 & 19.9$\pm$1.2  & 500$\pm$40     & 23.7$\pm$1.4 & 521$\pm$8     & 61$\pm$8     & 18$\pm$2     &     17$\pm$4      & 88.9$\pm$1.7   & 370  &                   &                \\
17504--3312 & 4.7$\pm$0.3   & 59.34$\pm$0.13 & 7.8$\pm$0.5  & 87.6$\pm$1.0  & 12.0$\pm$1.2 & 15$\pm$2     &                   &                & 43   &                   &                \\
17521--2938 & 4.3$\pm$0.3   & 69$\pm$13      & 5.4$\pm$0.3  & 49.8$\pm$0.7  & 8.2$\pm$1.7  & 7.0$\pm$1.1  &                   &                & 110  &                   &                \\
17545--3056 & 5.5$\pm$0.3   & 50$\pm$20      & 6.5$\pm$0.4  & 75.6$\pm$0.9  & 10.2$\pm$0.9 & 9.8$\pm$1.2  &     5.3$\pm$1.2   & $^{c}$45.56    & 57   & $^{c}$7.0         & $^{c}$0.25     \\
17545--3317 & 7.9$\pm$0.5   &                & 9.1$\pm$0.6  & 50.1$\pm$0.5  & 8.0$\pm$1.1  & 3.5$\pm$0.4  & $^{c}$1.7         & 14$\pm$2       & 34   &                   & $^{c}$0.42     \\
17583--3346 &               & 35.0$\pm$0.8   &              & 59.0$\pm$0.6  & 4.0$\pm$0.3  & 1.6$\pm$0.4  & $^{c}$1.1         & 13.0$\pm$1.6   & 26   & $^{c}$1.0         &                \\
17584--3147 & 5.1$\pm$0.3   & 99$\pm$17      & 6.6$\pm$0.4  & 167$\pm$4     & 18$\pm$3     & 11.0$\pm$1.3 &     6.82$\pm$0.13 & $^{c}$43.67    & 12   &                   &                \\
18019--3121 & 5.7$\pm$0.4   & 69.17$\pm$0.04 & 5.2$\pm$0.3  & 64.5$\pm$0.8  & 9$\pm$2      & 3.4$\pm$0.9  &                   &                & 62   &                   &                \\
18040--2726 & 6.5$\pm$0.4   & 130$\pm$20     & 6.4$\pm$0.4  & 137$\pm$2     & 17.0$\pm$1.4 & 9.9$\pm$1.2  &     8.8$\pm$0.4   & 69$\pm$3       & 66   & $^{c}$1.7         & $^{c}$1.1      \\
18040--2953 & 5.7$\pm$0.4   & 60$\pm$7       & 6.3$\pm$0.4  & 143$\pm$2     & 11$\pm$2     & 5.9$\pm$1.0  & $^{c}$2.9         & 25.01$\pm$0.02 & 42   &                   & $^{c}$1.7      \\
18091--2437 & 3.1$\pm$0.2   &                & 3.8$\pm$0.2  & 63.8$\pm$0.7  & 9.7$\pm$1.5  & 5.6$\pm$0.6  &                   &                & 86   &                   &                \\
18092--2347 & 17.7$\pm$1.1  &                & 18.1$\pm$1.1 & 245$\pm$5     & 26$\pm$2     & 22.4$\pm$3   &     13.2$\pm$0.3  & 115$\pm$7      & 110  &                   & $^{c}$4.7      \\
18092--2508 & 9.5$\pm$0.6   & 121.0$\pm$0.9  & 10.6$\pm$0.7 & 115$\pm$2     & 11.8$\pm$1.1 & 5.6$\pm$0.7  & $^{c}$4.7         & 26.9$\pm$0.9   & 90   & $^{c}$0.79        & $^{c}$2.0      \\
18195--2804 &               & 30.4$\pm$0.5   &              & 95.1$\pm$1.7  & 7.7$\pm$1.5  & 4.0$\pm$0.5  &     1.7$\pm$0.3   & 16$\pm$2       & 17   & $^{c}$0.6$\pm$0.8 & $^{c}$1.4      \\
18201--2549 &               & 64.9$\pm$0.4   &              & 133.7$\pm$1.1 & 10.2$\pm$1.4 & 2.6$\pm$0.2  & $^{c}$1.2         & 13$\pm$3       & 23   &                   & $^{c}$1.3      \\
  \noalign{\smallskip}
  \hline
  \end{tabular}
  \end{center}
   \begin{list}{}{}
     \item[Notes:] Data without errors are upper limits; $^{a}$ 2MASS
       data; $^{b}$ VVV-DR1 data; $^{c}$ No confirmed detection;
       $^{d}$ VVV-DR1 fluxes in Jy\,$\times$\,10$^{-5}$:
       F$_{z}$=761$\pm$2 and F$_{Y}$=3017$\pm$4.
  \end{list}
\end{minipage}
\end{sidewaystable*}


\subsection{Observed bolometric flux}
\label{obsflux}

The bulge sample of AGB stars is nowadays covered over a large part of
their SED by a multitude of infrared surveys. This allows us to
determine the bolometric flux with much higher accuracy compared to
earlier times, when (uncertain) bolometric corrections or (unsuitable)
black-body fitting had to be applied to cover the unobserved parts of
the SED \citep[e.g.][]{vanderVeen89,Ortiz02}. However, the photometric
data available is not complete for all sources so that a determination
of the bolometric flux by integration of an interpolated curve between
the available data points will be unsatisfactory. Also bolometric
corrections (although small ones) for the spectral ranges outside the
infrared are still required. We therefore decided to fit the observed
SEDs with models of stars obscured by a dusty radial symmetric CSE and
to determine the bolometric fluxes from the models. Because of the
large distance of the \GB, the stars may be observed through large
columns of intervening dust, such that extinction may have a
noticeable influence on the shape of the SEDs and on the observed
bolometric flux.

To de-redden the photometric data, we corrected the flux at 2.2
\mic\ ($K_S$ band) using the extinction maps towards the Galactic
bulge given by \cite{Gonzalez12}. The absolute extinctions towards the
sources of our sample are 0.13\,$\le$\,A$_{Ks}$\,$\le$\,2.45 mag (see
Table \ref{sample}). Two sources (IRAS~16582--3059 and
IRAS~17030--3053) are not covered by the maps and we assigned
$A_{Ks}$\,=\,0.15 mag to them. About half of the sources are also
covered by the extinction maps of \cite{Nidever12}. Their absolute
extinctions are 0.14\,$\pm$\,0.04 mag higher for $A_{Ks}$\,$<$\,1.0
mag than those of \cite{Gonzalez12}. Because the uncertainties given
in both papers are 0.1 mag, we considered this difference as not
significant. For higher extinctions, the scatter $\sigma$ between the
maps is larger with $\sigma$\,=\,0.3 mag.

To correct the other photometric bands, we determined the extinction
coefficients A$_\lambda$/A$_{Ks}$ from the extinction curves given by
\cite{Chen13} up to $\lambda$\,=\,8 \micron\ and from \cite{Gao09} up
to 24 \micron. The extinction coefficients beyond $\lambda$\,=\,24
\micron\ were determined by extrapolation using a power law with
$\beta$\,=\,--1.7.

The de-reddened SEDs were modelled using the radiation transport code
DUSTY \citep{Ivezic99}. Because the observed photometry taken at
different epochs is likely affected by variability, the photometric
data are expected to scatter around the mean SED. The relatively
sparse coverage of the SEDs with photometry and the variability
preclude the determination of accurate models. We therefore decided to
use a standard model, in which most of the parameters are fixed, and
only the optical depth of the shell was left as variable parameter.

Combining the information from the IRAS LRS classifications and the
maser detections, O-rich chemistry can be inferred for 28 objects. 
Carbon stars are extremely rare in the \GB\ \citep{Blanco89}. Hence,
we decided to assume O-rich chemistry for the remaining seven objects,
following
\cite{Lewis92}, who found that infrared sources without OH maser
detection but with IRAS colours similar to those of OH/IR stars are
most probably O-rich variable AGB stars. Consequently, we only used
models with O-rich chemistry.

The standard model assumes a central star with an effective
temperature T$_{eff}$\,=\,2500~K, and a dust condensation temperature
T$_{d}$\,=\,1000~K. We used the optical constants for amorphous cold
silicates from \cite{Ossenkopf92} and took the standard MRN
\citep{Mathis77} dust size distribution with
$n(a)$\,$\propto$\,$a^{-3.5}$, where $n$ is the number density and $a$
is the size of the grains. The grain sizes were limited to
0.005\,$\le$\,$a$\,$\le$\,0.25~\micron. The inner radius $R_{min}$ of
the CSE is determined by DUSTY from the dust condensation temperature
assumed. The outer radius of the shell was set to
$R_{max}$\,=\,100~$R_{min}$. Instead of describing the density
distribution in the shell by an analytic profile, the option to
calculate the wind structure from hydrodynamics was chosen, which
allows us to determine mass-loss rates from the models
\citep{Ivezic95}.  Models were calculated for optical depths
0.2\,$<$\,$\tau_{9.7\mu m}$\,$<$\,50 and the best model was found by
minimising the deviations from the de-reddened flux densities. In
several cases two models were fitted representing the star at
different variability states. Because of the requirement of a very red
IRAS colour, most stars possess a CSE with a high optical depth
$\tau_{9.7\mu m}$\,$>$\,10, and only four have 5\,$<$\,$\tau_{9.7\mu
  m}$\,$<$\,10. The bolometric fluxes determined vary by a factor 20
and are in the range $f_{bol}$\,=\,$10^{-12}$\,--\,$2 \times
10^{-11}$~Watt\,m$^{-2}$.

The results are given in Table \ref{DUSTY-results}. It lists the
object name in the first column. The best-fit optical depth
$\tau_{9.7\mu m}$, and the bolometric flux $f_{bol}$ delivered by
DUSTY are given in the two following columns. The last two columns
give luminosities and mass-loss rates discussed in the next Section.

The uncertainties of $\log f_{bol}$ are approximately 0.2~dex, which
we attribute mostly to variability. For the cases where models for
different variability states are given, the difference is
$\sim$\,0.3~dex. We conclude therefore that the peak-to-peak
luminosity variation of the large-amplitude variable objects is
$\ge$\,2. This number is probably a lower limit, as \cite{Suh04}
determined for two extremely obscured OH/IR stars in the Galactic disk
luminosity variations by a factor 3\,--\,3.6. The bolometric fluxes
determined here cannot be individually corrected for variability, but
in general we expect them to be close to the mean bolometric
fluxes. The deviations of the observed bolometric fluxes from the mean
fluxes is at most half of the peak-to-peak variations, and very likely
even lower in those cases where average bolometric fluxes are
available from fits at different variability states.

\begin{figure}
  \centering
  \includegraphics[width=\columnwidth]{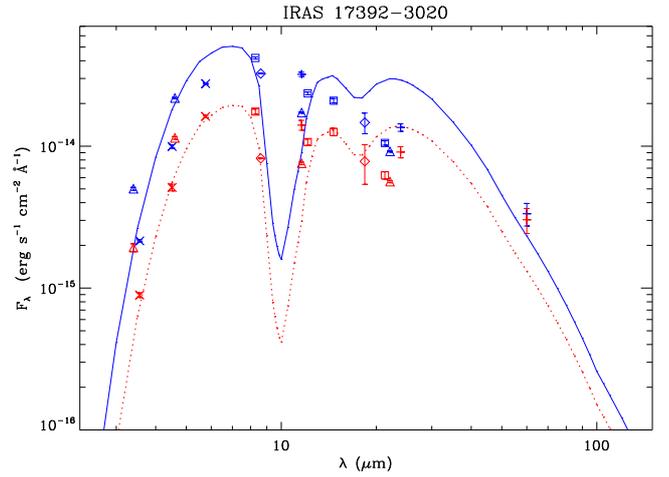}
  \caption{Observational photometry and model fits of
    IRAS\,17392$-$3020. For each photometric observation, the
    uncorrected flux density (lower red symbols) and the de-reddened
    flux density for A$_{Ks}$\,=\,1.8\,mag (upper blue symbols) are
    given. The best-fitting models are given as dotted (red) curve for
    the observed photometry and as solid (blue) curve for the
    de-reddened photometry. The de-reddened bolometric flux is a
    factor 2.2 enhanced. Symbols are as in Fig.\,7.}
  \label{extinction}	
\end{figure}

The de-reddened SEDs and the models are shown in
Fig.\,7. The observed SEDs are the result of the emission
from the circumstellar dust with high optical depths, which in
addition are reddened further by interstellar extinction. The removal
of the interstellar extinction makes the SEDs bluer and brighter than
the observed SEDs. As an example, the influence of extinction on
photometry and model fit is shown in Fig. \ref{extinction} for
IRAS\,17392$-$3020. We found that the increase in logarithmic
bolometric flux surpasses 0.2~dex only if A$_{Ks}$\,$>$\,1.0\,mag,
which applies for four sources in the sample. Therefore, the largest
contributions to the uncertainties of the $f_{bol}$ determination is
in general from the variability and not from interstellar extinction.

For several sources the DUSTY models predict less flux than observed
in the near infrared at $\lambda$\,$<$\,3~\micron\ and in the far
infrared at wavelengths 60\,$<$\,$\lambda$\,$<$\,160~\micron. These
sources are marked in Table \ref{sample} as `NIR-Exc' and
`FIR-Exc'. The observed excess emission makes only a minor
contribution to the overall luminosity and has been neglected for the
luminosity determination. 

The near-infrared excess presented by IRAS~17128--3528 and
IRAS~18092--2508 (see Fig.\,7) is evident from the
unexpectedly high brightness in the J, H, and K filters measured by
the VVV-DR1 survey. We verified that this is not due to
missidentification or contamination of neighbouring objects. These
stars may have developed deviations of their dust distribution from
spherical symmetry or their circumstellar shell may have been diluted
after the end of the strong AGB mass loss, so that the light of the
central star could escape to form the excess. They might currently be
starting the post-AGB evolution and may later develop double-peaked
SEDs as seen in the PPN candidates IRAS 17276--2846 and IRAS
17504--3312 (cf. Sect.\,\ref{SED}). Alternatively, they might belong
to binary systems, where the excess emission comes from the companion.
In any case, additional photometry is required to verify the near-IR
excesses for these sources.

The SEDs of IRAS~17292--2727, IRAS~17367--2722, IRAS~17367--3633, and
IRAS~17583--3346 (Fig.\,7) present a far-infrared 
excess. This far-infrared excess emission can not be attributed to
photometric errors or variability since consistent flux densities have
been measured at different times by more than one survey. Such excess
emission has been seen before in a number of objects, in which the
excess can be followed over a large wavelength interval extending up
to the submillimeter and millimeter ranges. To account for the excess
flux an additional dust component of larger cool grains is required
to model the SEDs of these stars \citep{SanchezContreras07}.

\begin{table*}
  \caption[]{Result of the DUSTY model fit.}
  \label{DUSTY-results}
  \begin{center}
  \begin{tabular}{lcccc}
    \hline
    \hline
    \noalign{\smallskip}
(1) & (2) & (3) & (4) & (5) \\
IRAS & $\tau_{9.7\mu m}$ & $-$log($f_{bol}$)$^{a}$ & L$^{b}$              & \Mdot    \\
name &                 &                       & $\times$$10^3$\,\Lsun & $\times$$10^{-5}$\,\Myr \\
    \noalign{\smallskip}    
    \hline                                                                                                                      
    \noalign{\smallskip}                                                                                                 
16582--3059       & 15.9           & 11.70            & ~~4.0          & 1.7 \\
17030--3053       & 10.9           & 11.75            & ~~3.5          & 1.0 \\
17107--3330$^{c}$  &~~6.2           & 11.64            & ~~4.5          & 1.0 \\
17128--3528$^{c}$  & 41.3           & 11.30            & ~~9.9          & 8.7 \\
17151--3642$^{c}$  & 23.4           & 11.25            &  11.1          & 6.0 \\
17171--2955$^{c}$  & 10.9           & 11.63            & ~~4.7          & 1.4 \\
17207--3632       & 34.2           & 11.00            &  19.8          & 12.8 \\
17251--2821       & 23.3\,--\,28.2 & 11.44\,--\,11.80 &   7.2\,--\,3.1 & 1.7\,--\,3.4 \\
17292--2727       & 15.9           & 11.05            &  17.7          & 5.7 \\
17316--3523       & 15.9           & 11.08            &  16.5          & 7.6 \\
17323--2424       & 28.2           & 11.67            & ~~4.2          & 2.7 \\
17341--3529$^{c}$  &  5.1\,--\,6.2  & 11.05\,--\,11.30 & 17.7\,--\,9.9  & 2.1\,--\,3.4 \\
17350--2413$^{c}$  &  9.0\,--\,10.9 & 11.50\,--\,11.95 & ~~6.3\,--\,2.2 & 0.7\,--\,1.7 \\
17351--3429       & 13.2           & 11.30            & ~~9.9          & 2.4 \\
17361--2358$^{c}$  & 15.9           & 11.57            & ~~5.3          & 2.1 \\
17367--2722       & 34.2           & 11.60            & ~~5.0          & 6.8 \\
17367--3633       & 10.9\,--\,13.2 & 10.80\,--\,11.15 & 31.4\,--\,14.0 & 4.7\,--\,9.3 \\
17368--3515       & 28.2           & 11.56            & ~~5.5          & 3.6 \\
17392--3020       & 41.3           & 11.01            &  19.4          & 11.3 \\
17418--2713       & 50.0\,--\,50.0 & 10.67\,--\,11.02 & 42.4\,--\,18.9 & 18.0\,--\,40.4 \\
17428--2438       & 15.9           & 11.40            & ~~7.9          & 3.2 \\
17495--2534       & 15.9\,--\,28.2 & 10.70\,--\,11.00 & 39.6\,--\,19.8 & 11.1\,--\,14.0 \\
17521--2938       & 41.3\,--\,50.0 & 11.60\,--\,11.88 &   4.9\,--\,2.6 & 2.3\,--\,3.7 \\
17545--3056       & 50.0           & 11.77            & ~~3.4          & 3.3 \\
17545--3317       & 15.9           & 11.50            & ~~6.3          & 2.6 \\
17583--3346       &~~5.1           & 11.40            & ~~7.9          & 1.6 \\
17584--3147       & 34.2           & 11.45            & ~~7.0          & 4.3 \\
18019--3121$^{c}$  & 28.2           & 11.71            & ~~3.9          & 2.5 \\
18040--2726       & 34.2\,--\,34.2 & 11.35\,--\,11.65 &   8.9\,--\,4.4 & 2.8\,--\,5.7 \\
18040--2953$^{c}$  & 19.3\,--\,23.3 & 11.30\,--\,11.65 &   9.9\,--\,4.4 & 2.4\,--\,4.6 \\
18091--2437       & 15.9\,--\,19.3 & 11.50\,--\,11.85 &   6.3\,--\,2.8 & 1.2\,--\,2.3 \\
18092--2347       & 50.0           & 11.19            &  12.8          & 10.6 \\
18092--2508       & 19.3           & 11.35            & ~~8.9          & 4.0 \\
18195--2804       & 15.9           & 11.55            & ~~5.6          & 1.9 \\
18201--2549$^{c}$  &~~5.1           & 11.25            &  11.2          & 2.2 \\
    \noalign{\smallskip}                                            
    \hline                                                          
  \end{tabular}                                                     
  \end{center}                                                      
  \begin{list}{}{}                                                  
     \item[Notes:] $^{a}$ $f_{bol}$ in Watt\,m$^{-2}$; $^{b}$
       Luminosity estimated using the distance of 8.0\,kpc to the
       Galactic centre. $^{c}$ v$_{exp}$\,=\,14\,km\,s$^{-1}$ was
       assumed. Otherwise v$_{exp}$ is derived from either OH or CO
       radio data.
  \end{list}                                                        
                                                                    
\end{table*}


\subsection{Luminosity}

We estimated the luminosity from the bolometric fluxes $f_{bol}$,
assuming a common distance to all of the sources in the sample
equivalent to the distance to the Galactic centre
(8.0\,$\pm$\,0.5\,kpc). The luminosity values obtained are tabulated
in the fourth column of Table\,\ref{DUSTY-results}. In case of two
fits, representing a brighter and a fainter phase, both luminosities
are given in the table.

To quantify the uncertainties of the luminosity determination, we take
the uncertainties of the distances and of the bolometric flux
determination into account. Taking the Galactic centre distance
uncertainty and the estimated size of the bulge ($\sim$\,2.8~kpc) into
account, the distance dependent uncertainty of the luminosity is
$\sim$\,35\%. This is of the same order as the uncertainty of the
bolometric flux due to variability and extinction (see
Sect.\,\ref{obsflux}). Therefore, luminosities of individual objects
are only accurate to within a factor of $\sim$\,2.

\begin{figure}
  \centering
  \includegraphics[width=\columnwidth]{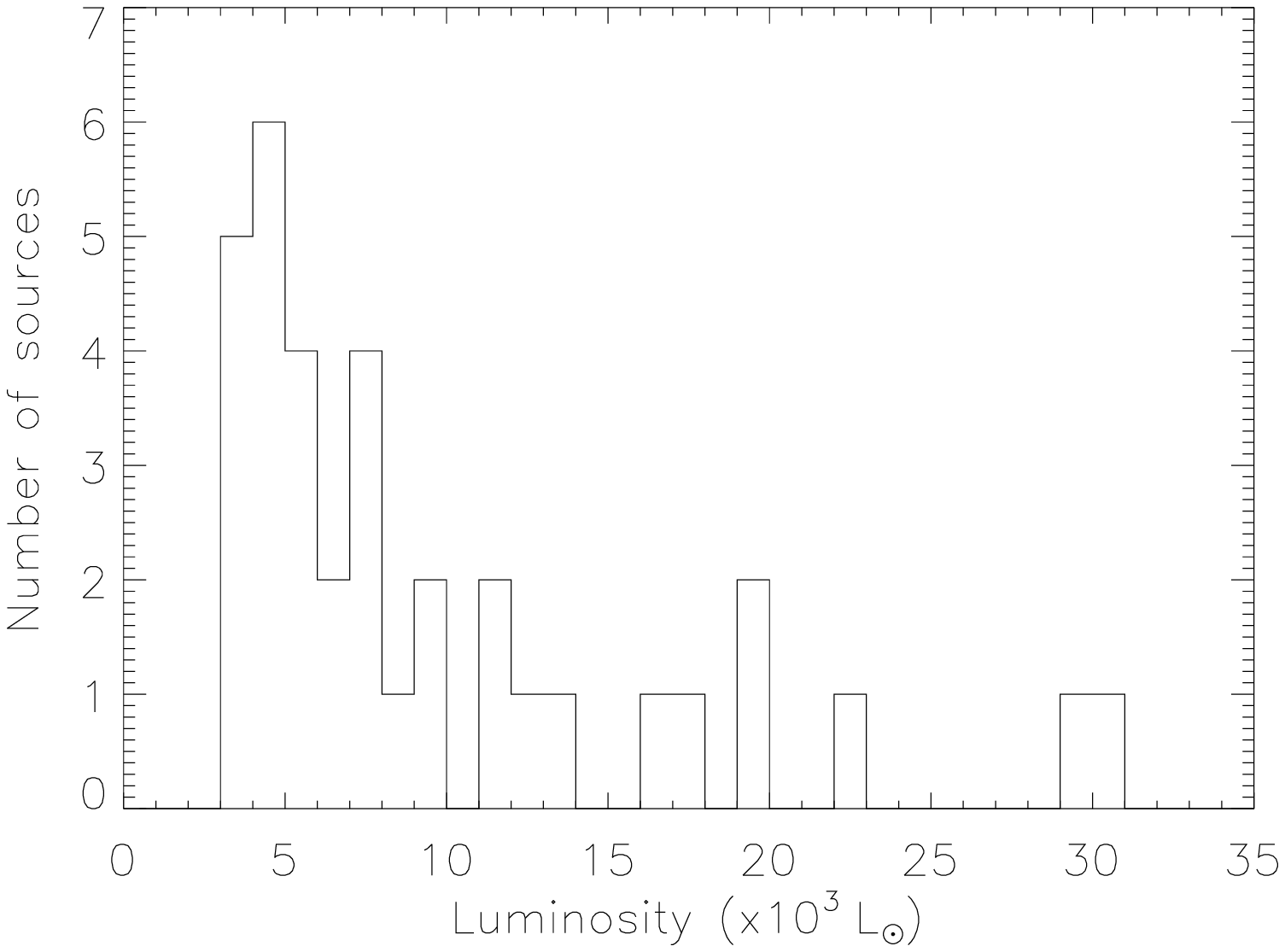}
  \caption{Luminosity distribution of the Galactic bulge sample.
    Luminosities were obtained from the model fitting of the SEDs. In
    the case of two fits for the same source, the average has been
    taken.}
  \label{fig:Labs}	
\end{figure}

Figure \ref{fig:Labs} shows a histogram of the luminosities in our
\GB\ sample. In the case of two model fits, we used the average
luminosity. The range of luminosities found goes from $\sim$\,3000 to
$\sim$\,30,000~\Lsun, peaking at about 4500~\Lsun.

As mentioned in Section 2, the minimum luminosity required by AGB
stars to be present in the current sample (e.g. $F_{25}$\,$>$\,7~Jy)
depends on the optical depth of the shells. Stars at a distance of 8.0
kpc and a SED represented by one of the DUSTY models would need for
$\tau_{9.7\mu m}$\,$>$\,20 merely a minimum luminosity of
2000\,--\,3000~\Lsun. For lower optical depths 5\,$<$\,$\tau_{9.7\mu
  m}$\,$<$\,20 this minimum luminosity rises to
$\sim$\,8500~\Lsun. Luminosities of the order of a few thousand solar
luminosities in Table \ref{DUSTY-results} are therefore found only for
high optical depths $ \tau_{9.7\mu m} > 10$. Stars with CSEs of lower
optical depths could enter the sample only with larger luminosities.

\subsection{Mass-loss rates}

The calculations of mass-loss rates by DUSTY assume that the mass loss
is radiatively driven. The model rates and terminal outflow velocities
are given for stars with a luminosity $10^4$~\Lsun, fixed gas-to-dust
ratio $r_{gd}=200$ and dust grain bulk density $\rho_s=3$
g\,cm$^{-3}$. Using the observed luminosities $L$ (Table
\ref{DUSTY-results}) and the expansion velocities v$_{exp}$ listed in
Table \ref{sample} as proxies for terminal outflow velocities, the
mass-loss rates were adjusted according to the scaling relations given
in the DUSTY manual \citep{Ivezic99}. For stars without OH maser or CO
observations v$_{exp}$\,=\,14 \kms\ was adopted. The mass-loss rates
\Mdot\ obtained, are in the range
$\sim$\,$10^{-5}$\,--\,3$\times$$10^{-4}$~\Myr. The rates for
individual stars are listed in Table \ref{DUSTY-results}. Because of
uncertainties inherent to the DUSTY code of 30\% and the uncertainties
of the luminosities, the error of individual mass-loss rates is also a
factor $\sim$\,2.

\begin{figure}
  \centering
  \includegraphics[width=\columnwidth]{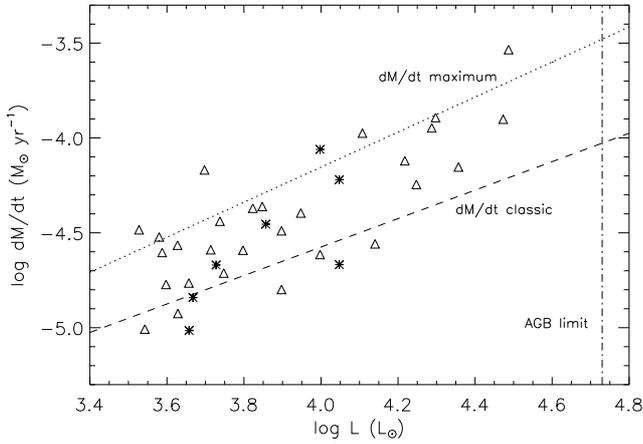}
  \caption{Mass-loss rates and luminosities derived from the DUSTY
    models fitting of the AGB star SEDs in our Galactic bulge
    sample. In cases where ranges of luminosities and mass-los rates
    are given in Table \ref{DUSTY-results}, we plot the average. Stars
    with envelopes of unknown chemistry are plotted with
    asterisks. The classical limit for mass-loss rates caused by
    radiation pressure on dust is shown by a dashed line, and the
    empirical mass-loss rate upper boundary for AGBs and red
    supergiants in the Large Magellanic Cloud found by
    \cite{vanLoon99} is plotted as a dotted line. Also the maximum
    luminosity achievable by AGB stars is marked by the vertical
    dash-dot line.}
  \label{fig:Mdot}	
\end{figure}

The scaling relations adjust the product $r_{gd} \cdot \rho_s$ to
accommodate the observed expansion velocities.  With the DUSTY values
for the gas-to-dust ratio $r_{gd}$ and dust grain bulk density
$\rho_s$, the model outflow velocities are generally only about half
as large as the observed velocities. To increase the model velocities,
the product needs to be decreased, as v$_{exp} \propto L^{0.25} \cdot
(r_{gd} \cdot \rho_s)^{-0.5}$. As $\rho_s=3$ g\,cm$^{-3}$ is already
considered as a conservatively low density for astronomical silicates
\citep{Draine84}, the gas-to-dust ratio is probably smaller.  For the
27 sources with observed expansion velocities (cf. Table
\ref{sample}), $r_{gd} = 44\pm20$ is required to bring the model and
observed expansion velocities into agreement. This value corresponds
to the lower end of the range $50 < r_{gd} < 180$ determined for OH/IR
stars in the Galactic disk by \cite{Justtanont06}.

The mass-loss rates in general are rather high. To check their
reliability, we compared them with those of \cite{Groenewegen06}, who
calculated mid- and far-infrared colours for mass losing AGB and
post-AGB stars based on own models. From the different dust
compositions \cite{Groenewegen06} employed, we used his pure silicate
models for comparison. The range of mid-IR colours he considered
covers only about 40\% (the blue part) of the colours of the GB
sample, but among these the mass-loss rates are in reasonable
agreement. The mean ratio between Groenewegen's and our mass-loss
rates is 0.9\,$\pm$\,0.4.

Figure \ref{fig:Mdot} plots the mass-loss rates against the
luminosities of the sample. Because of scaling relations the two
quantities are not independent from each other, and in the classical
case a relation $\dot{M}_{classic} = L \cdot (c\, v_{exp})^{-1}
\propto$\,$L^{0.75}$ \citep{vanLoon99,Elitzur01} is expected.  The
scatter of the mass-loss rates for a given luminosity is caused mainly
by the scatter of optical depths $\tau_{9.7\mu m}$ required to model
the SEDs and to a lesser degree by the variations of the expansion
velocities. In general, the luminosities and mass-loss rates cover the
range between the classical limit in the $(\dot{M},L)$-plane and the
empirical limit found by \cite{vanLoon99} in the Large Magellanic
Cloud (LMC). The classical limit is determined by the assumption that
the photons transfer momentum to the dust by a single scattering
event.  This limit is superseded however at high optical depths, where
photons experience multiple scattering events, before escaping from
the dust shell. Compliance with this empirical limit was found also
for the AGB stars in the central regions of M33, which have a near
solar metallicity comparable with the Galactic bulge \citep{Javadi13}.

Few stars exceed the empirical limit. For example, IRAS\,17418$-$2713
has the largest mass-loss rate in the sample of $3\times 10^{-4}$
\Myr, and exceeds the empirical limit by a factor $\sim1.5$. The very
steep rise of the SED in the near-infrared requires a high optical
depth for the model dust shell, which leads to a high mass-loss rate
in combination with the inferred high luminosity of
$\sim$\,30,000~\Lsun. However, given the uncertainties for the
luminosities and mass-loss rates, a systematic difference of the upper
limit of the $(\dot{M},L)$-relation of AGB stars in the bulge and the
LMC cannot be inferred.

\section{The nature of the sample}
\label{discussion}

\subsection{The distribution of luminosities and mass-loss rates}

Our sample is made up of the most reddened AGB stars of the \GB\ known
so far. We found a large range of luminosities with a peak in the
luminosity distribution at L\,$\sim$\,4500~\Lsun\ consistent with
low-mass progenitors, and with a tail extending to high luminosities.

This result is in good agreement with other studies of less reddened
AGB stars in the \GB. The luminosity distribution of
\cite{Blommaert98} for a sample of OH/IR stars located close to the
Galactic centre spans a similar range and has a median value of
$\sim$\,4600~\Lsun. The median luminosity of Galactic centre OH/IR
stars monitored by \cite{Wood98b} for variability is
$\sim$\,6400~\Lsun, and their range of luminosities extends to
L\,$>$\,30,000~\Lsun\ consistent with an upper limit for the mass
range $>6$ \Msun. Both studies relied on measurements in the
near-infrared (K, L-band), making the colours of their stars bluer
than in our sample.

Intermediate in colours between the Galactic centre OH/IR stars of the
just above mentioned studies and the colours of our sample are the
OH/IR stars observed by the ISOGAL survey \citep{Ortiz02} and of the
MSX-selected selected sample of AGB stars \citep{Ojha07}. Their
luminosity distributions peak around 8000 \Lsun, and objects with
higher luminosities are found in both samples.  The red IRAS sources
studied by \cite{vanderVeen90} are very similar to our sample. Their
sources cover the colour range 0.0\,$\le$\,$[12]-[25]$\,$\le$\,1.3,
which overlaps in its red part with the colour range
($[12]-[25]$\,$\ge$\,0.75\,mag) of the present sample.  Most of their
sources are bluer than ours, with only seven sources in common. They
find the luminosities to be strongly peaking at 5000--5500 \Lsun\ and
observed a tail of high luminosity sources extending to well above
20,000~\Lsun.

We therefore conclude that the luminosity distributions of the
different samples of AGB stars observed in the \GB\ are consistent
with our result in showing that the majority of the stars have
luminosities well below 10,000~\Lsun, but that stars with luminosities
up to the AGB limit are present as well. There is no evidence for
differences in the luminosity distribution with colour, which would
indicate the presence of mass segregation.

Mass-loss rates in the range from $10^{-7}$ to $10^{-4}$~\Myr\ for
\GB\ AGB stars were derived by \cite{Ojha07}, using the models of
\cite{Groenewegen06}. There are nine sources in common with our
sample. The deviations in the mass-loss rates are a factor
0.6\,--\,1.5, except for IRAS\,17418$-$2713, in which the rates differ
by a factor of 10. \cite{Ojha07} underestimate the rate in this case,
because of their adoption of a mean luminosity for all objects to
scale the model mass-loss rates. IRAS\,17418$-$2713 is actually a
factor $\sim$\,4 more luminous, which explains at least part of the
difference in the mass-loss rates.

\subsection{Predictions from AGB evolutionary models}

High mass-loss rates of the order $10^{-5}$ \Myr\ or higher (dubbed
{\it superwind} hereafter) imply that the stars lose most of their
mass in a relatively short time. Actually the mass-loss rate
determines the stellar lifetime on the AGB. According to the models,
during the evolution on the TP-AGB, lasting $\sim\,$0.25\,--\,2.2
million years, luminosities and mass-loss rates steadily increase,
until the last $\sim$\,60,000\,--\,120,000 years when the stars enter
the superwind phase \citep{Vassiliadis93}. These time ranges vary
slightly depending on the metallicity assumed and how the mass-loss
process is implemented in model calculations
\citep{Blocker95,Marigo07,Weiss09}. The stars of our sample have
mass-loss rates in the range from
($\sim$\,1\,--\,30)\,$\times$$10^{-5}$~\Myr, characteristic of the
superwind phase, and we assume that they will leave the AGB within the
next $\sim$\,$10^5$ years. Hence, a comparison with stellar
evolutionary models has to assign these stars to the low-temperature,
high-luminosity ends of the evolutionary tracks on the AGB for low-
and intermediate mass stars.

The final phases of these tracks on the AGB are characterised by
similar effective temperatures of 2500\,--\,3000~K, and models for
different zero-age main-sequence (ZAMS) masses are then basically
distinguished by the final luminosity reached and in part by their
chemical composition \citep{Marigo07,Weiss09}. While
\cite{Vassiliadis93} and \cite{Blocker95} do not discuss the
conversion of O-rich to C-rich chemistry in their AGB models due to
the third dredge-up, \cite{Marigo07} predict that stars in the mass
range $\sim$\,2.0\,--\,4.0~\Msun\ with solar metallicity
(Z\,=\,0.019), or $\sim$\,1.5\,--\,4.0~\Msun\ with typical metallicity
for the LMC, (Z\,=\,0.008), end as C stars. In \cite{Weiss09}, the
corresponding mass range for solar metallicity is
1.8\,--\,3.0~\Msun\ and, for Z\,=\,0.008, no model with
\Mms\,$<$\,5.0~\Msun\ ends O-rich. For most stars of our sample
(80\%), however, the O-rich chemistry is evident from the presence of
masers and/or a signature of silicates in the infrared. Hence, if
these stars are indeed close to the end of their TP-AGB evolution,
they should have evolved either from intermediate-mass stars with
\Mms\,$>$\,4.0~\Msun\ experiencing HBB, or from low-mass stars of at
least solar metallicity. This conclusion would be avoided by the
suppression of carbon star formation for the rest of the masses, which
may occur for higher than solar metallicity or for increased helium
content \citep{Karakas14}.


\begin{figure}
  \centering
  \includegraphics[width=\columnwidth,angle=0]{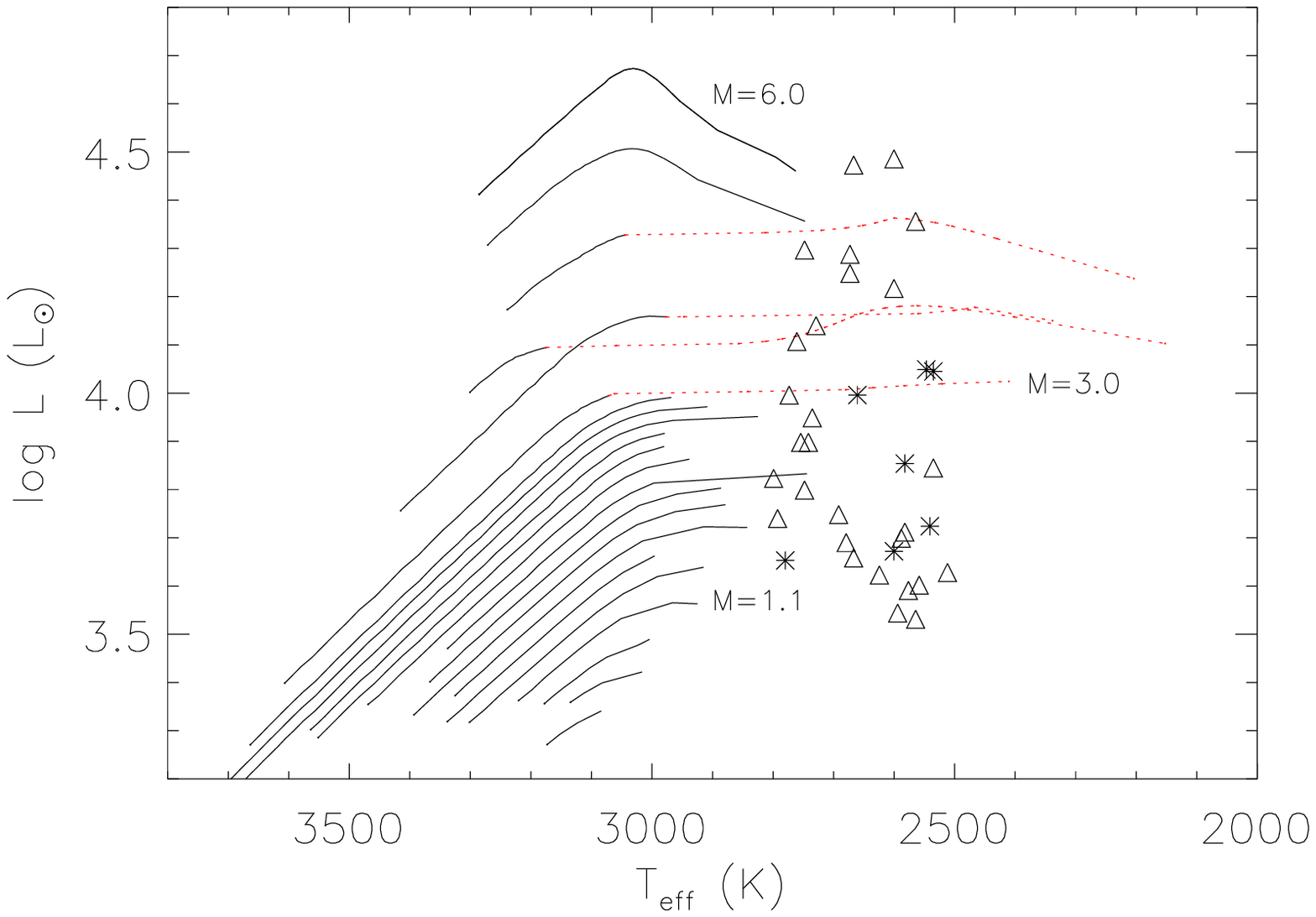}
  \caption{Location of the GB sample stars in the Hertzsprung-Russell
    diagram overlayed on evolutionary tracks on the AGB from
    \cite{Bertelli08,Bertelli09} for solar composition. Tracks are for
    main-sequence masses 0.7\,$\le$\,\Mms\,$\le$\,6.0 \Msun. Segments
    of the tracks, where the photospheric chemical composition is
    C-rich, are shown in red (dotted). Stars with envelopes of unknown
    chemistry are plotted with asterisks. Average luminosities of the
    GB sample stars are taken from Table\,\ref{DUSTY-results}. For the
    unknown effective temperatures random values between 2500 and
    2800~K were adopted.}
  \label{fig:Padova}	
\end{figure}


Figure \ref{fig:Padova} compares the locations of the GB sample stars
in the Hertzsprung-Russell diagram with the models of
\cite{Bertelli08,Bertelli09}, describing the TP-AGB stellar evolution
for ZAMS masses 0.7\,$\le$\,\Mms\,$\le$\,6.0~\Msun, assuming solar
composition (Z\,=\,0.017, Y\,=\,0.26)\footnote{Available at
  http://stev.oapd.inaf.it/YZVAR/}. The effective temperatures
$T_{eff}$ of the stars are not known, but we assume that they are
$<$\,3000~K. To avoid crowding in the figure, we assigned random
effective temperatures to the stars in the range
2500\,$\le$\,$T_{eff}$\,$\le$\,2800~K. Contrary to the model
predictions, the mass range 1.1\,$\la$\,\Mms\,$\la$\,6.0~\Msun\ is
continuously covered by the sample, including the range in which the
stars should end as C-rich. Most of the stars, in which the chemistry of
the envelope is unknown, have relatively low luminosity in line with
our assumption that they are all oxygen rich.

\subsection{The high-luminosity group}

The luminosity distribution of the GB sample (Fig. \ref{fig:Labs}) has
a steep increase at low luminosities, peaks at $\sim$\,4500~\Lsun, and
falls off less steeply towards higher luminosities. For the following
discussion, we name sources with L\,$\ge$\,10,000~\Lsun\ the
`high-luminosity group' and those with L\,$\le$\,7000~\Lsun\ the
`low-luminosity group'.

In the `high-luminosity group' two sources (IRAS~17418$-$2713 and
IRAS~17495--2534) have the highest luminosities
($\sim$\,30,000\,$\pm$\,10,000~\Lsun; Table
\ref{DUSTY-results}). Apart from the possibility that they could be
foreground sources and hence are actually less luminous, their
luminosities are compatible with intermediate-mass stars
(\Mms\,$\sim$\,4\,--\,6~\Msun) experiencing HBB, preventing them from
becoming C stars. Both show the absorption feature from amorphous
silicates at 10\,\micron\ (Table \ref{sample}), confirming their
O-rich chemical composition. Furthermore, IRAS~17418--2713 shows
crystalline silicate emission features in the infrared
\citep{Garcia-Hernandez07}, while IRAS~17495--2534 is the only source
known so far showing these features in absorption
\citep{Speck08}. Speck et al. argue that the rarity of such a
distinctive feature is due to their emergence only in massive AGB
stars, which is consistent with our findings.

The rest of the group, making up about one quarter of the whole GB
sample, has luminosities between 10,000\,--\,25,000~\Lsun. These
luminosities are predicted for the mass range, where for solar
composition, stars end as C-rich. For lower metallicities, this group
will even increase, as conversion to C-rich chemistry also occurs for
smaller masses. However, masers and silicate features in the infrared
are observed in most stars from this group, ruling out a C-rich
chemistry. It is unlikely that a C-rich central star coexists with OH
maser emission in all these stars, because such a configuration can
last only for $\sim$\,1000 years, which is the time needed to replace
the O-rich material in the OH maser shell at
$R_{OH}$\,$\approx$\,5\,$\times$$10^{16}$~cm, assuming typical outfow
velocities of 15~\kms. As the effective temperatures are unknown,
these stars in principle could have temperatures
$T_{eff}$\,$>$\,3000~K, assigning them to an earlier evolutionary
phase, where the chemical composition is still O-rich, and the
luminosities are almost the same (cf. Fig. \ref{fig:Padova}). However,
none of the model calculations
\citep{Vassiliadis93,Blocker95,Marigo07,Weiss09} predict the observed
high mass-loss rates at less advanced phases of TP-AGB evolution.

The presence of O-rich objects at the tip of the AGB with
luminosities, corresponding to masses where the stars should be
already converted to C stars, may have two possibility. Either phases
of very high mass-rates can occur, before the O\,$\longrightarrow$\,C
transition is experienced, or for some of the stars from this mass
range the transition might be delayed or avoided. The first
possibility is unlikely except in the case of higher mass stars,
because it is likely that very little mass is left over after a
superwind phase to continue the evolution with further thermal pulses,
allowing an O\,$\longrightarrow$\,C transition before the stars leave
the AGB.  Therefore, the second possibility may be more rewarding to
follow. \cite{Weiss09} find that models with super-solar metallicities
can avoid the conversion to C/O\,$>$\,1 before the end of the AGB
simply because they start with smaller C/O ratios, and therefore more
thermal pulses are needed to reach C/O\,$\sim$\,1. Also,
\cite{Karakas14} finds that at a super-solar metallicity Z=0.03, no
carbon stars are formed except in a narrow mass range of 3.25--4
\Msun. Therefore, in environments where a mix of metallicities is
present, a mix of stars originating from similar ZAMS masses, but
with different chemistries, might also be present. It would imply that most
of the objects from the GB sample were born in bulge regions with
higher than solar metallicities. A similar line of argument was
forwarded by \cite{Jura93} to explain the differences in the period
distribution of Mira variables in the solar neighbourhood and the
Galactic centre.  They suggested that Mira variables have higher
metallicity in the Galactic centre and are therefore oxygen rich,
while stars of the same mass in the solar neighbourhood are less
metal rich and already converted into carbon stars.


\begin{figure}
  \centering
  \includegraphics[width=\columnwidth,angle=0]{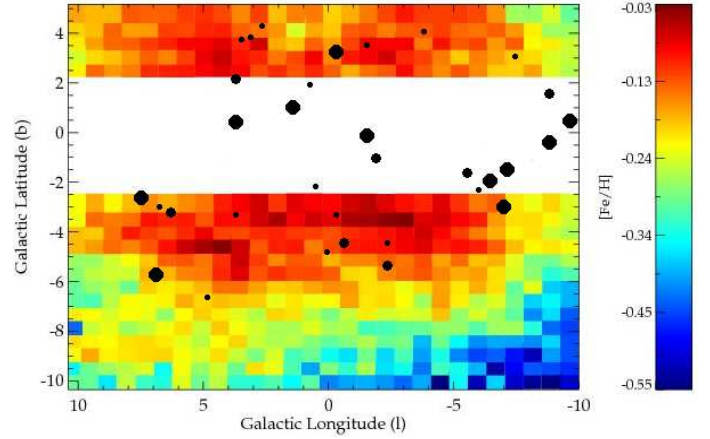}
  \caption{Distribution of the GB sample stars overlayed on the
    distribution of mean metallicities in the \GB\ of
    \cite{Gonzalez13}. The different symbol sizes mark the members of
    the low- and high-luminosity group and the stars with luminosities
    in between.}
  \label{fig:metallicities}	
\end{figure}


The metallicity argument can be tested for consistency in comparison
with the spatial distribution of the GB sample with the global
metallicity distribution in the \GB. Fig. \ref{fig:metallicities}
shows the metallicity map of \cite{Gonzalez13}, which clearly confirms
the vertical metallicity gradient with metal-rich stars dominating the
inner bulge in regions closer to the Galactic plane ($\mid b\mid <
5^{\circ}$). The GB sample members are overplotted. They are
completely confined to the more metal rich part of the bulge. Although
the mean metallicities in the $30^\prime \times 45^\prime$ bins of the
metallicity map do not reach super-solar metallicities, higher
metallicities may occur on smaller spatial scales, where the GB sample
members may be located. Therefore, at least this comparison does not
contradict the model predictions that the chemistry of
intermediate-mass stars at the termination of TP-AGB evolution depends
on the metal content of the environment in which they were born.

\subsection{The low-luminosity group}

Almost half of the sample has luminosities L\,$<$\,7000
\Lsun\ (log\,L\,$<$\,3.85) and mass-loss rates
(1\,--\,5)\,$\times$$10^{-5}$~\Myr. These kinds of combinations of
relatively low luminosities and very high mass-loss rates are not
predicted by the models either, at least as long the stars are
considered to be on the TP-AGB and in between two thermal pulses (the
interpulse phase). Assuming that the stars are in the final phase of
their AGB evolution (the last couple of interpulse periods) and making
the requirement that the luminosities do not surpass 7000~\Lsun, these
stars must have descended from
1.1\,$\la$\,\Mms\,$\la$\,1.8~\Msun\ main-sequence stars
(Fig. \ref{fig:Padova}).

In models of TP-AGB evolution, mass-loss rates exceeding
$10^{-7}$~\Msun\ are not assumed to occur before large amplitude
pulsations of the stellar envelope have been established. Most members
of the GB sample have not been monitored for variability, except the
few sources in common with \cite{vanderVeen90}, for which they report
large amplitude variability with periods P$>750$ days. The other
sources usually have high IRAS variability index ($>95$, see Table
\ref{sample}), indicating that they are also, in the majority, large
amplitude variable stars. We therefore exclude the remote possibility
that the members of the low-luminosity group are intermediate-mass
stars (\Mms\,>\,3.0~\Msun) on the early AGB preceding the TP-AGB,
although the observed luminosities (L\,$<$\,7000~\Lsun) are achieved
by models during this phase \citep{Bertelli08}.

During the TP-AGB, the mass-loss rates steadily increase, modulated by
the intervening thermal pulses, which affect the surface luminosity
and radius of the stars. During the interpulse phase, the stellar
luminosity reaches its maximum shortly before the next thermal pulse
happens and is referred here to as quiescent luminosity. The models of
\cite{Vassiliadis93} provide the mass-loss rates ($>$\,$10^{-5}$~\Myr)
observed for \Mms\,$>$\,2.0~\Msun\ during the last couple of
interpulse periods, but the correspondent model luminosities are too
high. \cite{Blocker95} presents models with \Mms\,=\,1, 3, 4, 5, and
7~\Msun. While the 1~\Msun\ model does not reach the TP-AGB phase, the
\Mms\,=\,3 \Msun\ model has quiescent luminosities in the required
range, but does not reach mass-loss rates surpassing
$10^{-5}$~\Myr. Models for larger masses do reach the required
mass-loss rates, but are again too luminous. \cite{Marigo07} and
\cite{Weiss09} also arrive at the same result: the high observed
mass-loss rates are not predicted for the mass ranges corresponding to
the observed quiescent luminosities. Given the short times the thermal
pulses last ($\sim$ several hundred years, \citealt{Vassiliadis93}), it
is unlikely that all these stars are currently experiencing a boost of
the mass-loss rate following the brief increases of the luminosity.

A possible explanation for the discrepancies between the observed
properties and the model predictions may originate in the assumptions
concerning the relation between mass-loss rates and other properties
in the TP-AGB models. As no complete mass-loss theory for AGB stars is
available, mass-loss rates are empirically coupled to pulsational
period. The periods are theoretically determined from relations
involving mass and radius, which are provided by the models. While
these relations provide resonable values in general, there might be
brief phases where the predictions are insufficient. At least for the
low luminosity part of our sample, which had main-sequence masses
\Mms\,$<$\,2~\Msun, the models predict mass-loss rates that are too
low in the brief final evolutionary phase.


\begin{figure}
  \centering
  \includegraphics[width=\columnwidth,angle=0]{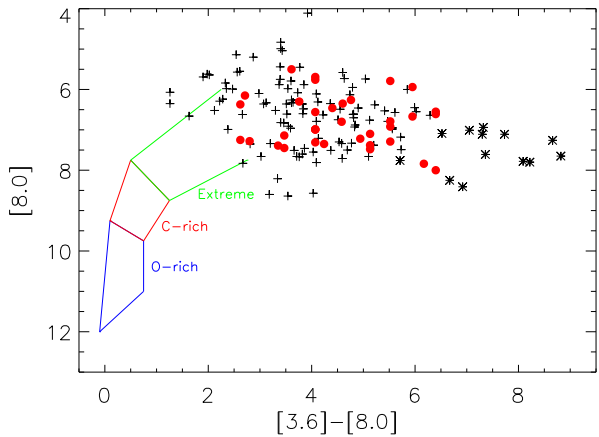}
  \caption{Location of the GB sample stars (red filled circles) in the
    [3.6]--[8.0]\,vs.\,[8.0] colour-magnitude diagram assuming a
    distance corresponding to the LMC. The boundaries of the locations
    of the different type of AGB stars (O-rich, C-rich and extreme) as
    defined by the SAGE project are taken from \cite{Blum06}. In
    addition, the `evolved stars' in the LMC (plus signs) identified
    by \cite{Gruendl09}, and the spectroscopically confirmed C stars
    (asterisks) from \cite{Gruendl08} are overplotted.}
  \label{fig:SAGE-CMD}	
\end{figure}


\subsection{Comparison with the AGB population in the LMC}
 
In the LMC, the AGB population has been studied extensively by the
SAGE project \citep{Blum06}. The chemistry of the population was
determined from photometry using basically the J--[3.6] vs. [3.6]
colour-magnitude diagram (CMD). In Fig. \ref{fig:SAGE-CMD} the
boundaries for different chemistries are delineated in the
[3.6]--[8.0] vs. [8.0] CMD, which is more appropriate for very
obscured stars. While C stars are in general found to be redder than
O-rich stars, Blum et al.  also define a group of `extreme stars' with
even redder colours ($J-[3.6]$\,$>$\,3.1 mag), for which the chemistry
could not be determined. The few known stars of this group, which have
been observed spectroscopically, are predominantly C-rich. Along with
the model predictions that the lowest effective temperatures
and hence the largest mass-loss rates are achieved when the stars
become C-rich, the `extreme stars' are uniformly classified as
predominantly C-rich \citep{Srinivasan11,Riebel12}.  However, a
minority population of O-rich stars is not excluded and contamination
with red supergiants and young stellar objects is probably also
present \citep{Srinivasan09,Blum14}.

We determined synthetic apparant magnitudes in the [3.6] and [8.0]
bands of the GLIMPSE photometric system \citep{Churchwell09} for all
stars of the GB sample using the DUSTY model fits. Comparison with the
actually observed magnitudes by GLIMPSE gives an uncertainty of
$\sim$\,1~mag. To compare the location of the sample in the CMD of
\cite{Blum06}, a distance modulus of 4.0 mag was used to account for
the distance difference between the LMC and the Galactic bulge. As
shown in Fig. \ref{fig:SAGE-CMD}, the GB sample stars are located in a
sparsely populated region extending the branch populated by the
`extreme stars' towards redder colours $[3.6]-[8.0]$\,$>$\,2.5
mag. This region of the CMD was also found by \cite{Gruendl09} to be
populated by `evolved stars'. The reddest of them (dubbed EROS for
`Extremely Red ObjectS') were spectroscopically observed by
\cite{Gruendl08}, showing C-rich chemistry. It follows that while the
SAGE `extreme stars' and the even redder `evolved stars' of
\cite{Gruendl09} in the LMC are predominantly C-rich, the
corresponding population in the Galactic bulge represented by the GB
sample seem to be O-rich.

\cite{Riebel12} determined dust mass-loss rates for the SAGE stars
using model
SEDs. Almost all `extreme stars' were matched with C-rich models. Most
dust mass-loss rates for O-rich stars are in the range
$10^{-11}$\,--\,$10^{-9}$~\Myr, and for C-rich stars $>$\,$10^{-10}$
\Myr. With a gas-to-dust ratio of $\sim$\,200 their total mass-loss
rates are well below the rates achieved during the superwind. The
separation of O-rich and C-rich sources in the SAGE CMDs is very
likely valid only as long as mass-loss rates are low, the CSEs are
optically thin and therefore the stellar temperature is able to
influence the infrared colours. As soon as the mass-loss rates
increase, and especially during superwind phases, the stars migrate in
CMDs towards redder colours, independent of their current
chemistry. \cite{Gruendl08} find a mean luminosity of 7100~\Lsun\ for
their EROS sources, and conclude that these stars must have descended
from 1.5\,--\,2.5 \Msun\ main-sequence stars. The total mass-loss
rates derived are in the range of
(4\,--\,23)\,$\times$\,$10^{-5}$~\Myr, much larger than derived for
the SAGE AGB stars. The EROS luminosities, main-sequence masses, and
mass-loss rates are in perfect agreement with those derived here for
the GB sample. The main difference is the chemical composition. As we
previously concluded for the GB sample, \cite{Gruendl08} noted
that the mass-loss rates of the EROS sources typically exceed the
maximum expected for both O-rich and C-rich AGB stars with
luminosities in the observed range.

\section{Conclusions}
\label{conclusions}

The reddest bright 25~\micron\ sources towards the Galactic bulge with
IRAS colours expected for AGB stars were found to span a luminosity
range from $\sim$\,3000 to $\sim$\,30,000~\Lsun\ and mass-loss rates
between $\sim$\,$10^{-5}$\,--\,3\,$\times$\,$10^{-4}$~\Myr. For the
lower luminosity group with L\,$<$\,7000~\Lsun\ the main-sequence
progenitor masses are in the range 1.1\,--\,1.8~\Msun, assuming solar
metallicity.  With mass-loss rates
(1\,--\,5)\,$\times$\,$10^{-5}$~\Myr\ their stellar envelopes are lost
in mere 10,000\,--\,200,000 years and AGB evolution is terminated.
One concludes that the heavily obscured AGB stars do not necessarily
descend from massive AGB stars (\Mms\,$>$\,4~\Msun) only. At higher
luminosities (10,000\,--\,25,000~\Lsun), AGB evolutionary models
predict that for metallicities similar to the Sun or less these
high-mass loss stars should already be converted to C-rich
chemistry. On the contrary, in the bulge these stars are mostly
O-rich. The models predict that the conversion to C-rich chemistry can
be delayed or inhibited for higher than solar metallicities. Our
sources might have been born in regions of the bulge, where several
generations of stars have leftover processed matter with higher than
solar metallicity. Their spatial distribution favourably coincides with
regions of higher mean metallicities in the \GB.


\begin{acknowledgements}
This paper has benefitted from discussions in depth with the referee
J. Th. van Loon. This work was partially funded by the Spanish MICINN
under the Consolider-Ingenio 2010 Program grant CSD2006-00070 First
Science with the GTC\footnote{\tt
  http://www.iac.es/consolider-ingenio-gtc}, and through grants
AyA2011-24052. This work has been supported by the CoSADIE
Coordination Action (FP7, Call INFRA-2012-3.3 Research
Infrastructures, project 312559). FJE acknowledges financial support
from the ARCHES project (7th Framework of the European Union, n
313146). Support was also granted by Deutsche Forschungsgemeinschaft
through grant EN~176/30. \\

\end{acknowledgements}

\bibliographystyle{aa} 
\bibliography{/pcdisk/muller/fran/RESEARCH/bibliography/references} 


\begin{figure}
\label{fig:SEDs}
\resizebox{\hsize}{!}{\includegraphics{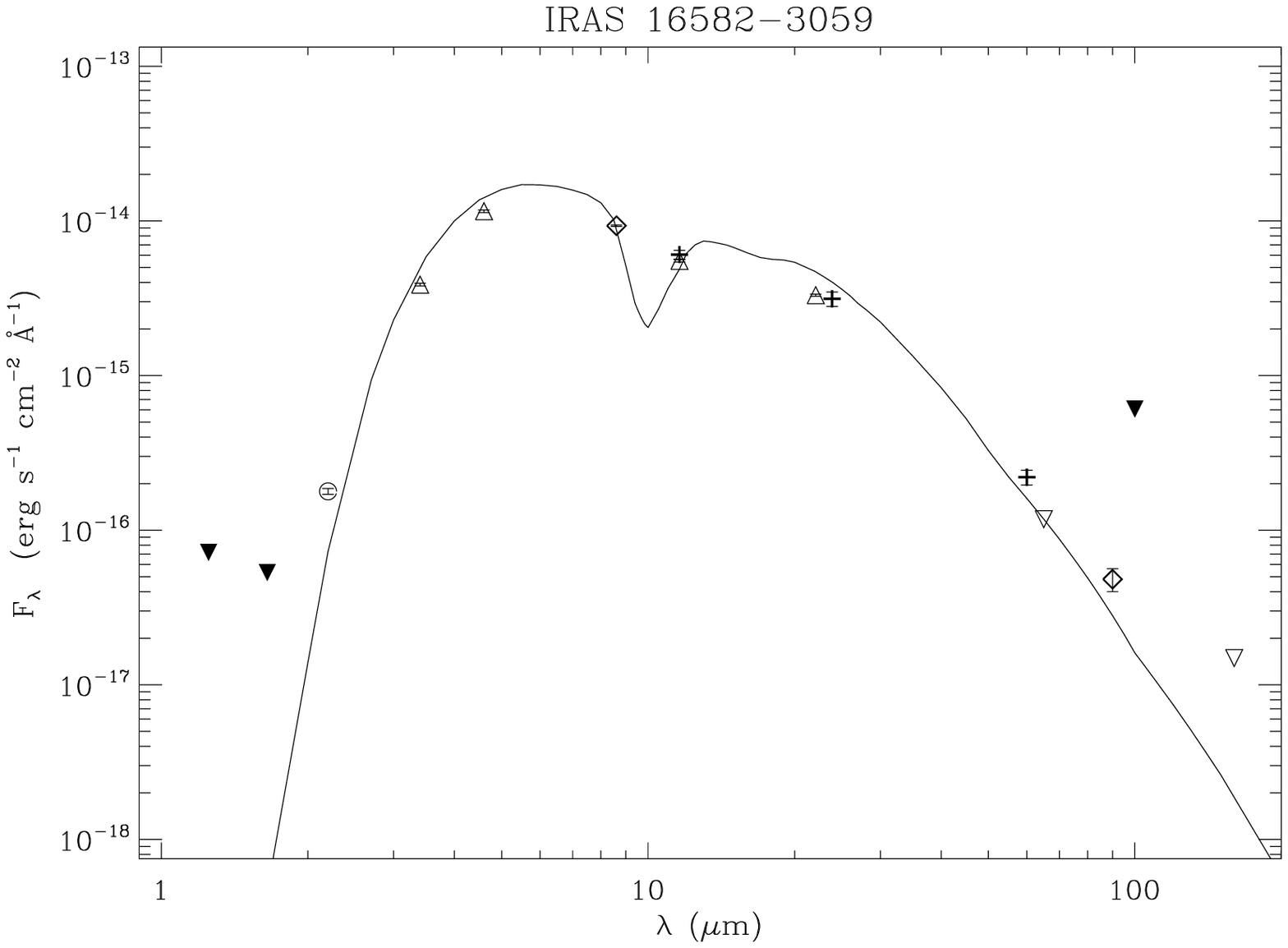}}
\resizebox{\hsize}{!}{\includegraphics{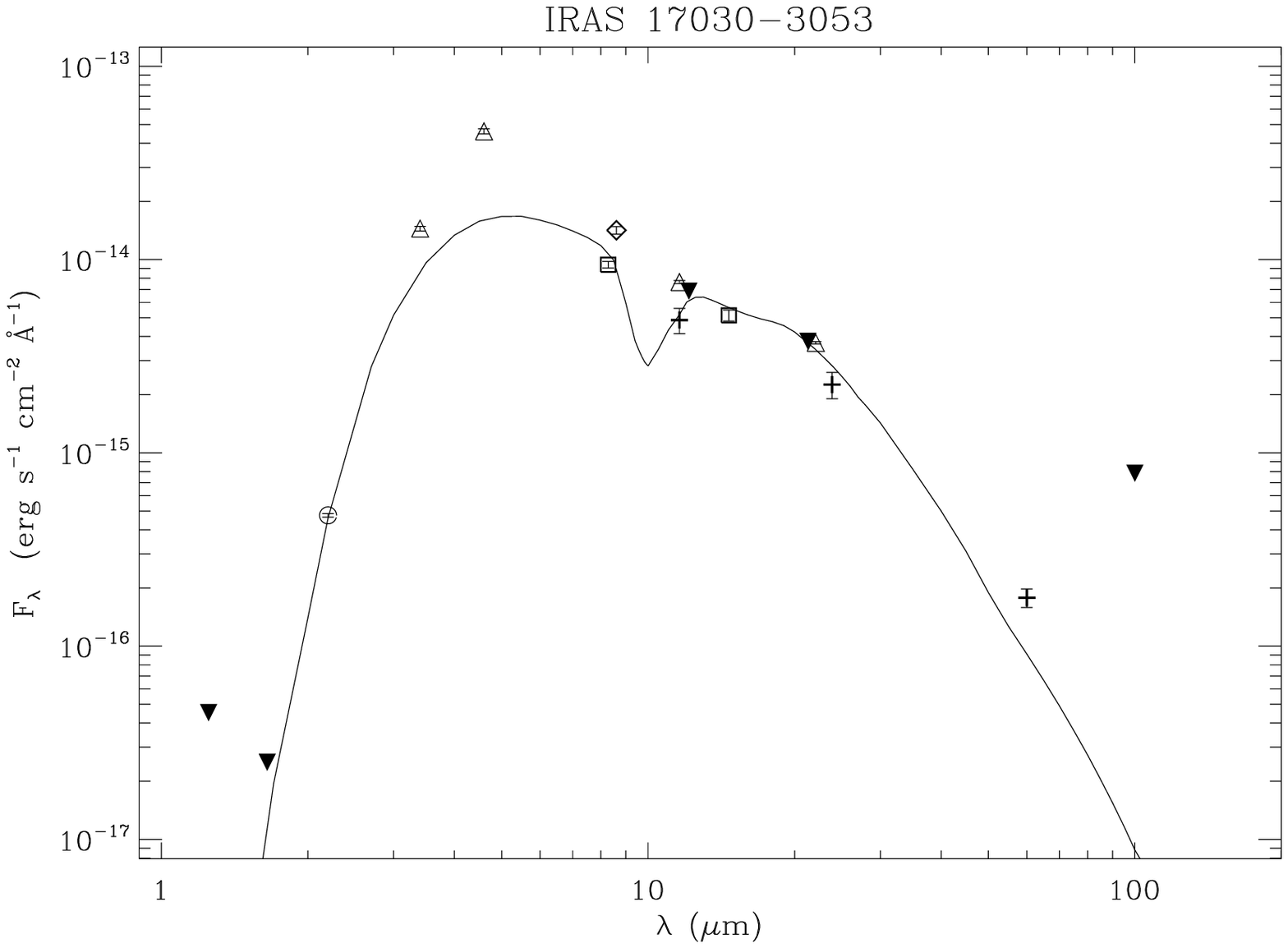}}
\resizebox{\hsize}{!}{\includegraphics{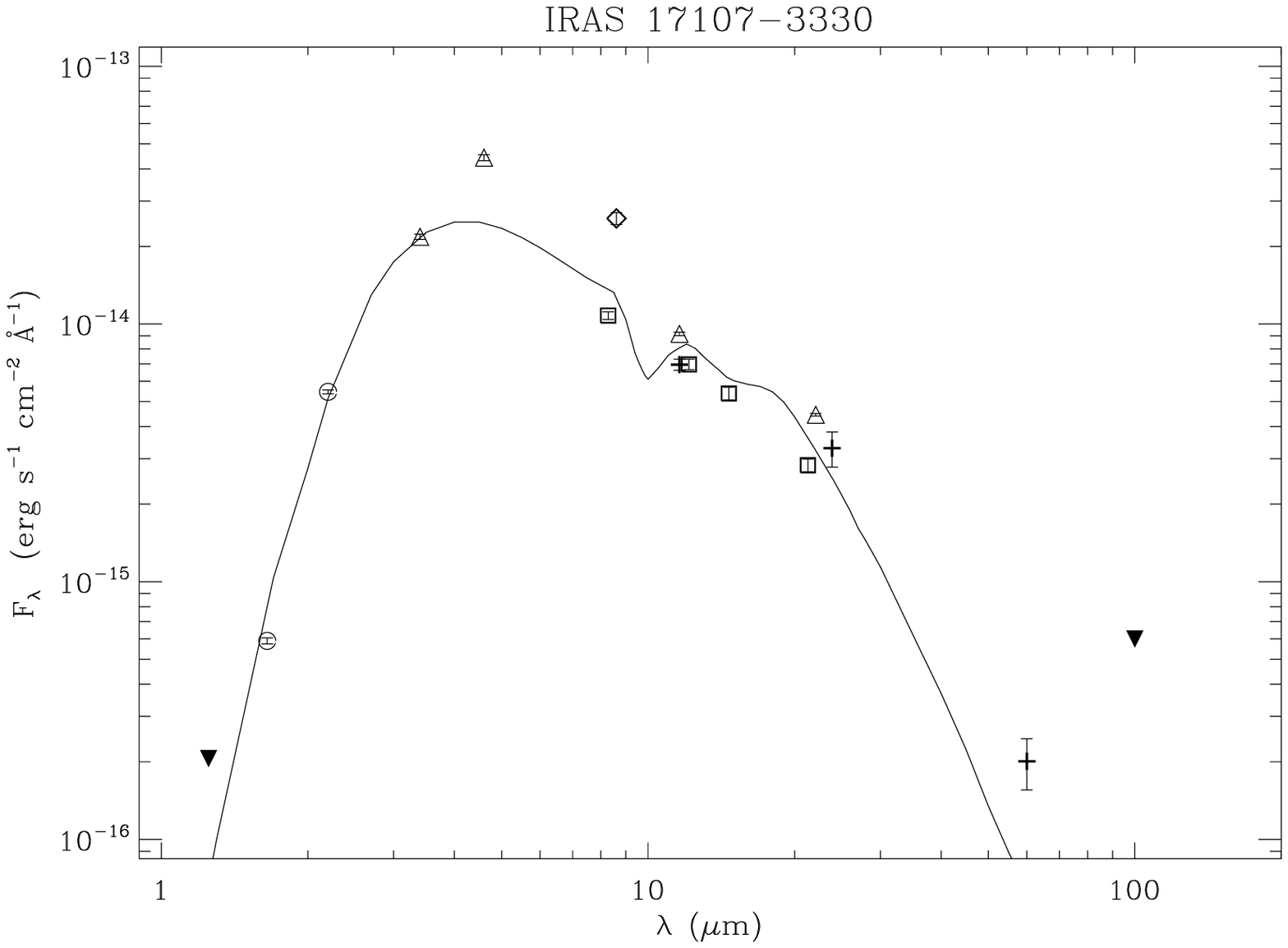}}
          \caption{Observation and model fits after correcting for
            interstellar extinction. 2MASS: circles; VVV-DR1: stars;
            WISE: triangles pointing up; GLIMPSE: sails; MSX: squares;
            AKARI: diamonds; IRAS: crosses; Upper limits: filled
            triangles pointing down; Unconfirmed detection: open
            triangles pointing down; Model: solid line (in case of two
            models, the solid line represents the brighter and the
            pointed line the fainter phase (see Section
            \ref{obsflux})).}
\end{figure}
\addtocounter{figure}{-1}
\begin{figure}
\resizebox{\hsize}{!}{\includegraphics{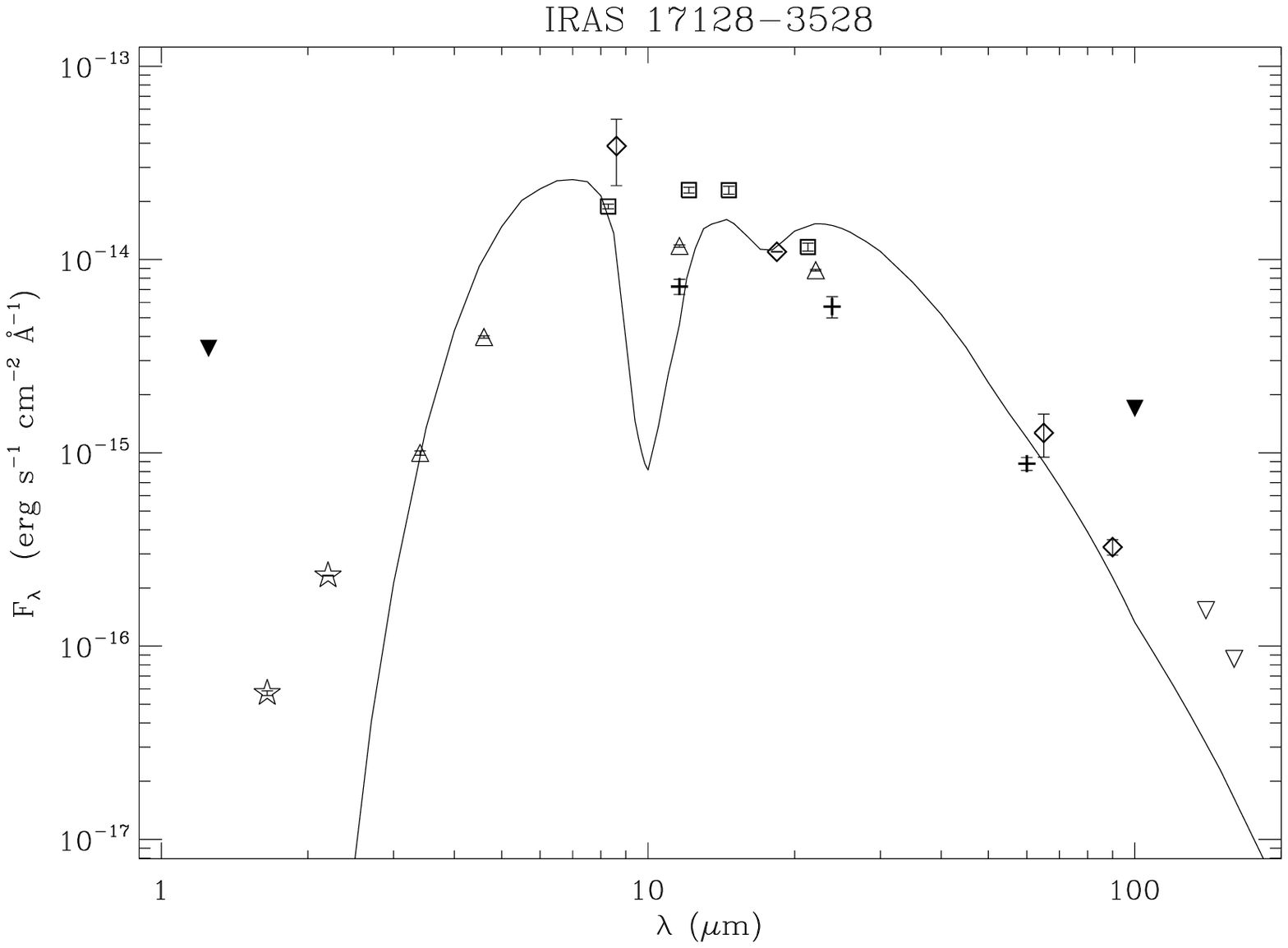}}
\resizebox{\hsize}{!}{\includegraphics{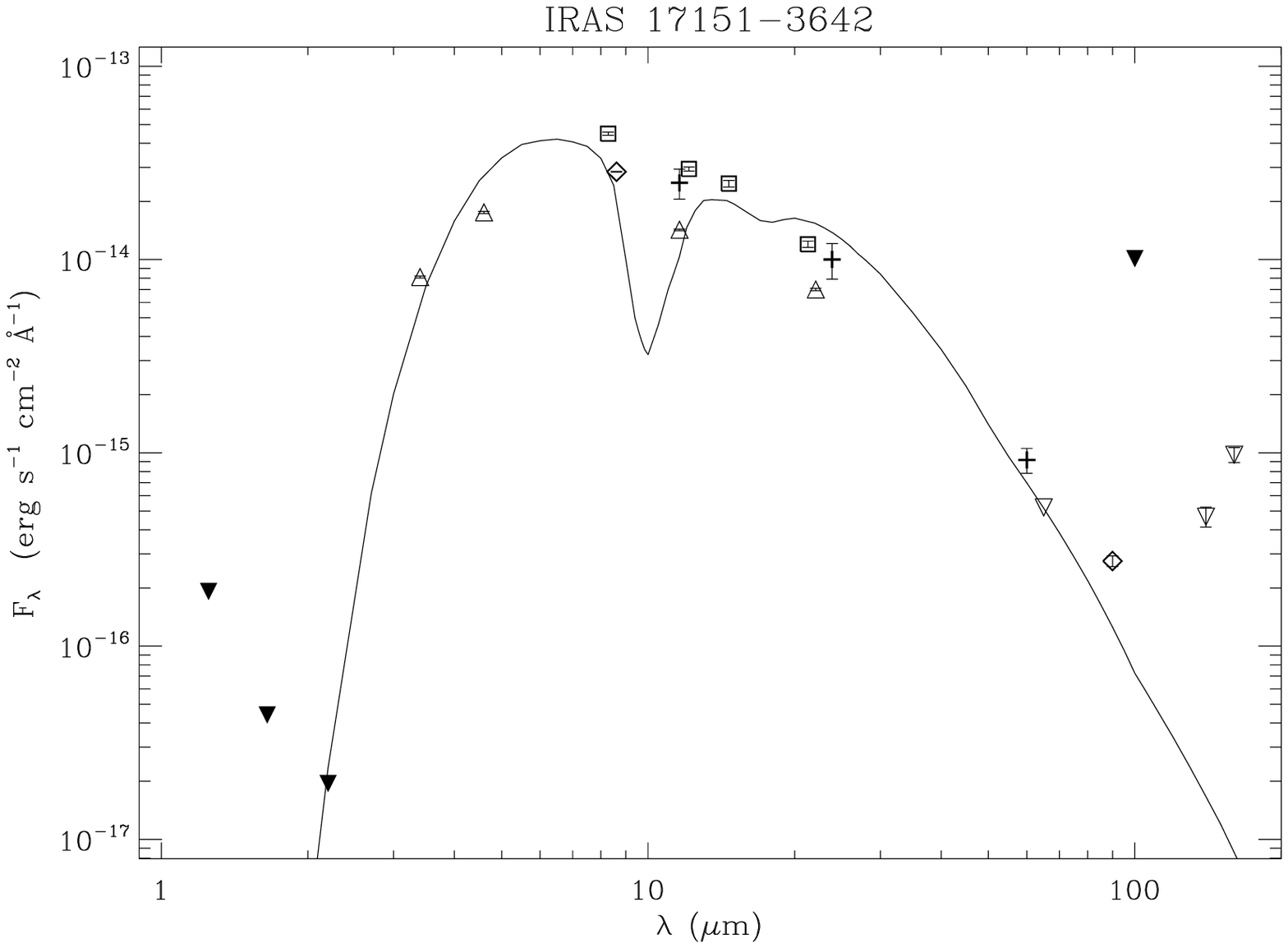}}
\resizebox{\hsize}{!}{\includegraphics{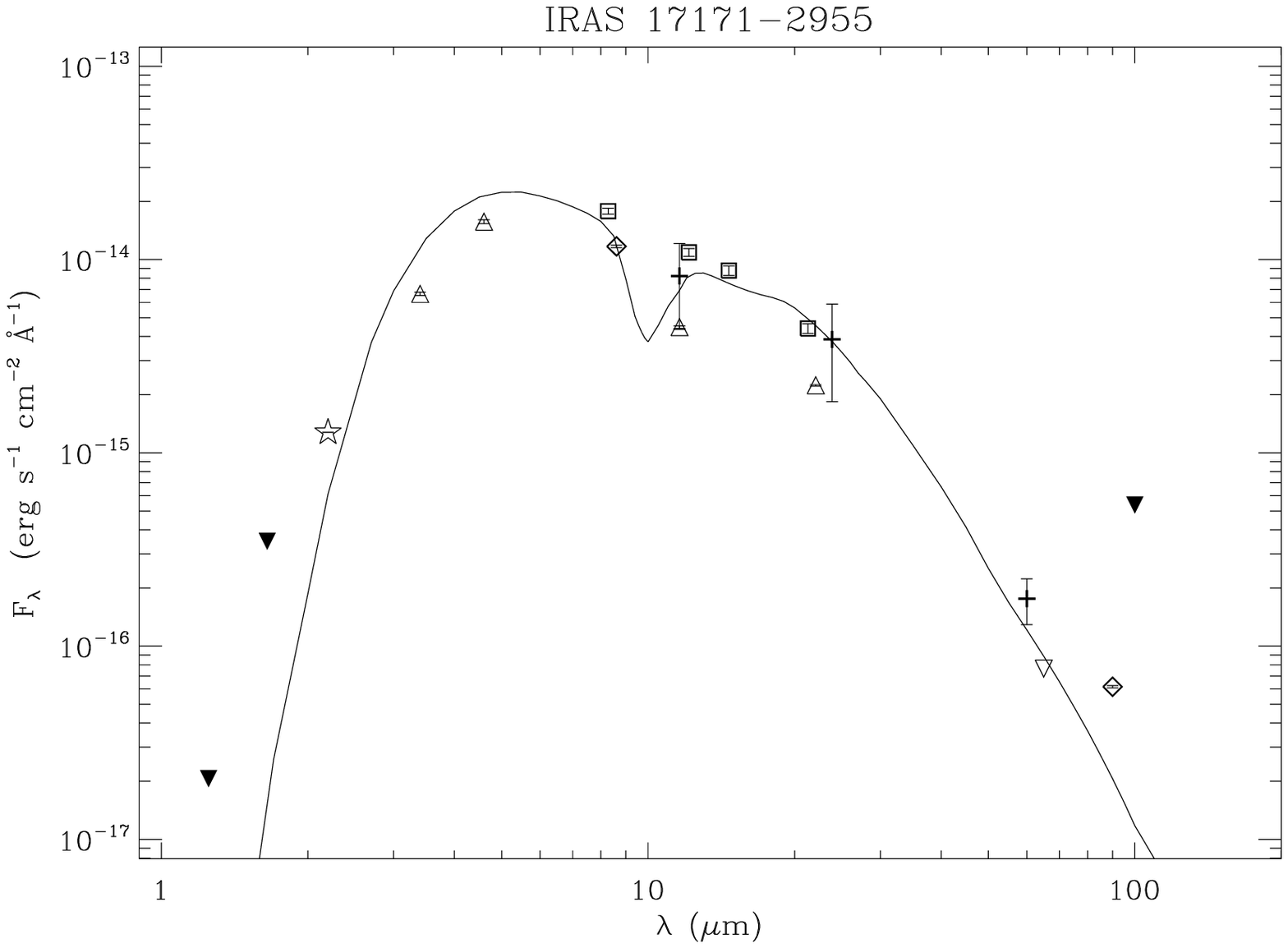}}
\resizebox{\hsize}{!}{\includegraphics{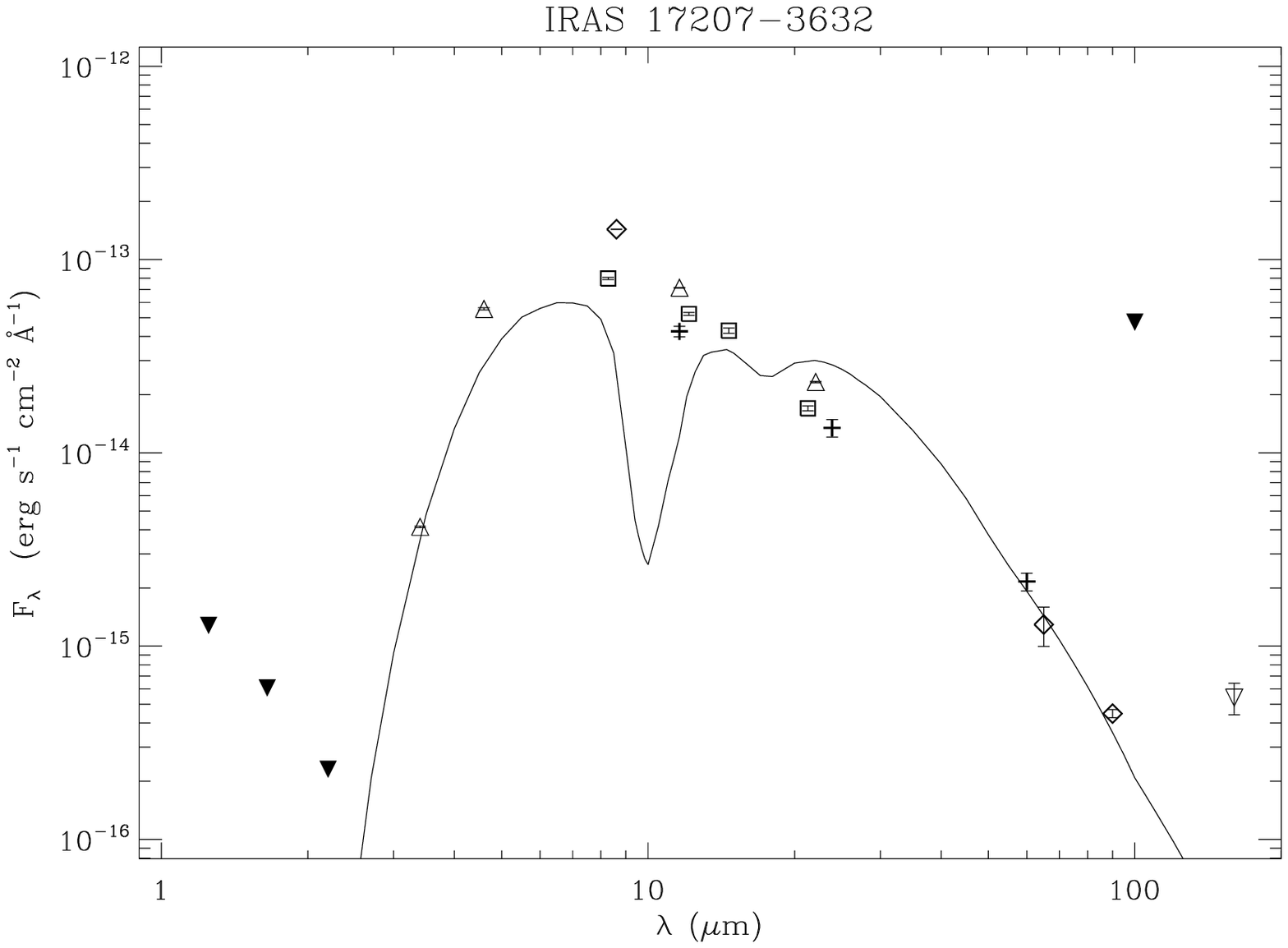}}
          \caption{continued.}
\end{figure}

\addtocounter{figure}{-1}
\begin{figure}
\resizebox{\hsize}{!}{\includegraphics{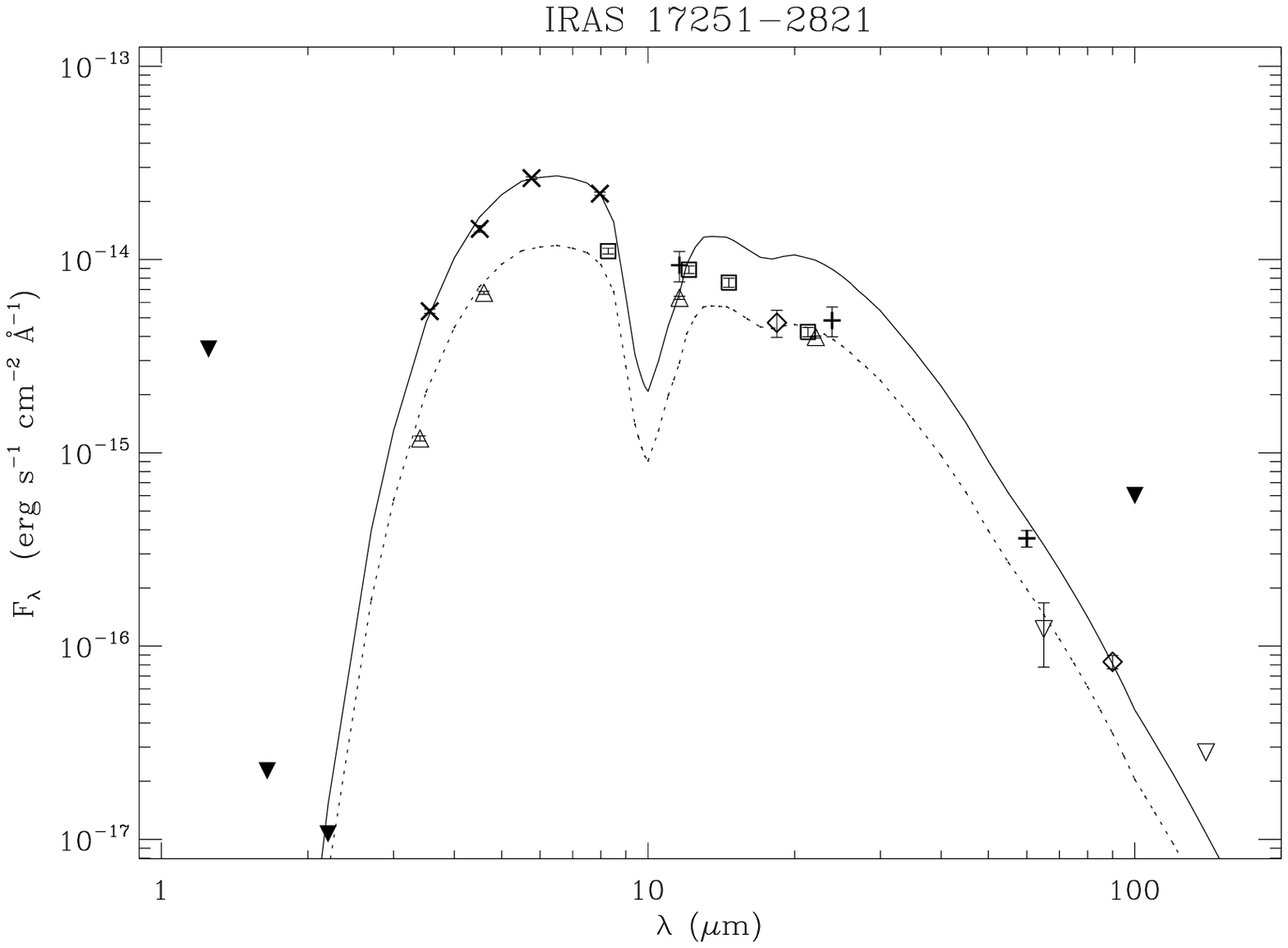}}
\resizebox{\hsize}{!}{\includegraphics{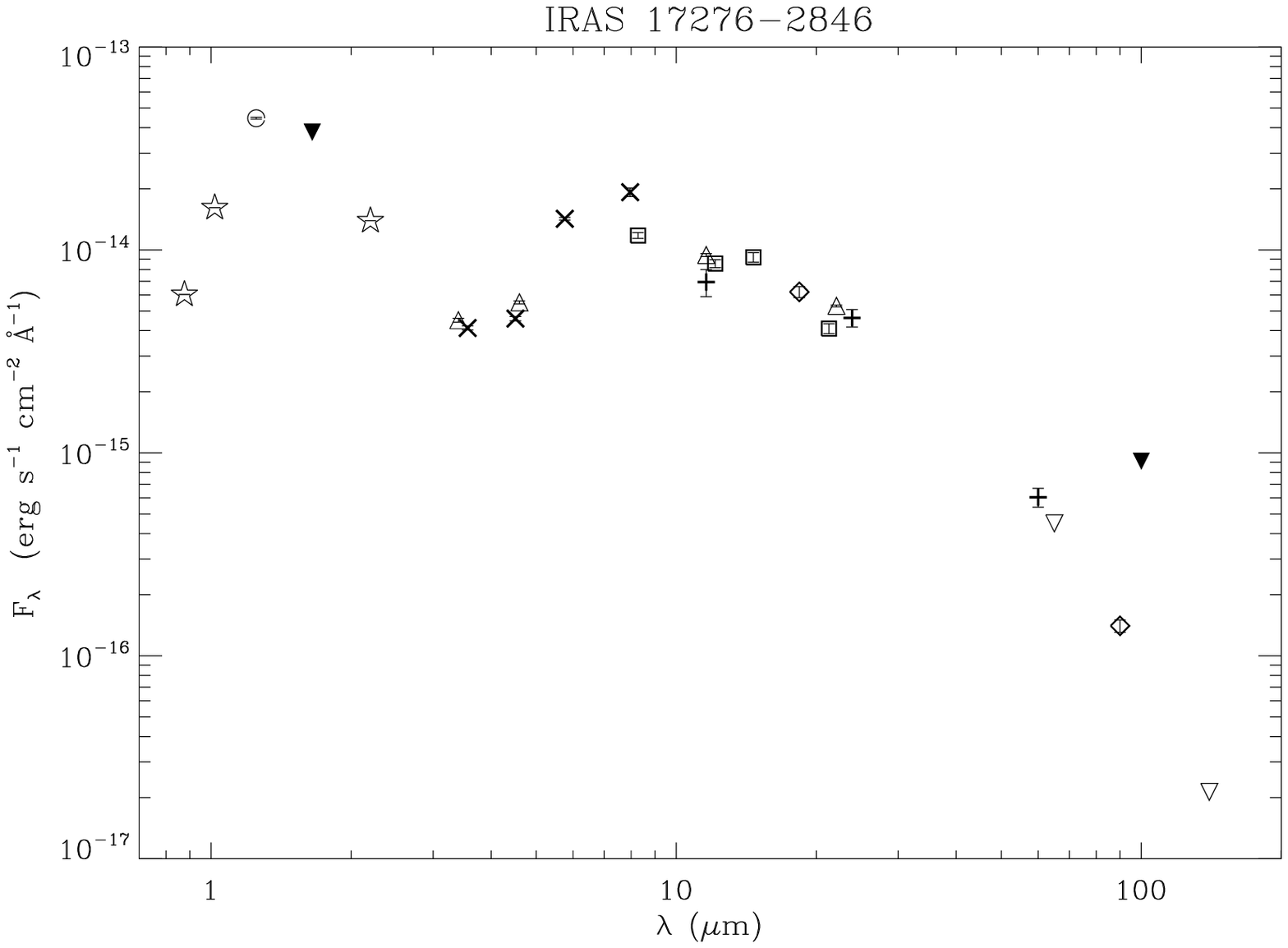}}
\resizebox{\hsize}{!}{\includegraphics{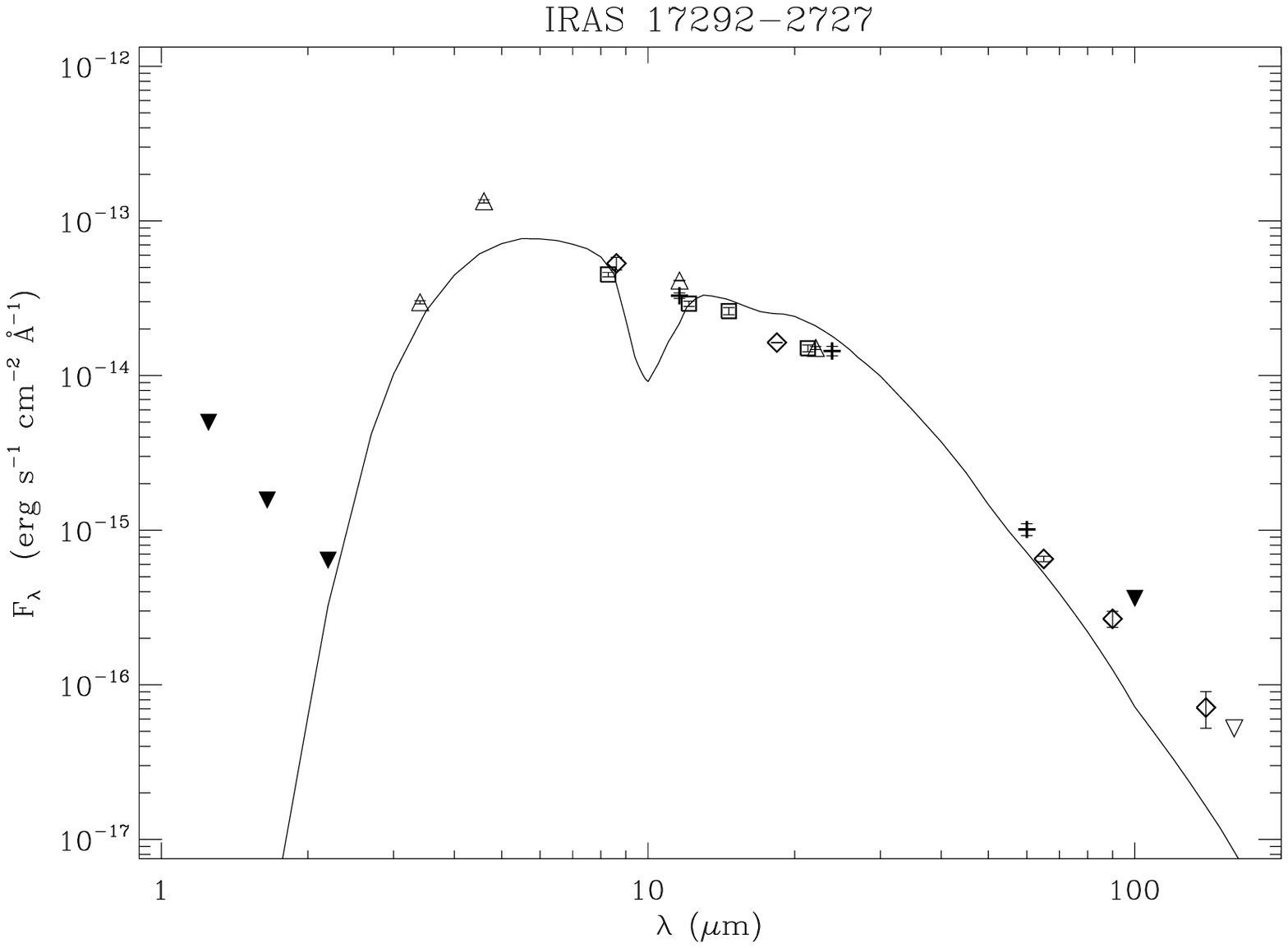}}
\resizebox{\hsize}{!}{\includegraphics{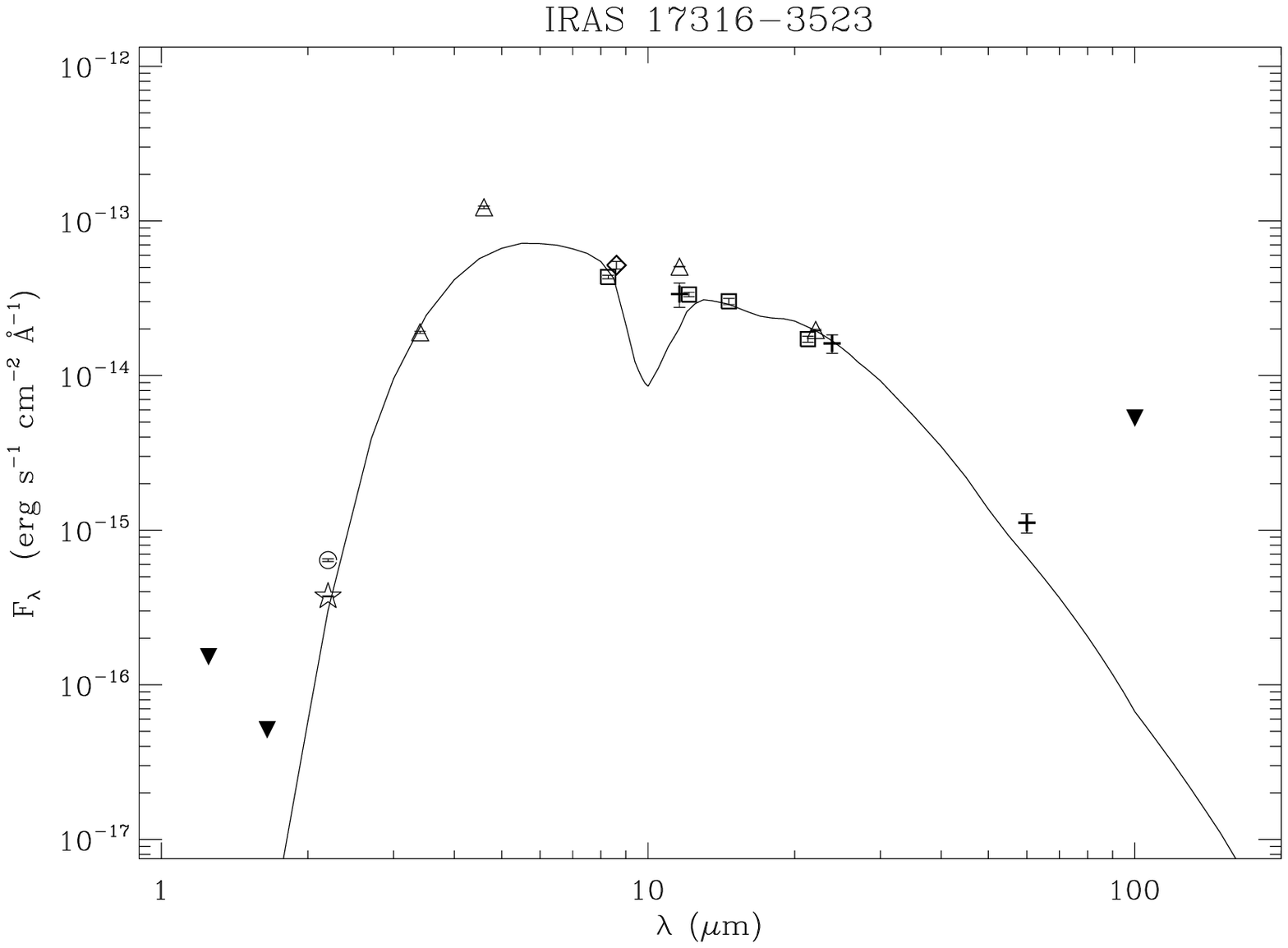}}
          \caption{continued.}
\end{figure}

\addtocounter{figure}{-1}
\begin{figure}
\resizebox{\hsize}{!}{\includegraphics{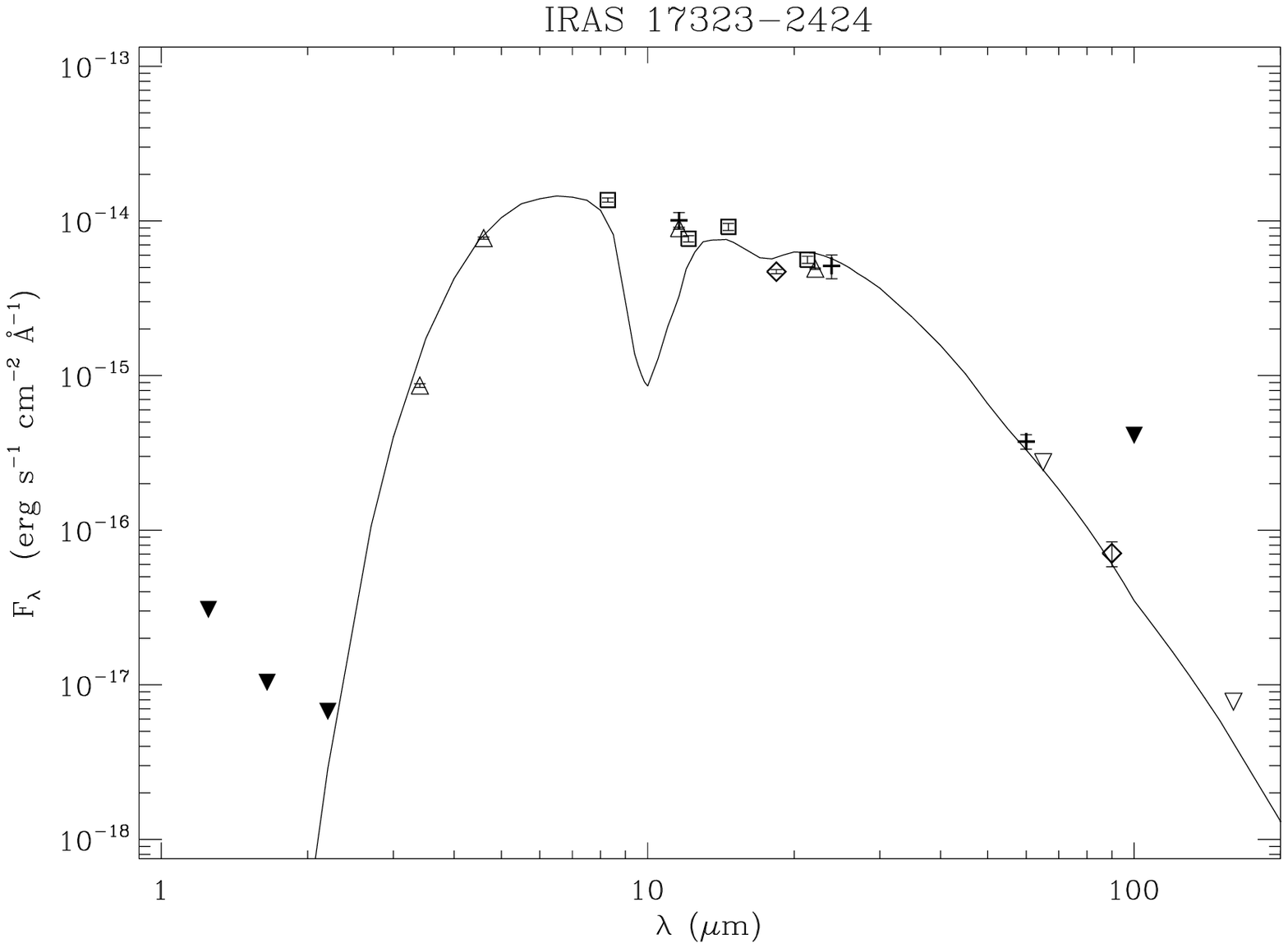}}
\resizebox{\hsize}{!}{\includegraphics{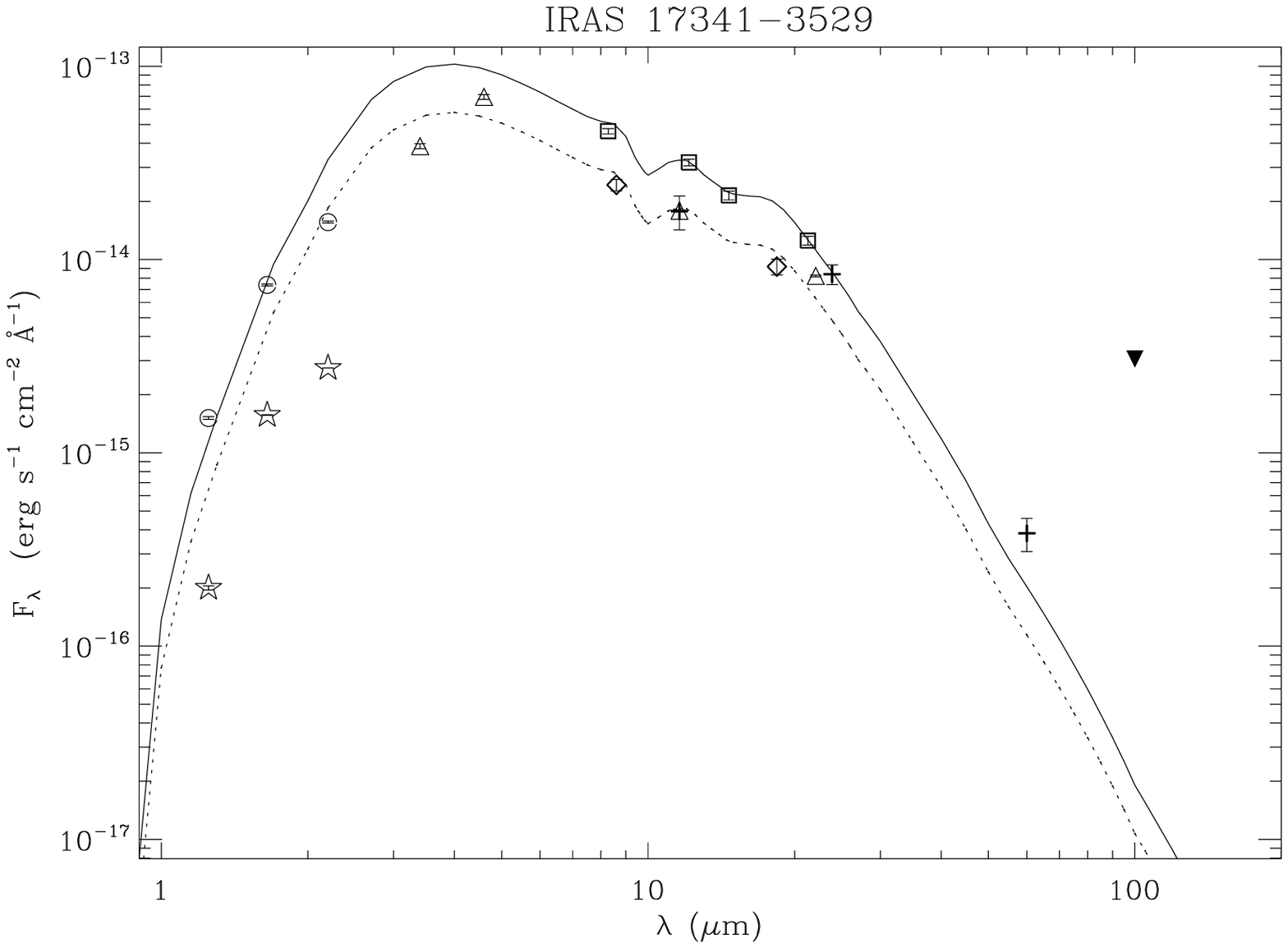}}
\resizebox{\hsize}{!}{\includegraphics{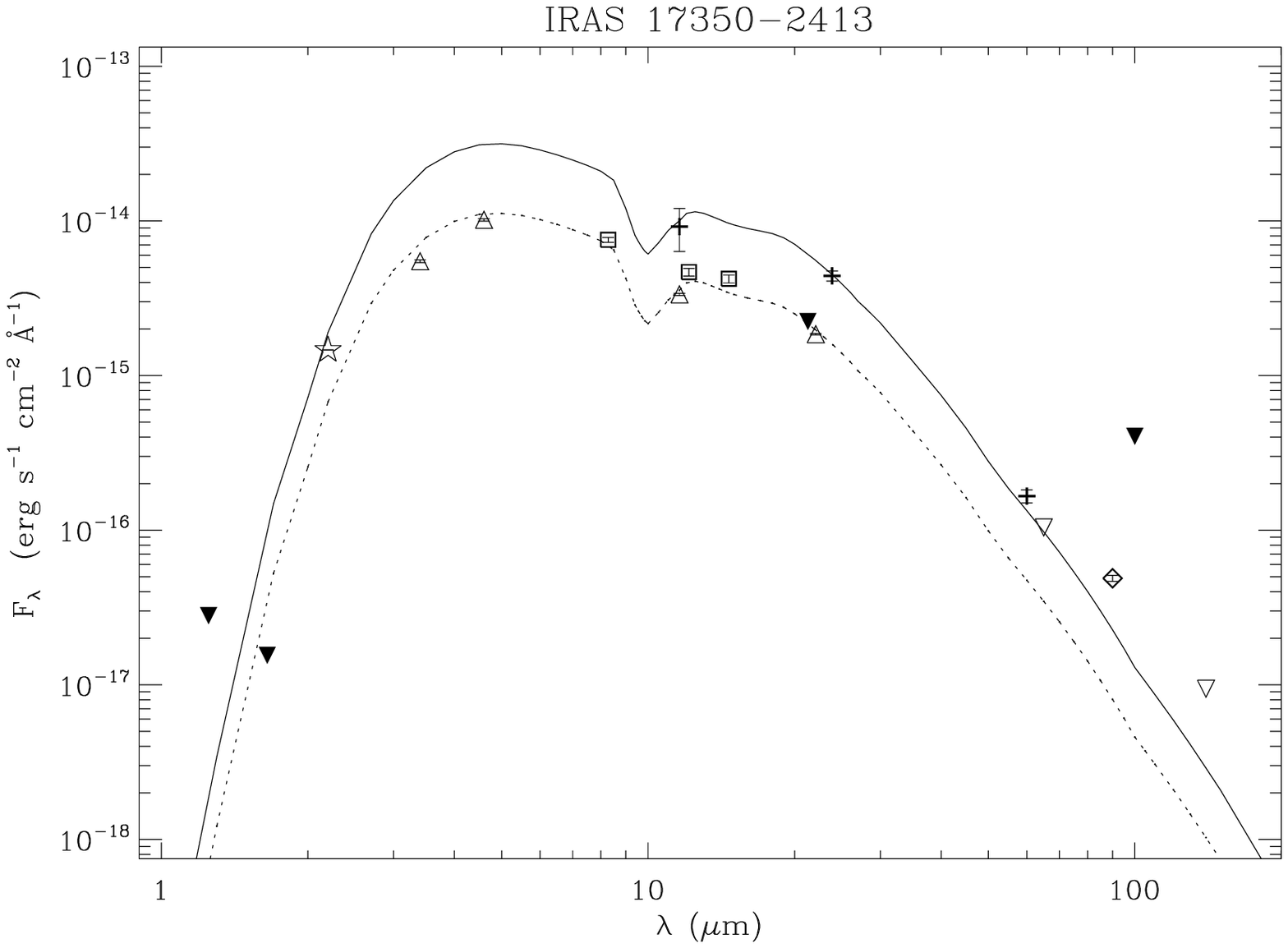}}
\resizebox{\hsize}{!}{\includegraphics{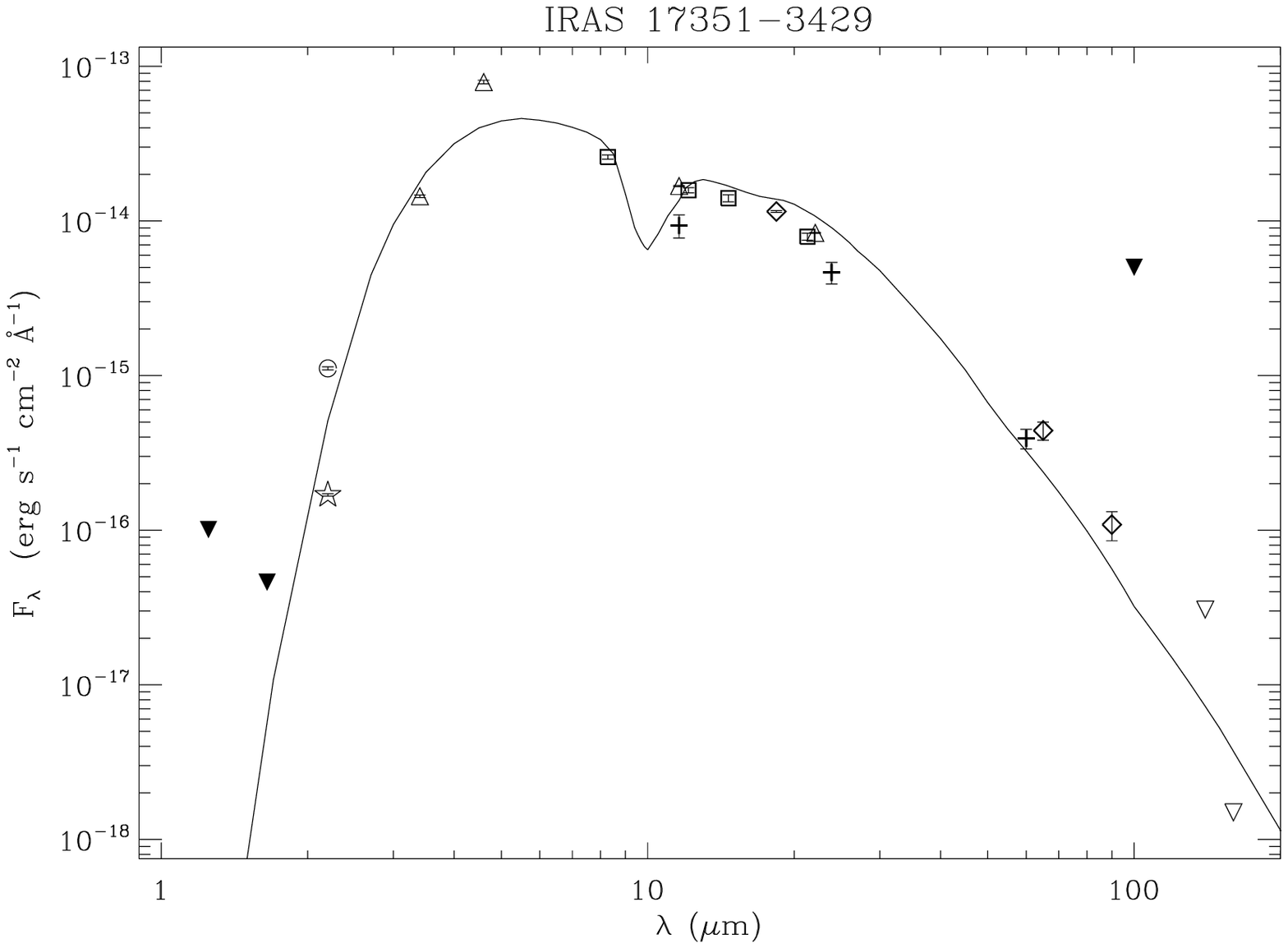}}
          \caption{continued.}
\end{figure}

\addtocounter{figure}{-1}
\begin{figure}
\resizebox{\hsize}{!}{\includegraphics{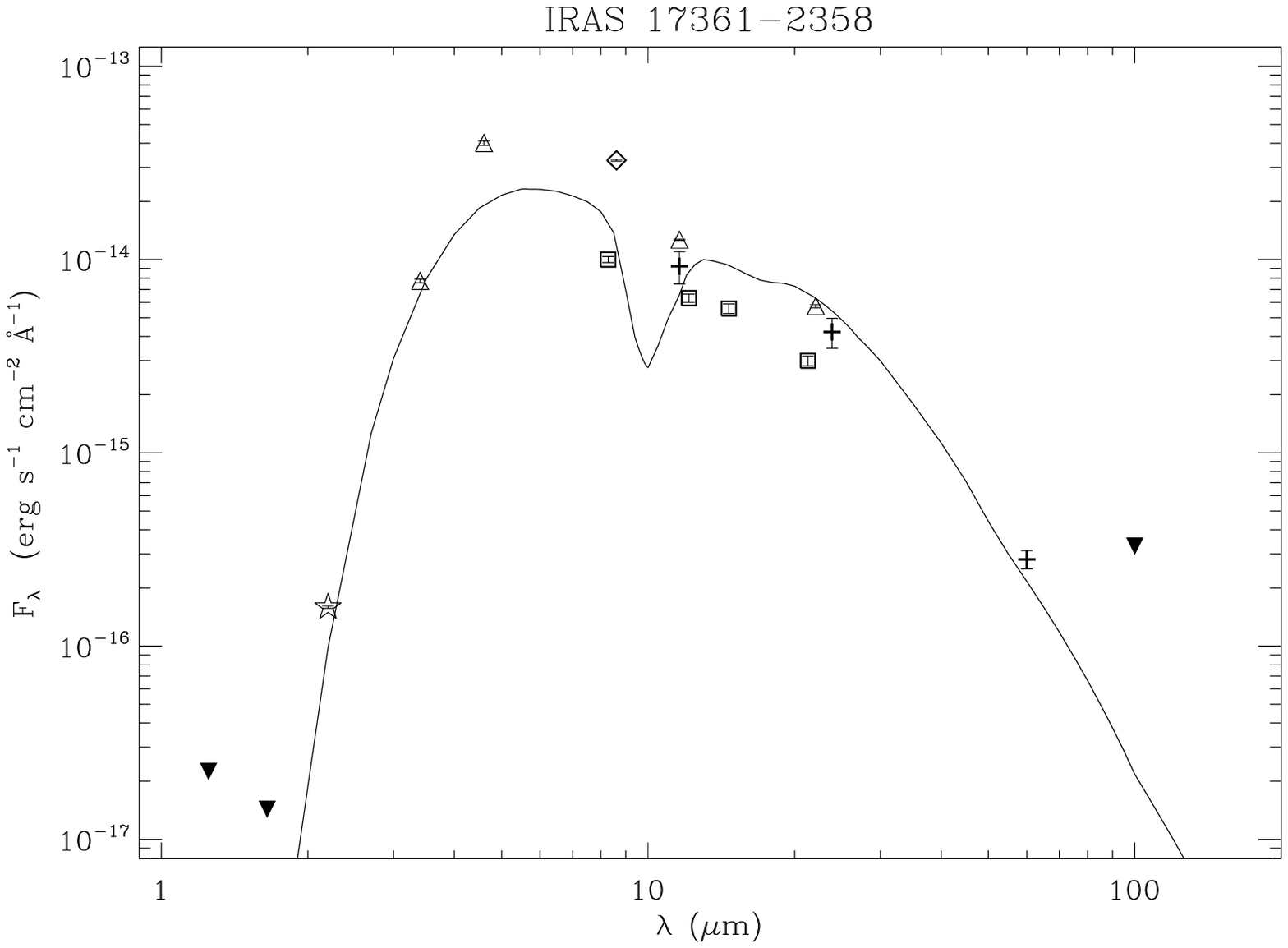}}
\resizebox{\hsize}{!}{\includegraphics{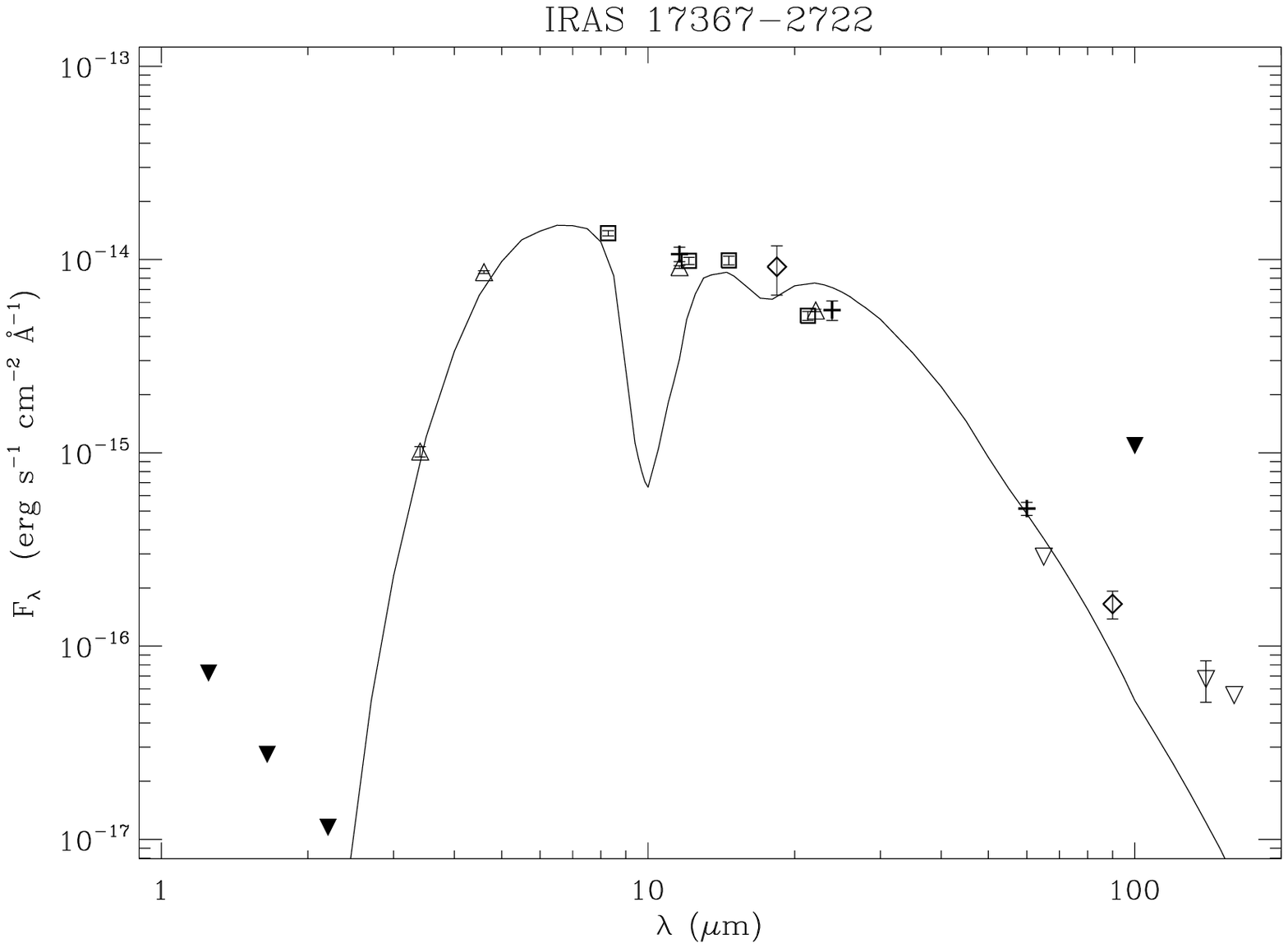}}
\resizebox{\hsize}{!}{\includegraphics{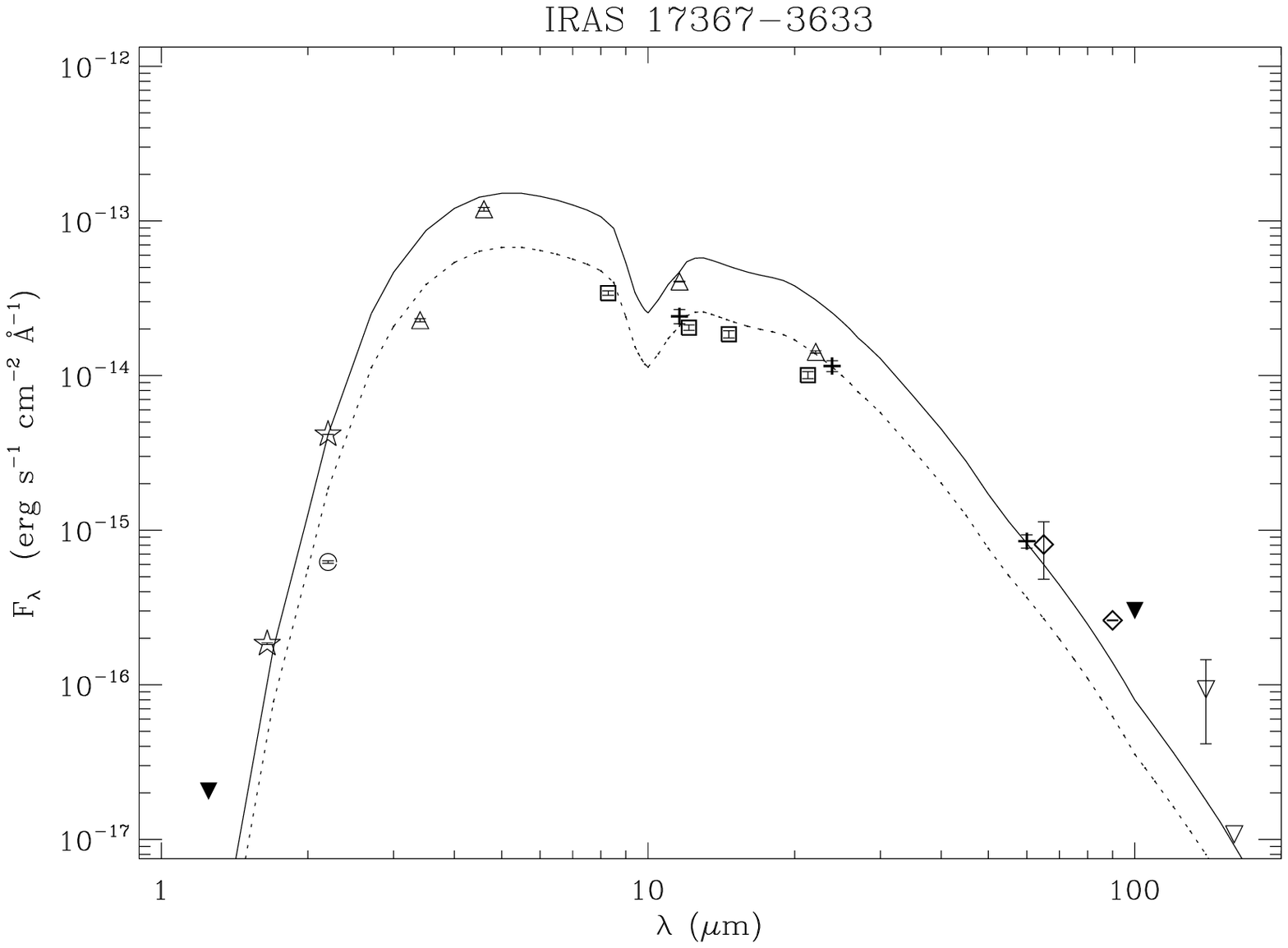}}
\resizebox{\hsize}{!}{\includegraphics{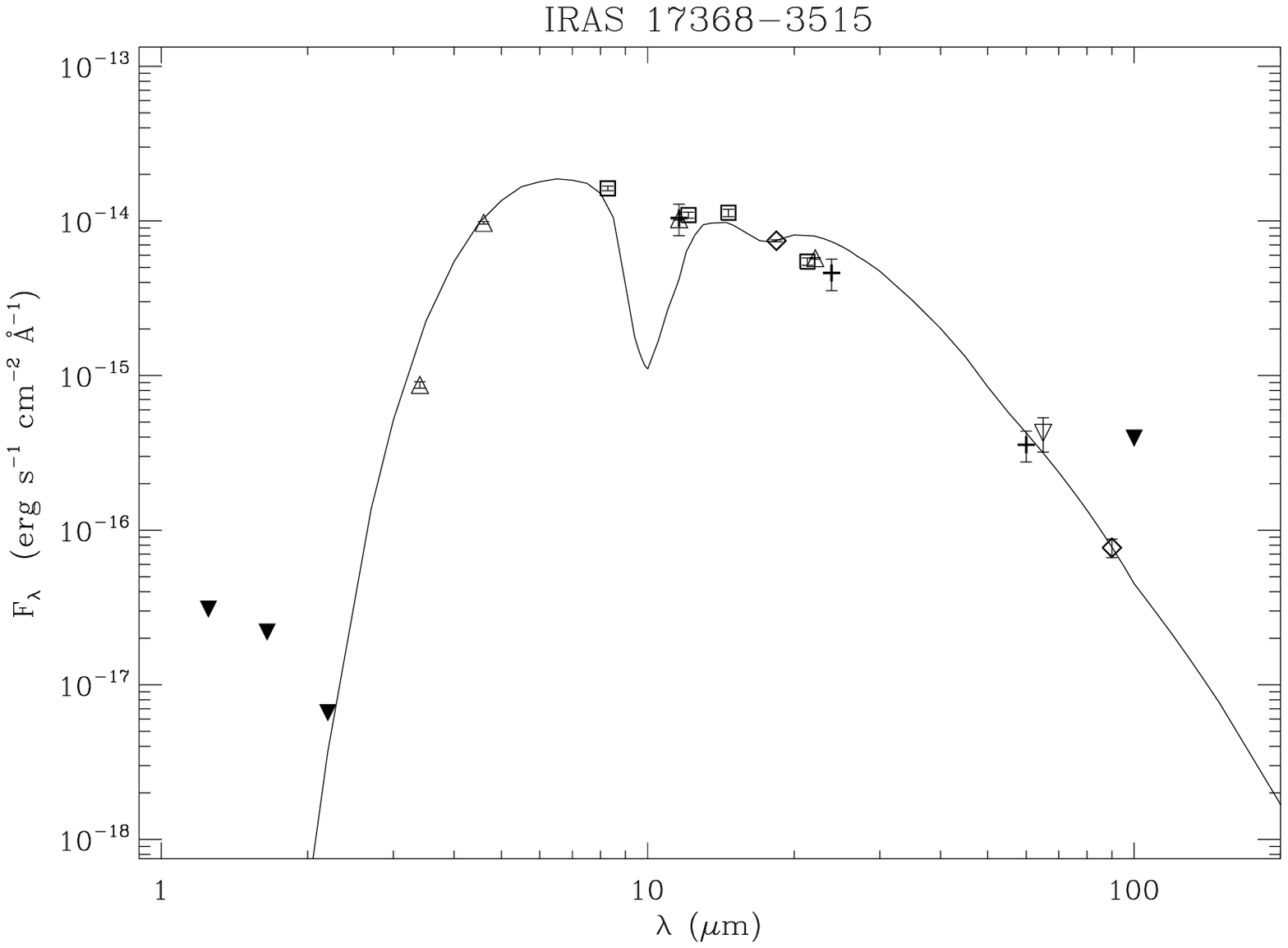}}
          \caption{continued.}
\end{figure}

\addtocounter{figure}{-1}
\begin{figure}
\resizebox{\hsize}{!}{\includegraphics{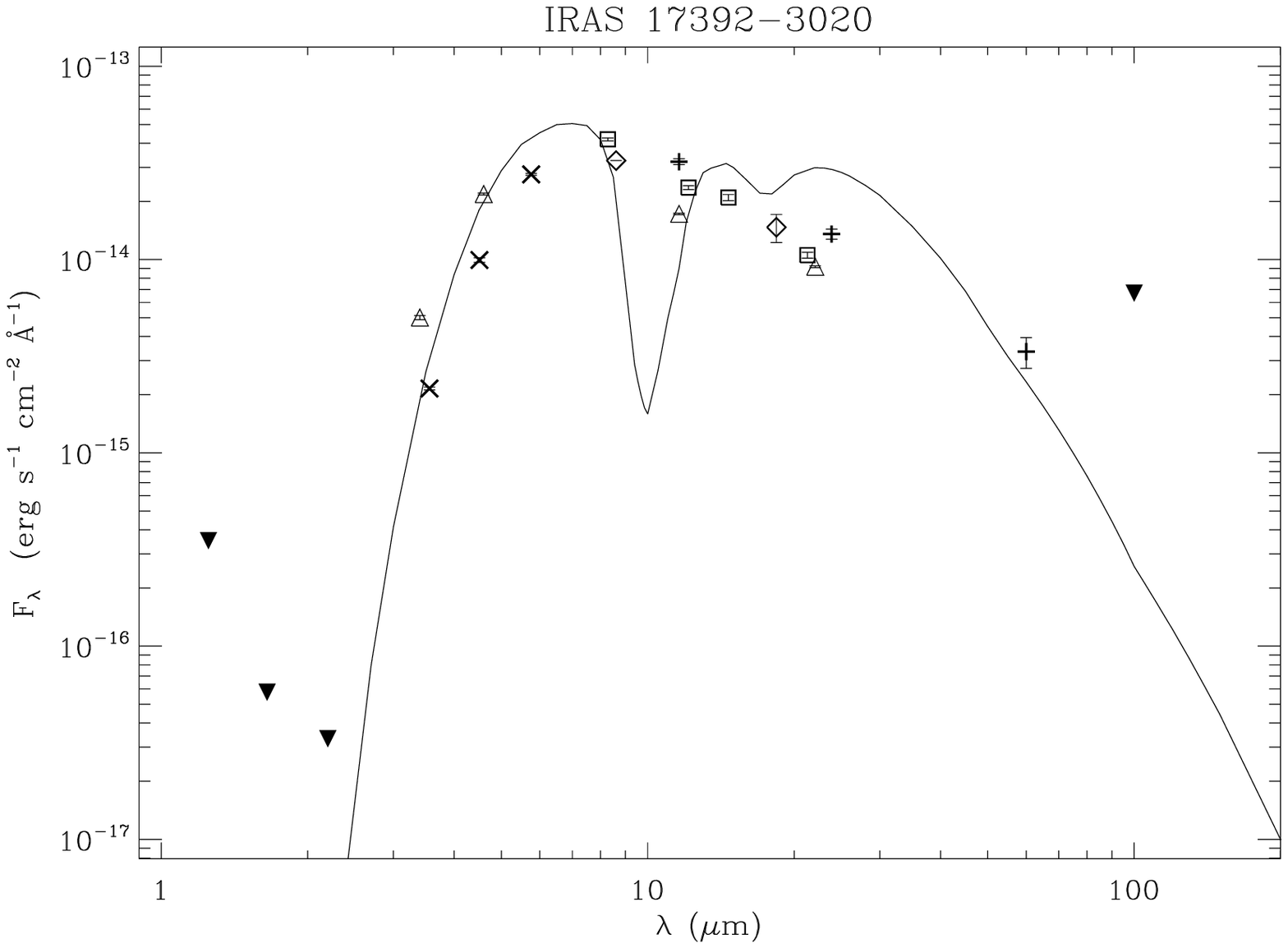}}
\resizebox{\hsize}{!}{\includegraphics{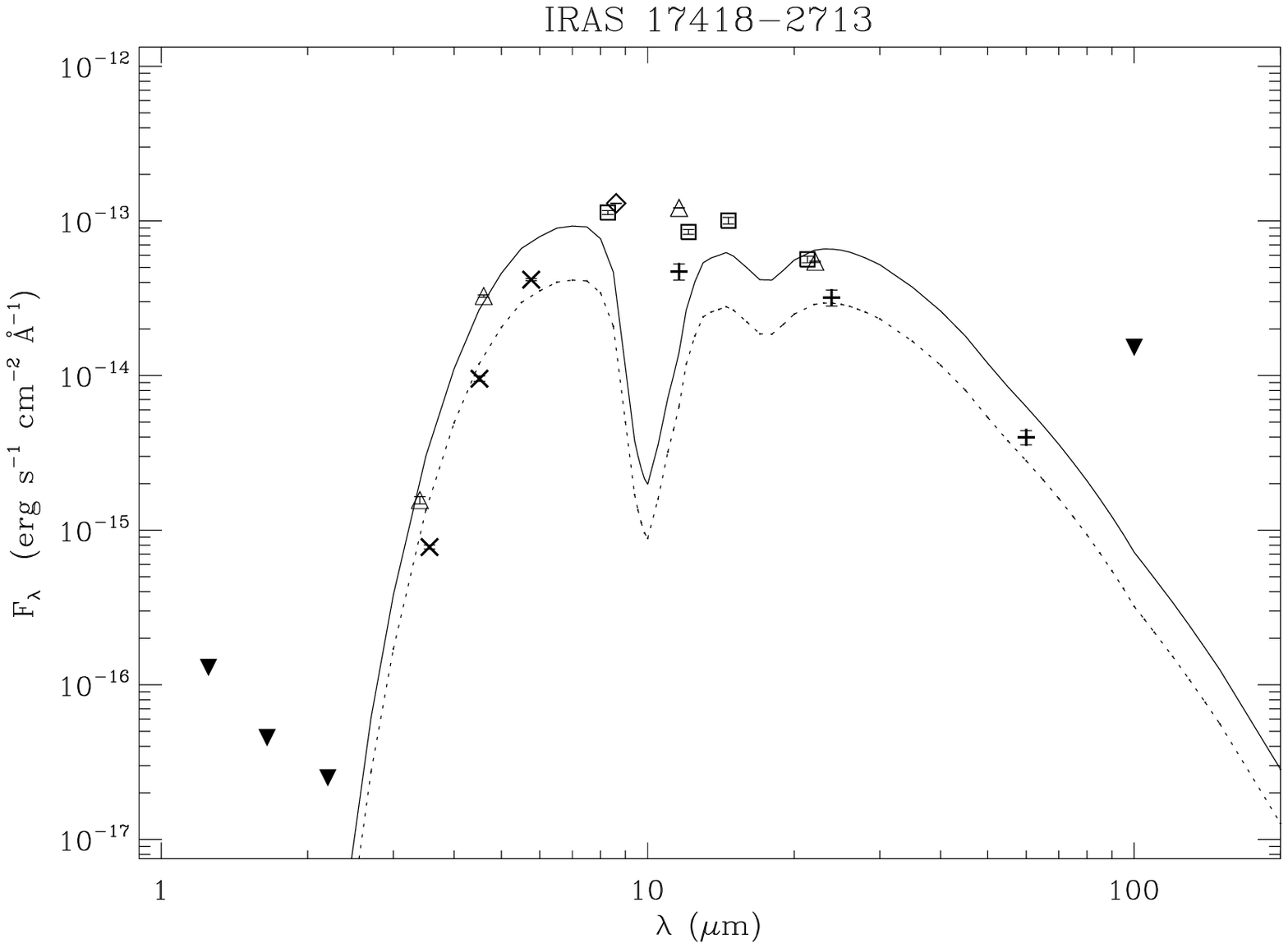}}
\resizebox{\hsize}{!}{\includegraphics{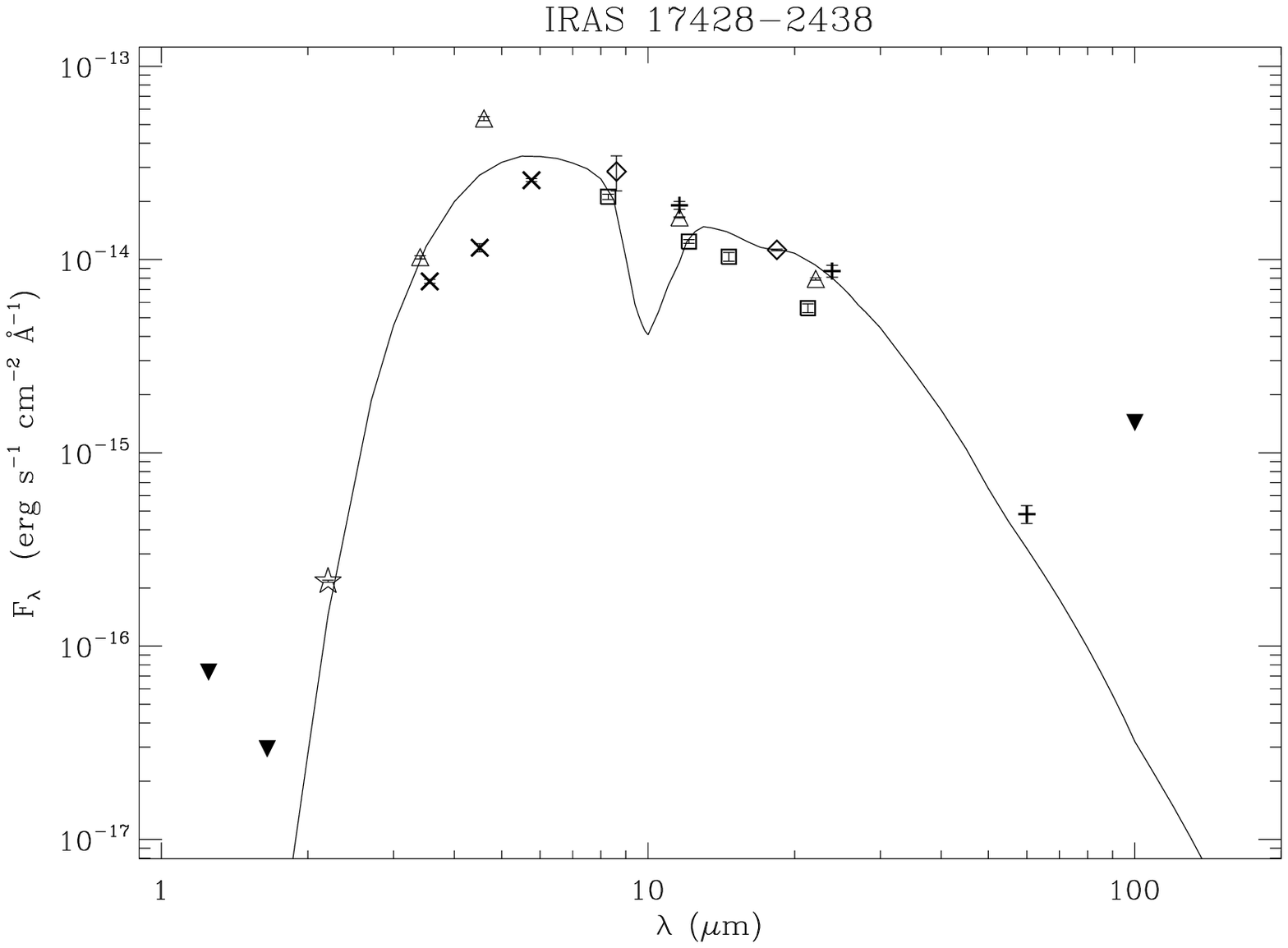}}
\resizebox{\hsize}{!}{\includegraphics{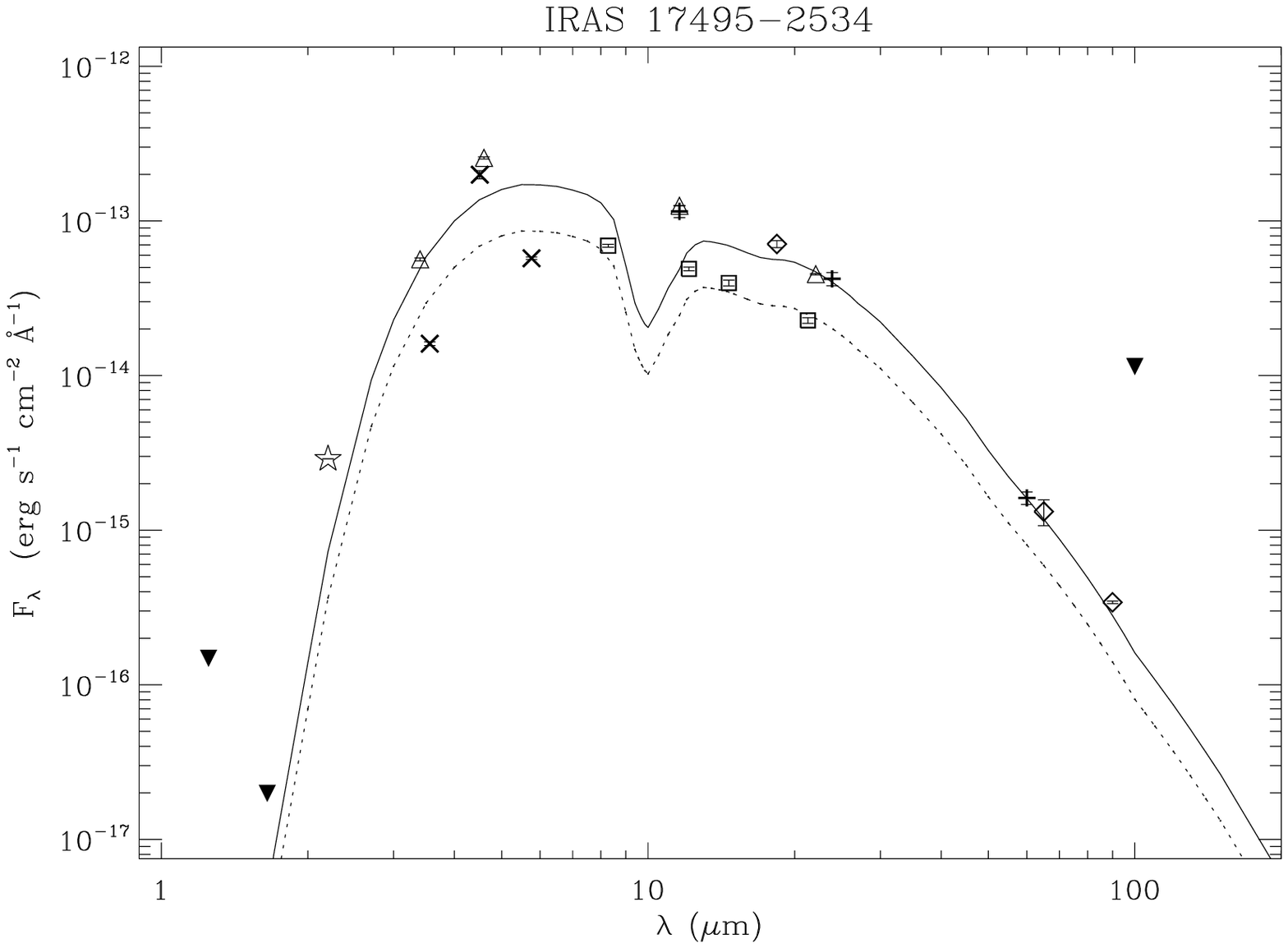}}
          \caption{continued.}
\end{figure}

\addtocounter{figure}{-1}
\begin{figure}
\resizebox{\hsize}{!}{\includegraphics{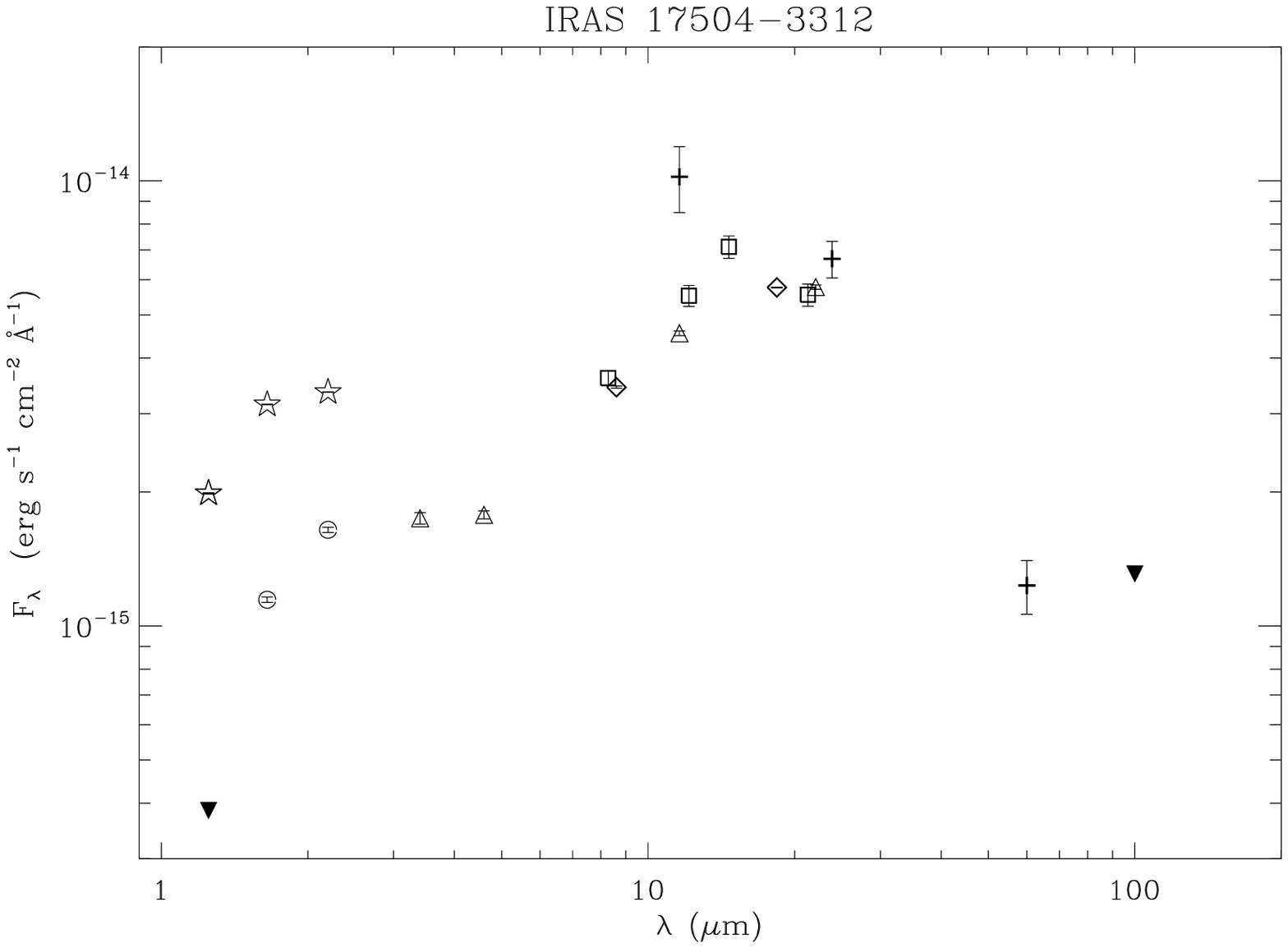}}
\resizebox{\hsize}{!}{\includegraphics{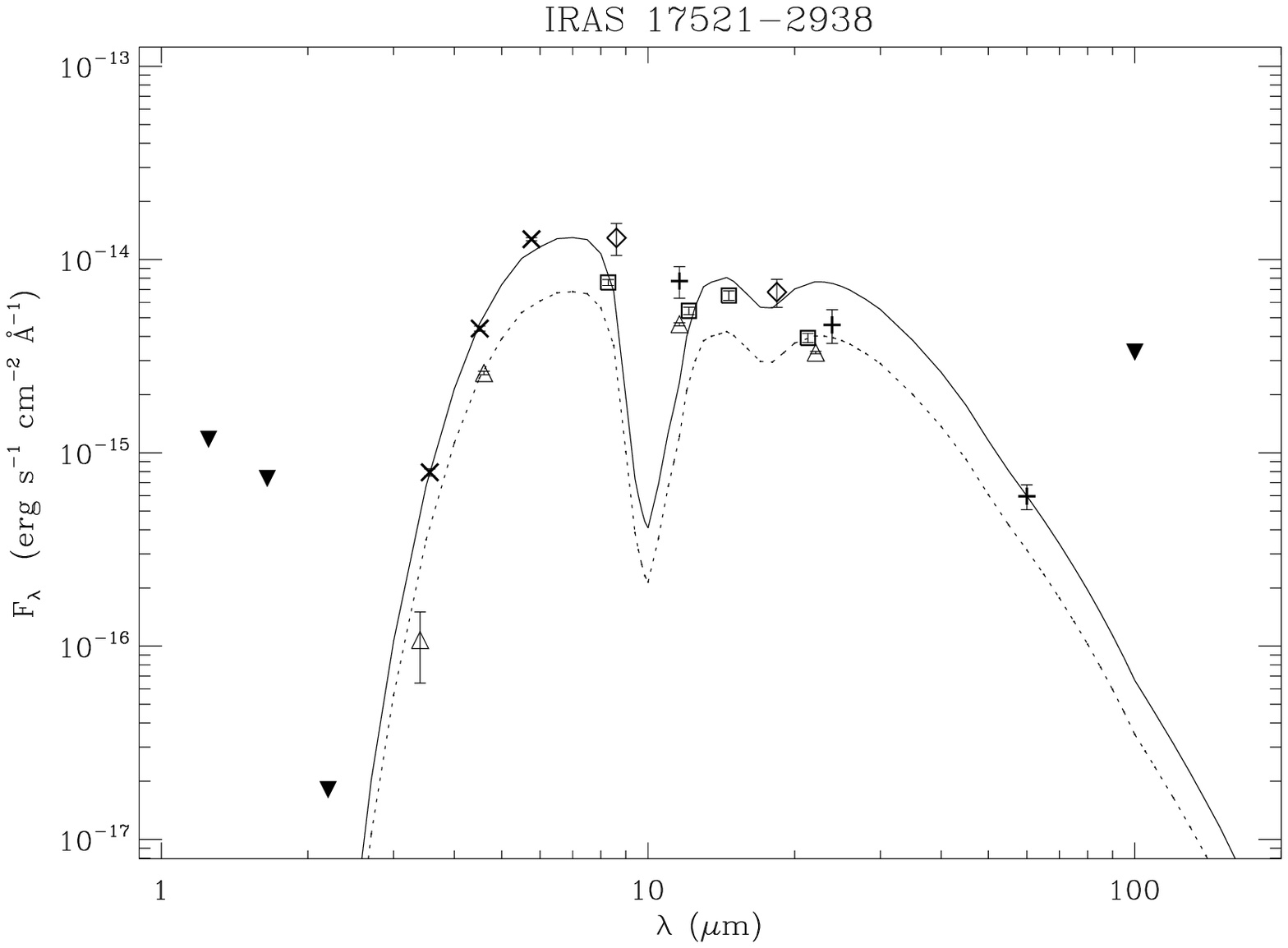}}
\resizebox{\hsize}{!}{\includegraphics{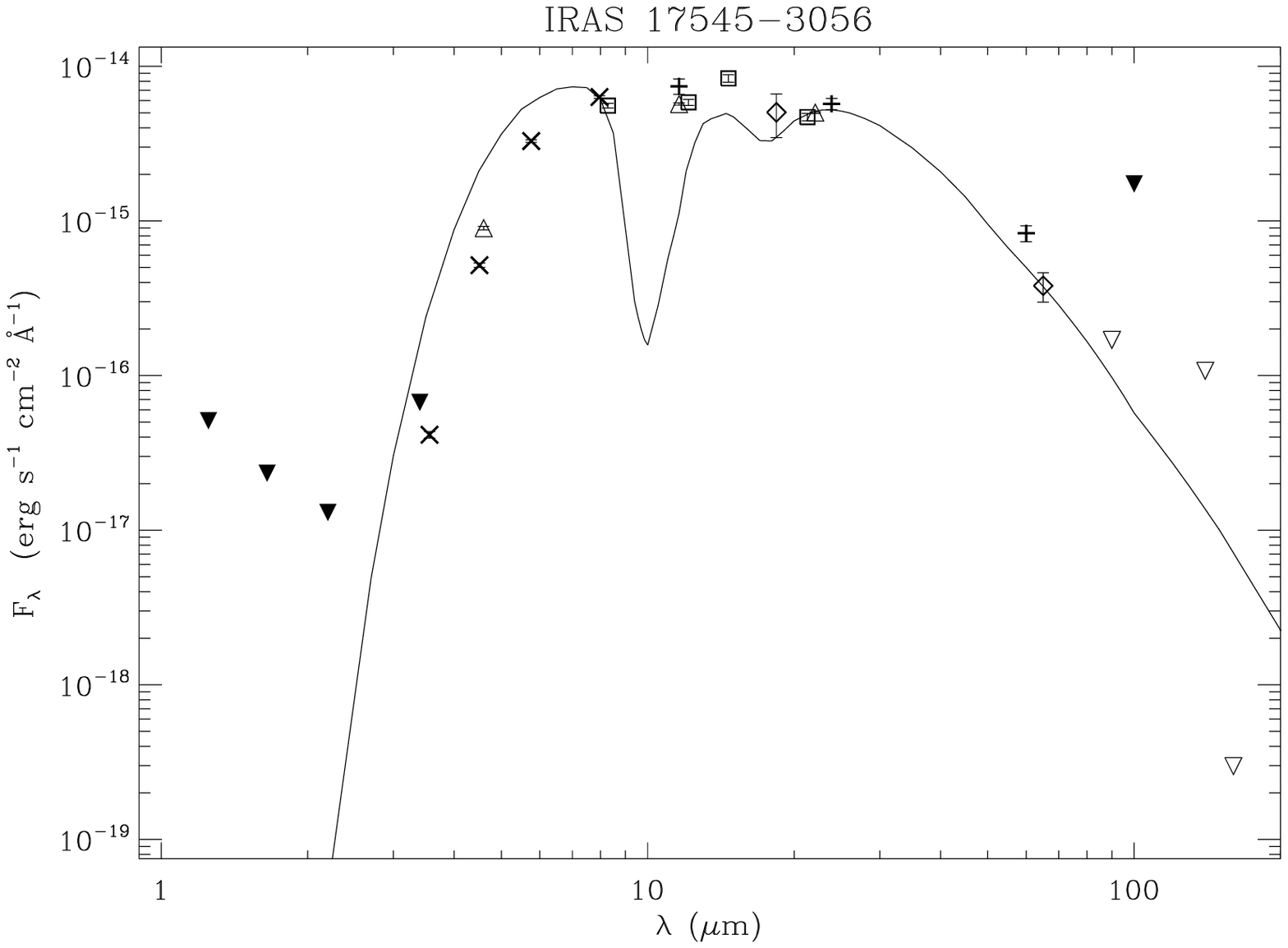}}
\resizebox{\hsize}{!}{\includegraphics{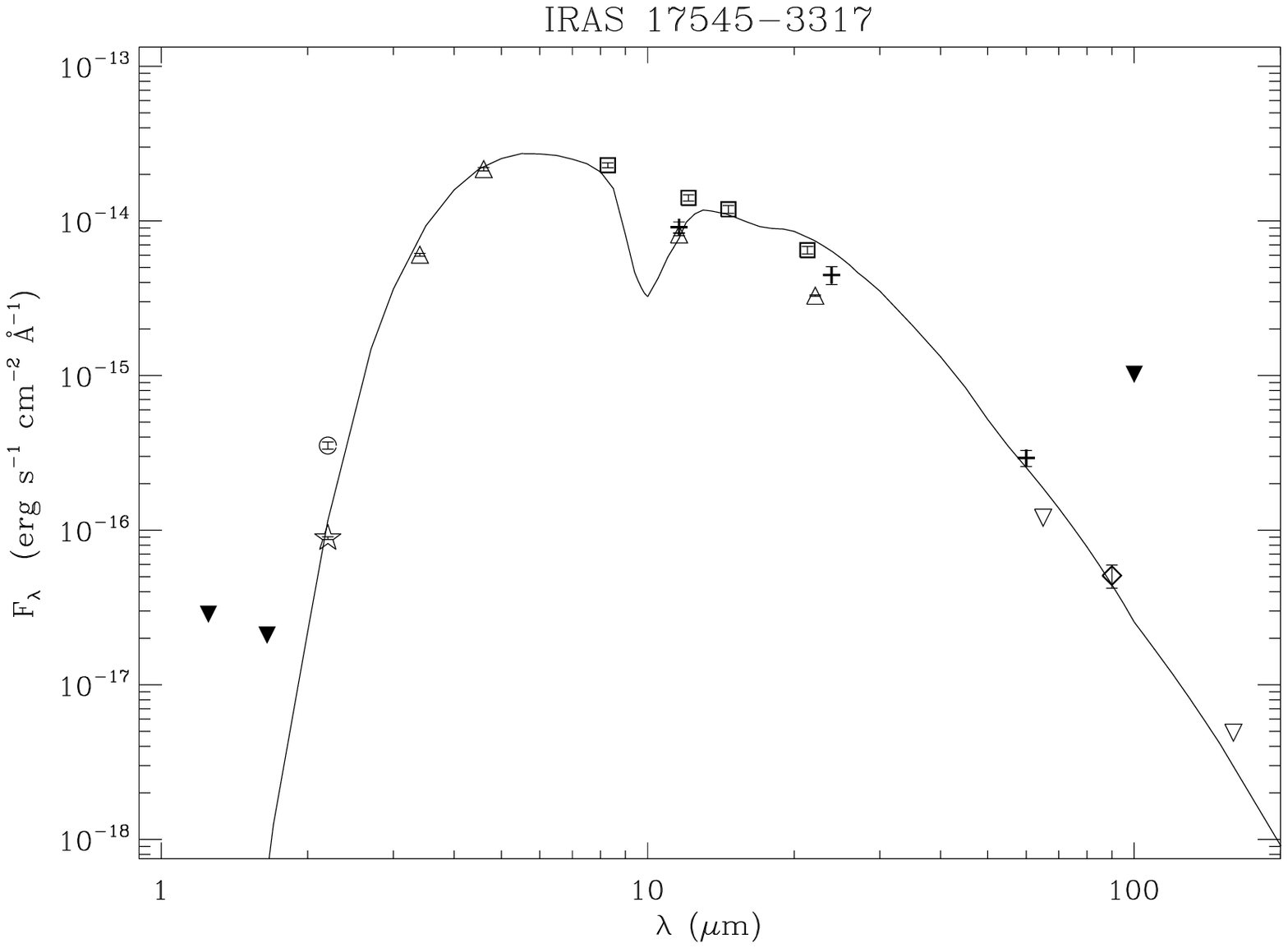}}
          \caption{continued.}
\end{figure}

\addtocounter{figure}{-1}
\begin{figure}
\resizebox{\hsize}{!}{\includegraphics{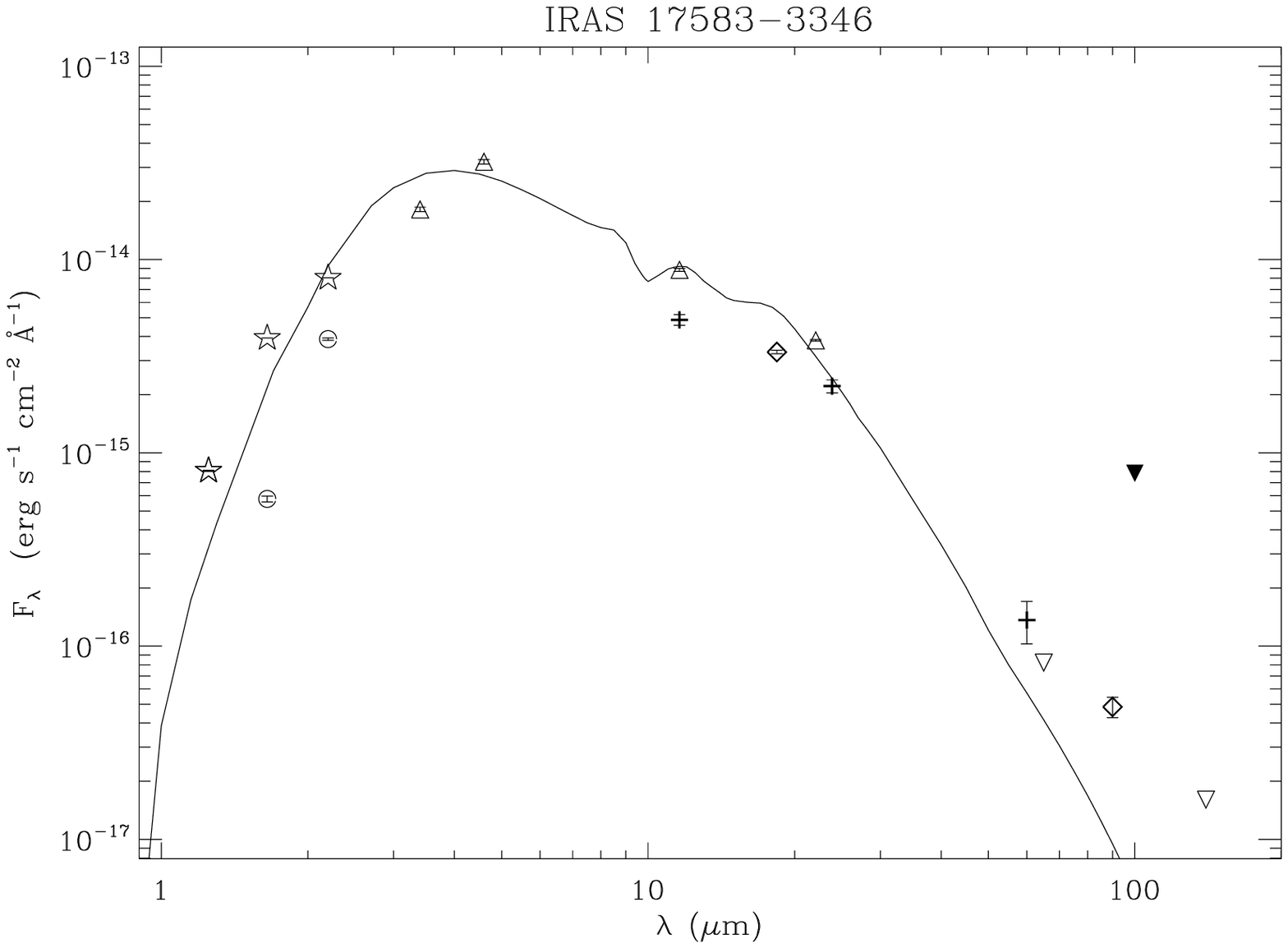}}
\resizebox{\hsize}{!}{\includegraphics{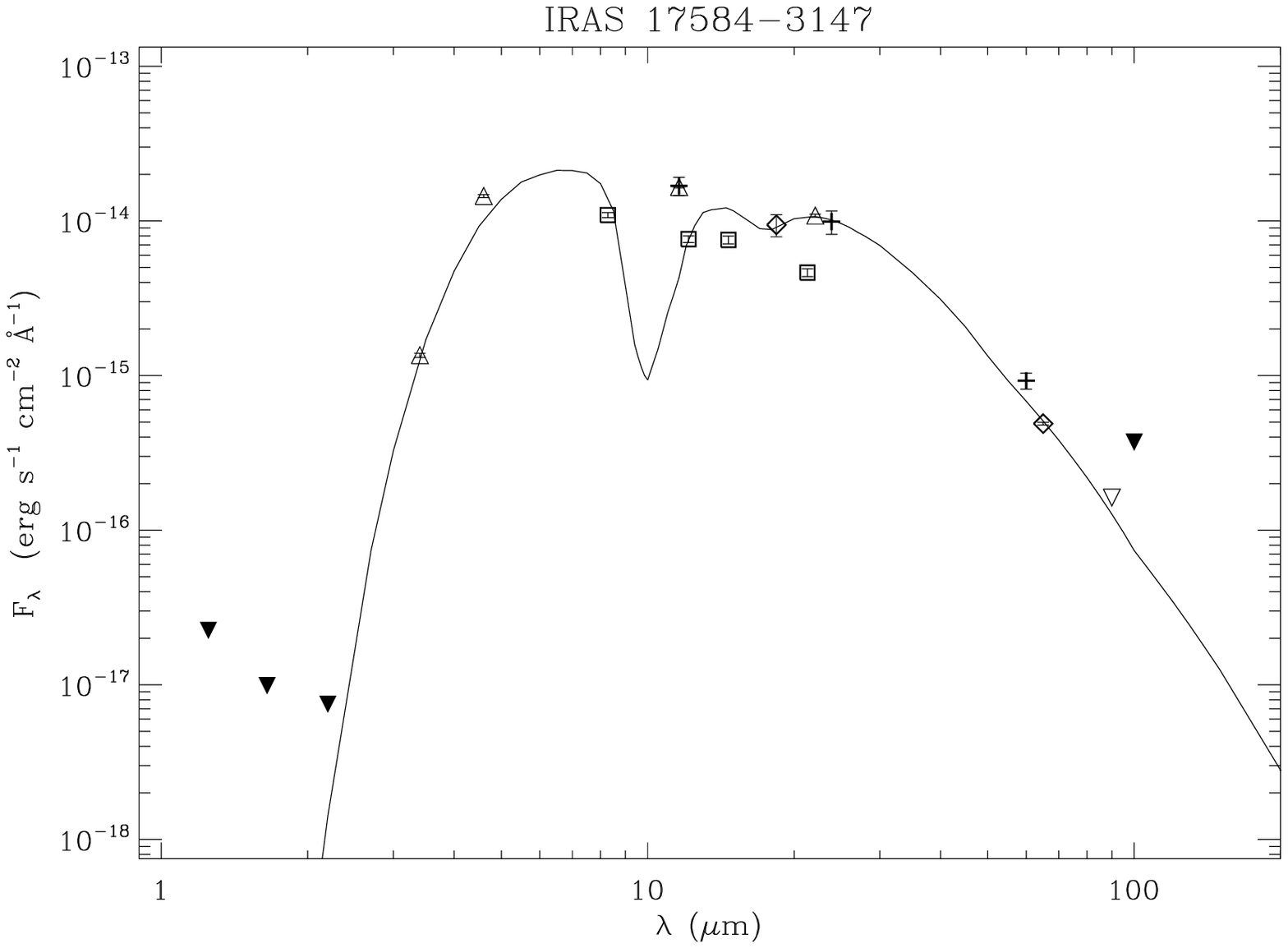}}
\resizebox{\hsize}{!}{\includegraphics{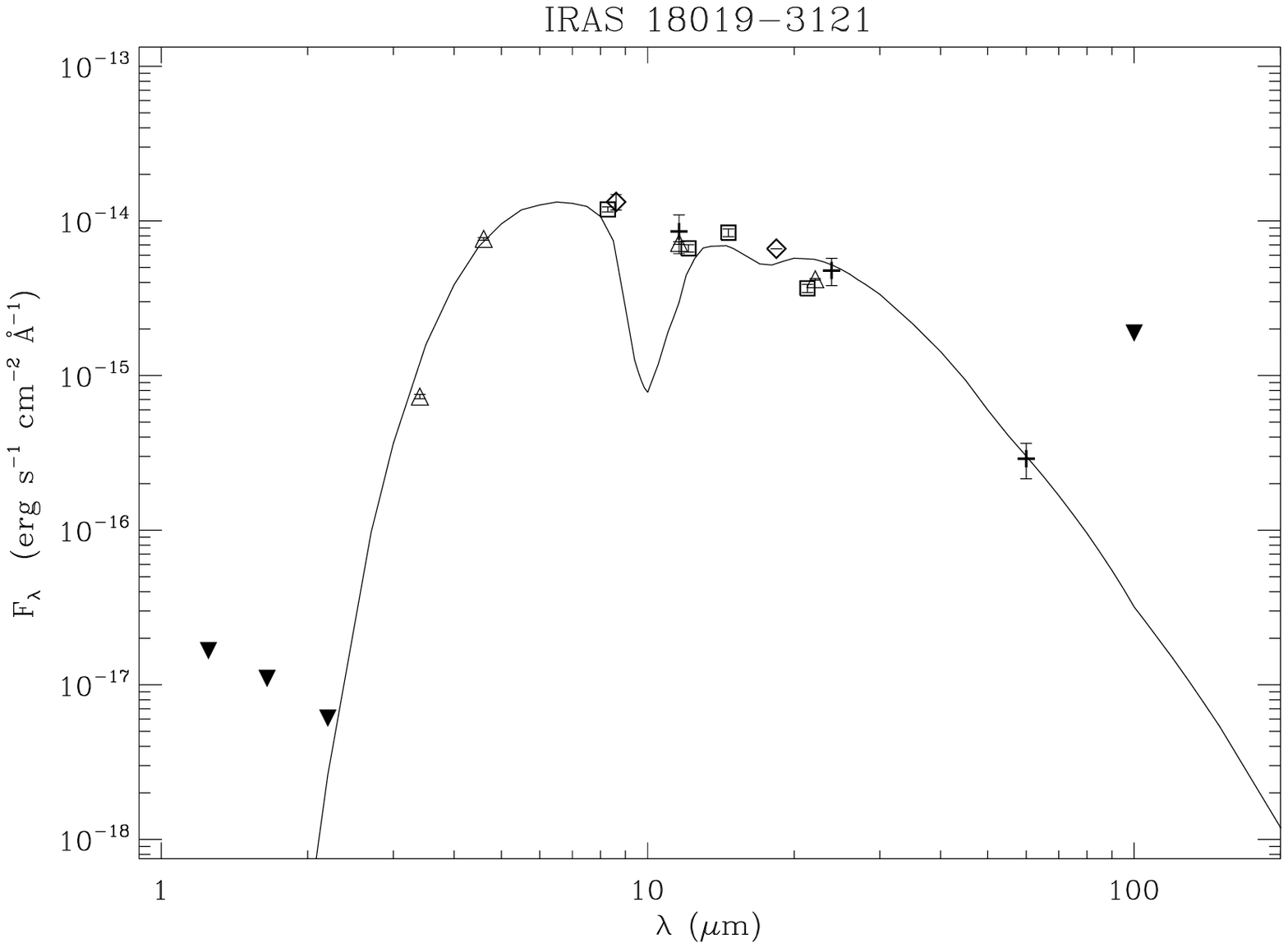}}
\resizebox{\hsize}{!}{\includegraphics{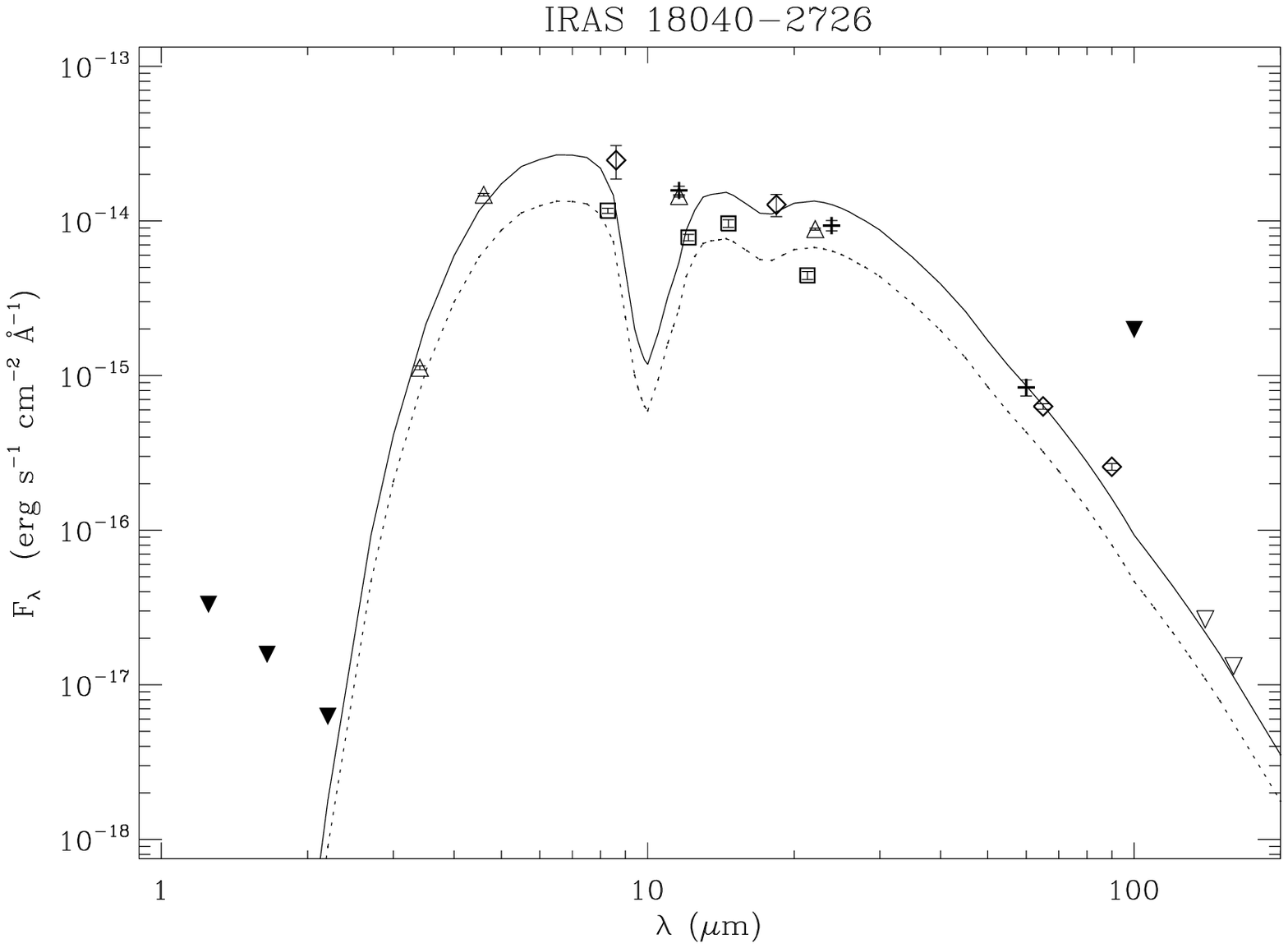}}
          \caption{continued.}
\end{figure}

\addtocounter{figure}{-1}
\begin{figure}
\resizebox{\hsize}{!}{\includegraphics{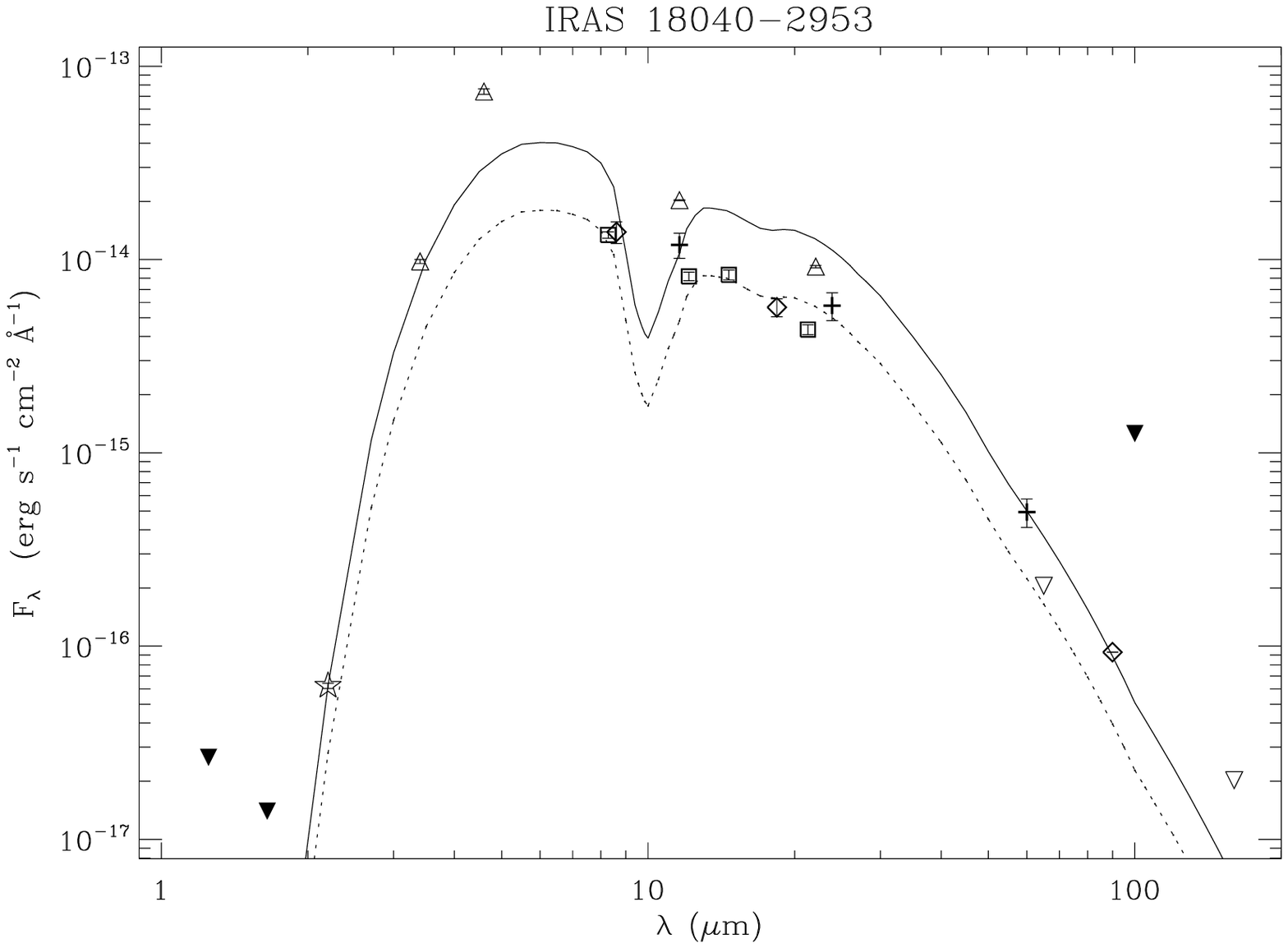}}
\resizebox{\hsize}{!}{\includegraphics{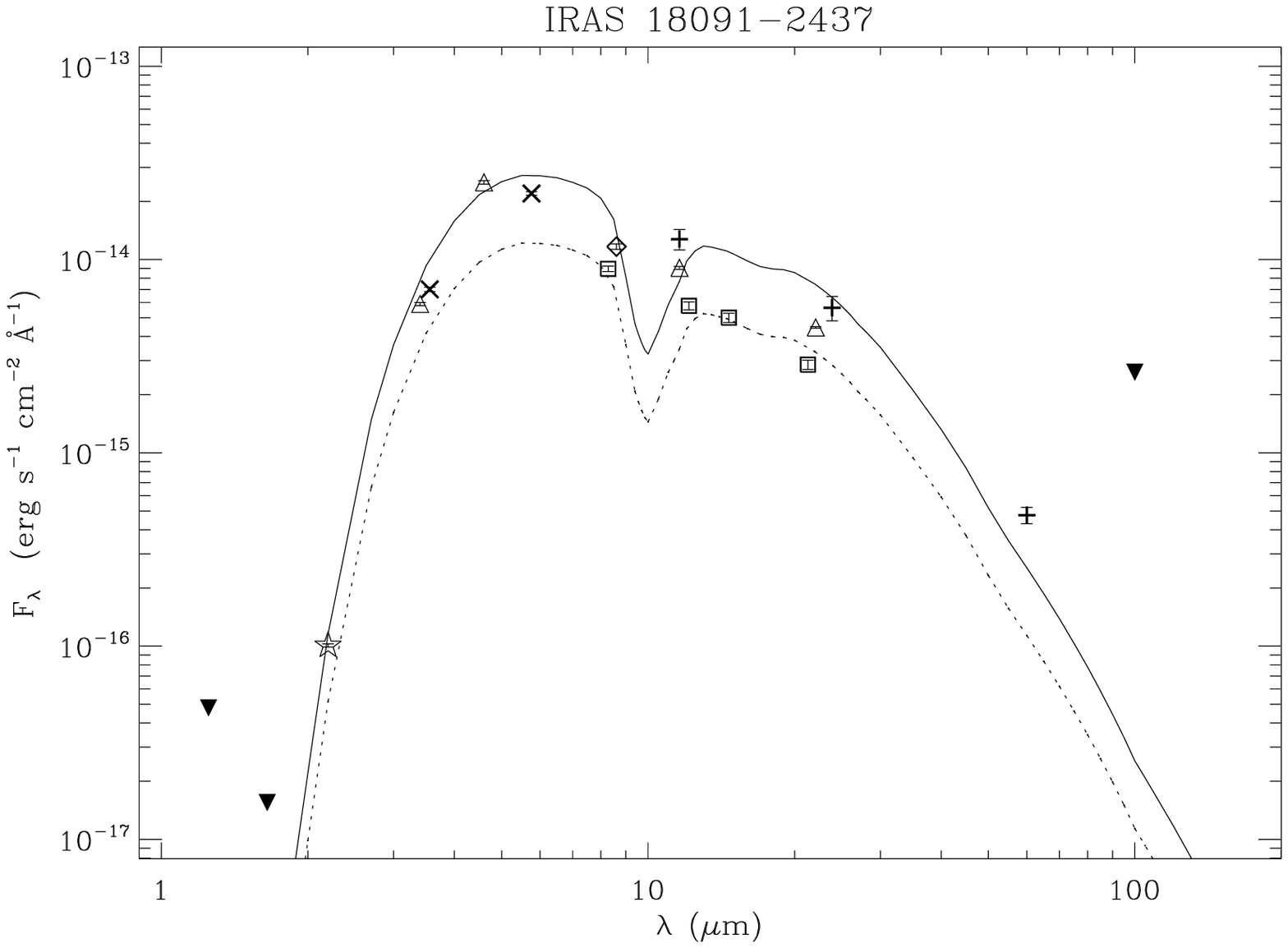}}
\resizebox{\hsize}{!}{\includegraphics{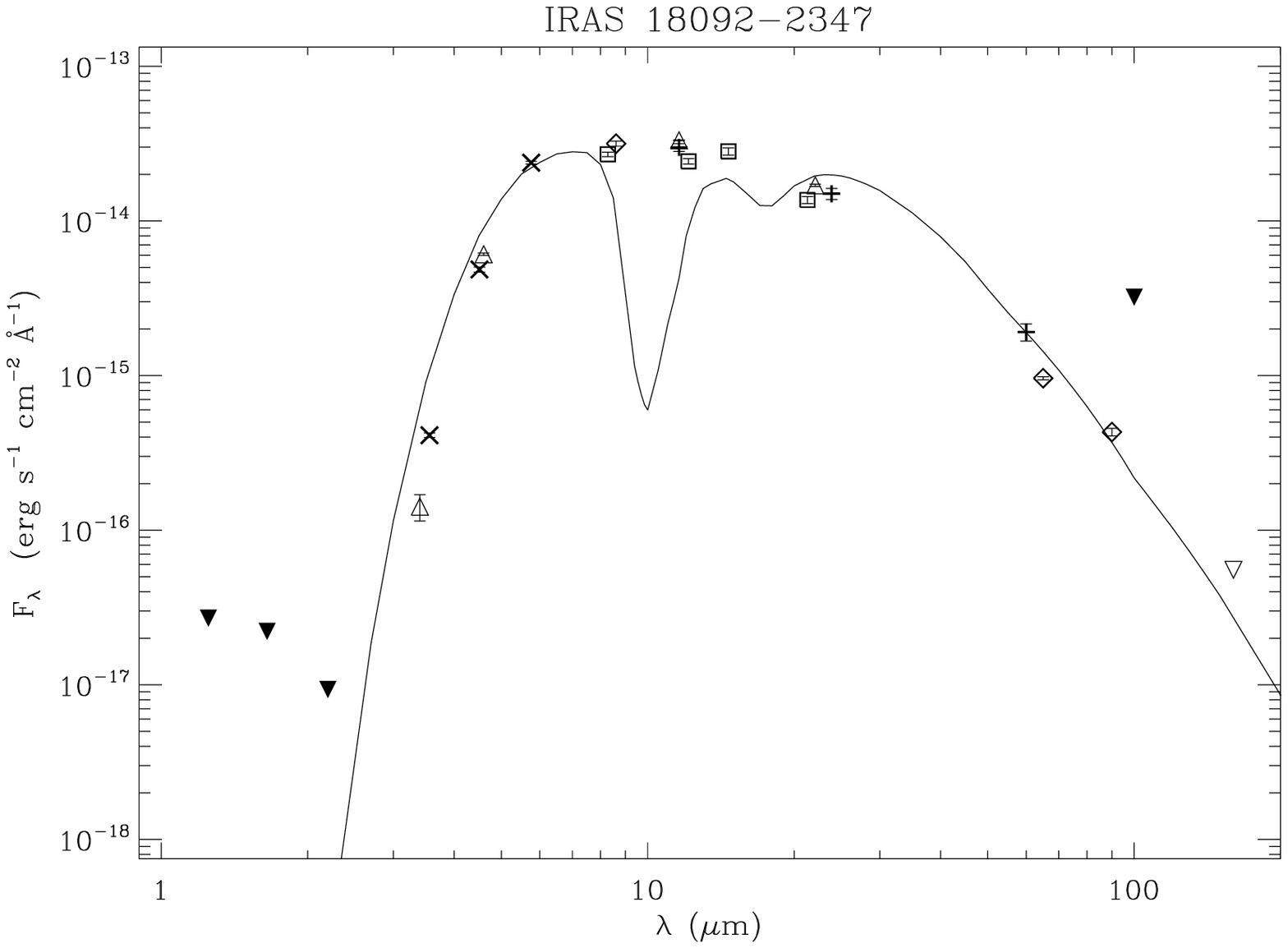}}
\resizebox{\hsize}{!}{\includegraphics{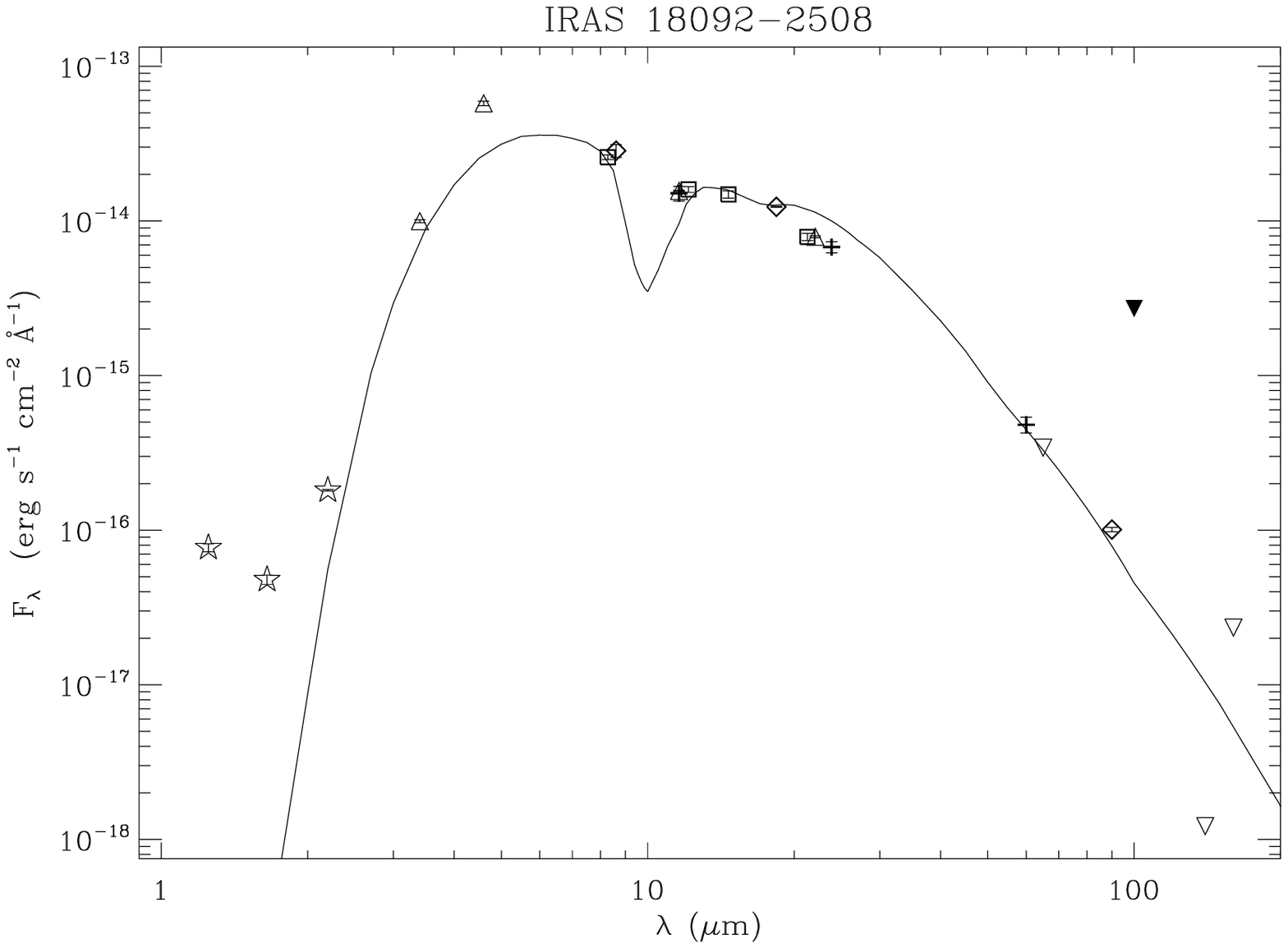}}
          \caption{continued.}
\end{figure}

\addtocounter{figure}{-1}
\begin{figure}
\resizebox{\hsize}{!}{\includegraphics{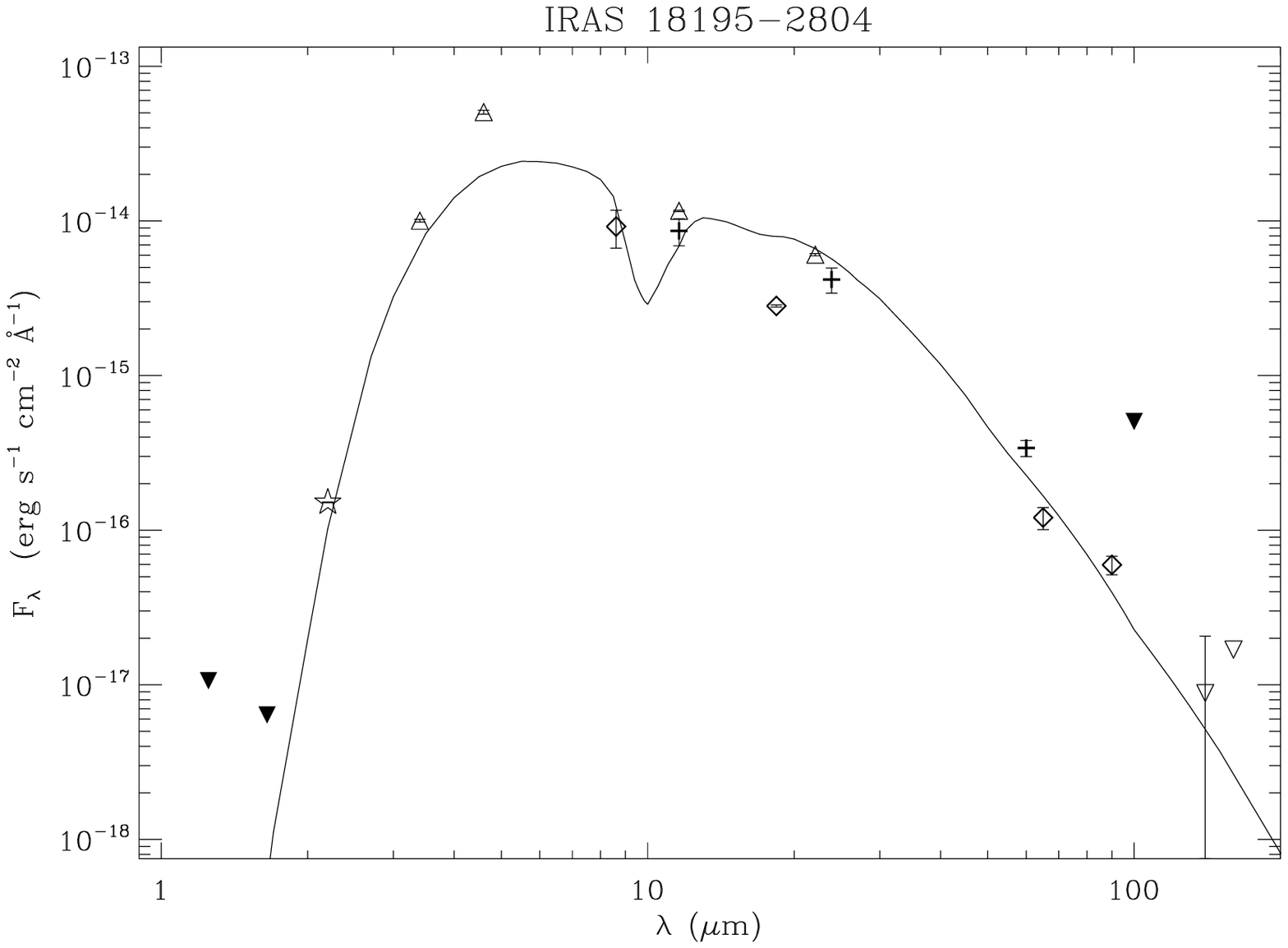}}
\resizebox{\hsize}{!}{\includegraphics{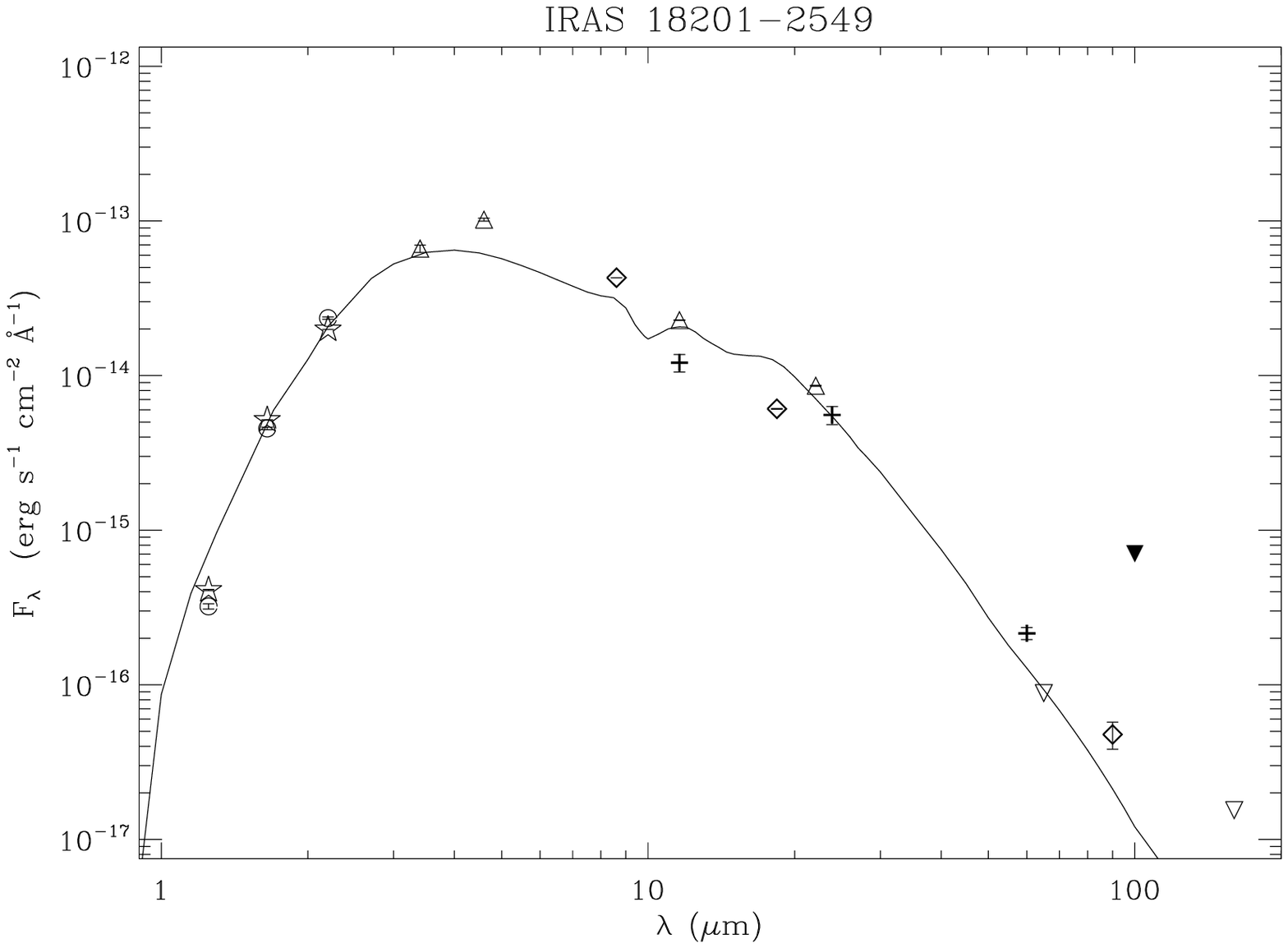}}
          \caption{continued.}
\end{figure}

\end{document}